\documentclass[astrosymb,trackchanges]{aastex7}

\usepackage{comment}
\usepackage{multirow}

\usepackage{tabularx}
\usepackage{arydshln}
\usepackage{ulem}
\usepackage[export]{adjustbox}

\newcommand\T{\rule{0pt}{2.6ex}}       
\newcommand\B{\rule[-1.2ex]{0pt}{0pt}}

\usepackage [english]{babel}
\usepackage [autostyle, english = american]{csquotes}
\MakeOuterQuote{"}

\def\simgt{\lower.5ex\hbox{$\; \buildrel > \over \sim \;$}}
\def\simlt{\lower.5ex\hbox{$\; \buildrel < \over \sim \;$}}

\newcommand\chandra{{\it Chandra}}
\newcommand\xmm{{\it XMM-Newton}}

\newcommand\nustar{{\it NuSTAR}}
\newcommand\suzaku{{\it Suzaku}}

\newcommand\xrism{{\it XRISM}}

\newcommand\src{{MAXI J1744-294}}
\newcommand\axj{{AX J1745.6-2901}}

\newcommand{\Fexxv}{Fe~{\sc xxv}}

\newcommand{\Fexxvi}{Fe~{\sc xxvi}}
\newcommand{\Caxix}{Ca~{\sc xix}}

\newcommand{\Sxvi}{S~{\sc xvi}}
\newcommand{\Sxv}{S~{\sc xv}}

\newcommand{\Nixxvii}{Ni~{\sc xxvii}}

\newcommand{\Fei}{Fe~{\sc i}}

\shorttitle{XRISM spectroscopy of the Galactic center I. The sources}
\shortauthors{Parra et al.}
\setwatermarkfontsize{170pt}
\graphicspath{{./}{figures/}}

\begin{document}

\title{XRISM spectroscopy of a crowded Galactic center region - I.\\
Disentangling the sources in the field of view}


\correspondingauthor{Maxime Parra}
\email{maxime.parrastro@gmail.com}

\correspondingauthor{Shifra Mandel}
\email{ss5018@columbia.edu}

\author[0009-0003-8610-853X]{Maxime Parra}
\affiliation{Department of Physics, Ehime University, 2-5, Bunkyocho, Matsuyama, Ehime 790-8577, Japan}
\email{maxime.parrastro@gmail.com}

\author[0009-0003-0653-2913]{Kai Matsunaga}
\affiliation{Department of Physics, Graduate School of Science, Kyoto University, Kitashirakawa Oiwake-cho, Sakyo-ku, Kyoto 606-8502, Japan}
\email{matsunaga.kai.i47@kyoto-u.jp}

\author[0000-0002-6126-7409]{Shifra Mandel}
\affiliation{Columbia Astrophysics Laboratory, Columbia University, New York, NY 10027, USA}
\email{ss5018@columbia.edu}

\author[0000-0002-9709-5389]{Kaya Mori} 
\affiliation{Columbia Astrophysics Laboratory, Columbia University, New York, NY 10027, USA}
\email{km211@columbia.edu}

\author[0000-0003-4580-4021]{Hideki Uchiyama}
\affiliation{Faculty of Education, Shizuoka University, 836 Ohya, Suruga-ku, Shizuoka, Shizuoka 422-8529, Japan}
\email{uchiyama.hideki@shizuoka.ac.jp}

\author[0000-0003-1130-5363]{Masayoshi Nobukawa}
\affiliation{Faculty of Education, Nara University of Education, Nara, 630-8502, Japan}
\email{nobukawa@cc.nara-edu.ac.jp}

\author{Tahir Yaqoob}
\affiliation{NASA/Goddard Space Flight Center, Greenbelt, MD 20771, USA}
\affiliation{Center for Research and Exploration in Space Science and Technology, NASA/GSFC (CRESST II), Greenbelt, MD 20771, USA}
\affiliation{Center for Space Science and Technology, University of Maryland, Baltimore County (UMBC), 1000 Hilltop Circle, Baltimore, MD 21250, USA}
\email{tahir@umbc.edu}

\author[0000-0001-6665-2499]{Takayuki Hayashi}
\affiliation{Department of Physics, Graduate School of Science, Kyoto University, Kitashirakawa Oiwake-cho, Sakyo-ku, Kyoto 606-8502, Japan}
\email{hayashi.takayuki.3f@kyoto-u.ac.jp}

\author[orcid=0000-0003-2161-0361]{Misaki Mizumoto}
\affiliation{Science Education Research Unit, University of Teacher Education Fukuoka, Munakata, Fukuoka 811-4192, Japan}
\email{mizumoto-m@fukuoka-edu.ac.jp}

\author[orcid=0000-0003-4808-893X]{Shinya Yamada}
\affiliation{Department of Physics, Rikkyo University, 3-34-1 Nishi Ikebukuro, Toshima-ku, Tokyo 171-8501, Japan}
\email{syamada@rikkyo.ac.jp}

\author[0000-0001-8195-6546]{Megumi Shidatsu} 
\affiliation{Department of Physics, Ehime University, 2-5, Bunkyocho, Matsuyama, Ehime 790-8577, Japan}
\email{shidatsu.megumi.wr@ehime-u.ac.jp}

\author[0000-0002-2218-2306]{Paul A. Draghis}
\affiliation{MIT Kavli Institute for Astrophysics and Space Research, Massachusetts Institute of Technology, Cambridge, MA 02139, USA}
\email{pdraghis@mit.edu}

\author[0000-0002-3252-9633]{Efrain Gatuzz}
\affiliation{Max-Planck-Institut f\"ur extraterrestrische Physik, Gie{\ss}enbachstra{\ss}e 1, 85748 Garching, Germany}
\email{egatuzz@mpe.mpg.de}

\author[0000-0001-5506-9855]{John A. Tomsick}
\affiliation{Space Sciences Laboratory, 7 Gauss Way, University of California, Berkeley, CA 94720-7450, USA}
\email{jtomsick@berkeley.edu}

\author{Charles~J.~Hailey}
\affiliation{Columbia Astrophysics Laboratory, Columbia University, New York, NY 10027, USA}
\email{chuckh@astro.columbia.edu}

\author[0000-0002-2006-1615]{Chichuan Jin} 
\affiliation{National Astronomical Observatories, Chinese Academy of Sciences, Beijing 100101, China}
\affiliation{School of Astronomy and Space Science, University of Chinese Academy of Sciences, Beijing 100049, China}
\affiliation{Institute for Frontier in Astronomy and Astrophysics, Beijing Normal University, Beijing 102206, China}
\email{ccjin@bao.ac.cn}

\author[0009-0008-1132-7494]{Benjamin Levin} 
\affiliation{Columbia Astrophysics Laboratory, Columbia University, New York, NY 10027, USA}
\email{bsl2134@columbia.edu}

\author[0000-0003-0293-3608]{Gabriele Ponti} 
\affiliation{INAF - Osservatorio Astronomico di Brera, Merate, Italy}
\affiliation{Max-Planck-Institut f\"ur extraterrestrische Physik, Gie{\ss}enbachstra{\ss}e 1, 85748 Garching, Germany}
\email{gabriele.ponti@inaf.it}

\author[0000-0003-1621-9392]{Mark Reynolds} 
\affiliation{Department of Astronomy, Ohio State University, 140 West 18th Ave., Columbus, OH 43210}
\affiliation{Department of Astronomy, University of Michigan, 1085 S. University Ave., Ann Arbor, MI 48109}
\email{reynolds.1362@osu.edu}

\begin{abstract}
The Galactic center is a complex and crowded region hosting the supermassive black hole Sgr A*, numerous accreting compact objects, and diffuse X-ray emission. This paper presents the first in a series of studies analyzing the \xrism\ observation of the X-ray transient \src/Swift J174540.2$-$290037
, located $\sim18$\arcsec\ from Sgr A*. The observation, conducted 
in March 2025, along with \xmm{} and \nustar{} coverage, aimed to investigate the Fe emission features of \src\ during its outburst. However, the region surrounding the source is heavily contaminated by X-ray emission from various diffuse and point sources, including strong line contributions from 
the supernova remnant Sgr A East and the Galactic center X-ray emission (GCXE).
Additionally, the nearby neutron star low-mass X-ray binary (NS-LMXB) AX J1745.6$-$2901 was also in outburst during the \xrism\ observation, further complicating the spectral analysis. This study focuses on disentangling the contributions of these overlapping sources by robustly modeling the background contamination and spatial-spectral mixing. We describe the methodologies, region selection, and data reduction techniques applied to the different instruments. Two complementary approaches -- empirical and physical modeling -- are employed to characterize diffuse emission and point-source contributions. The results provide a foundation for the detailed spectral analysis of \src, \axj, and the surrounding interstellar medium (ISM), which will be presented in subsequent papers. This study highlights the challenges and robust solutions for analyzing \xrism/Resolve data from crowded regions in conjunction with other X-ray telescope data. 

\end{abstract}

\keywords{\uat{X-ray transient sources}{1852} --- \uat{Galactic center}{565} --- \uat{Low-mass x-ray binary stars}{939} --- \uat{Stellar mass black holes}{1611} --- \uat{Neutron stars}{1108} --- \uat{Accretion}{14} --- \uat{High Energy astrophysics}{739}}

\section{Introduction} \label{sec:intro}

Since the advent of the first X-ray telescopes nearly six decades ago, the X-ray emission originating from the center of our Galaxy has been a source of fascination and mystique \citep{Giacconi1962, Bowyer1965, Kellogg1971, Koyama1989}.  The full complexity of the many different X-ray emitters in the Galactic center (GC) was revealed by the \chandra\ observatory, which identified thousands of point sources in addition to several distinct regions of diffuse emission within the GC region (\citealt{Wang2002, Muno2009, Zhu2018}).
More recently, multiple studies by \suzaku\ \citep{Koyama2007, Yamauchi2009, Uchiyama2011, Uchiyama2013, Nishiyama2013, Nobukawa2016} and \xmm\ \citep{Ponti2015b, Anastasopoulou2023, Stel2025, Anastasopoulou2025} have focused on the Galactic center diffuse emission (GCDE), particularly the Fe-K emission lines  
along the Galactic center and ridge. Although these studies often differed in their interpretation for its origin and composition, they consistently found that (1) the Fe XXV emission in the Galactic center shows higher emissivity compared to the Galactic ridge \citep{Uchiyama2011, Heard2013, Nishiyama2013, Nobukawa2016}, and (2) the neutral and lightly ionized Fe-I K$\alpha$ emission predominantly originates from X-ray reflection off of dense molecular clouds \citep{Nobukawa2016, Anastasopoulou2025, Stel2025}.

The Galactic center, in addition to hosting the supermassive black hole (SMBH) Sgr A*, is also the site of the highest concentration of accreting compact objects in our Galaxy \citep{Degenaar2012, Mori2013, Hailey2016, Hailey2018, Mori2021, Mori2022, Mandel2025a}.  In particular, a large number of low-mass X-ray binaries (LMXBs) have been identified through transient outbursts in the central $\sim50$ pc \citep{Muno2005, Degenaar2012, Degenaar2015, Mori2019, Mori2021}.  A new bright outbursting black hole (BH) candidate, initially named MAXI J1744-294\footnote{ To avoid confusion with the many other Swift transients detected in the Galactic center region, we will adopt the "new" name of MAXI J1744-294 in all of our studies.} and later confirmed to be a repeat outburst of Swift J174540.2-290037 (T37) \citep{Kudo2025, Mandel2026_AtelT37}, was detected in early 2025, only $\sim18\arcsec$ from Sgr A*. This discovery prompted a large multi-wavelength campaign to determine the properties and follow the evolution of the source \citep{Mandel2026_ApJ}. As part of that campaign, a \xrism\ observation of the transient was performed on March 3, 2025 \citep{Mandel2025c}, along with simultaneous  \xmm{} and \nustar{} coverage, with the primary goal of studying \src{}'s Fe emission features. 

\begin{figure*}[t!]
    \includegraphics[clip,trim=0cm 0.4cm 0cm 1.2cm,width=1.0\textwidth]{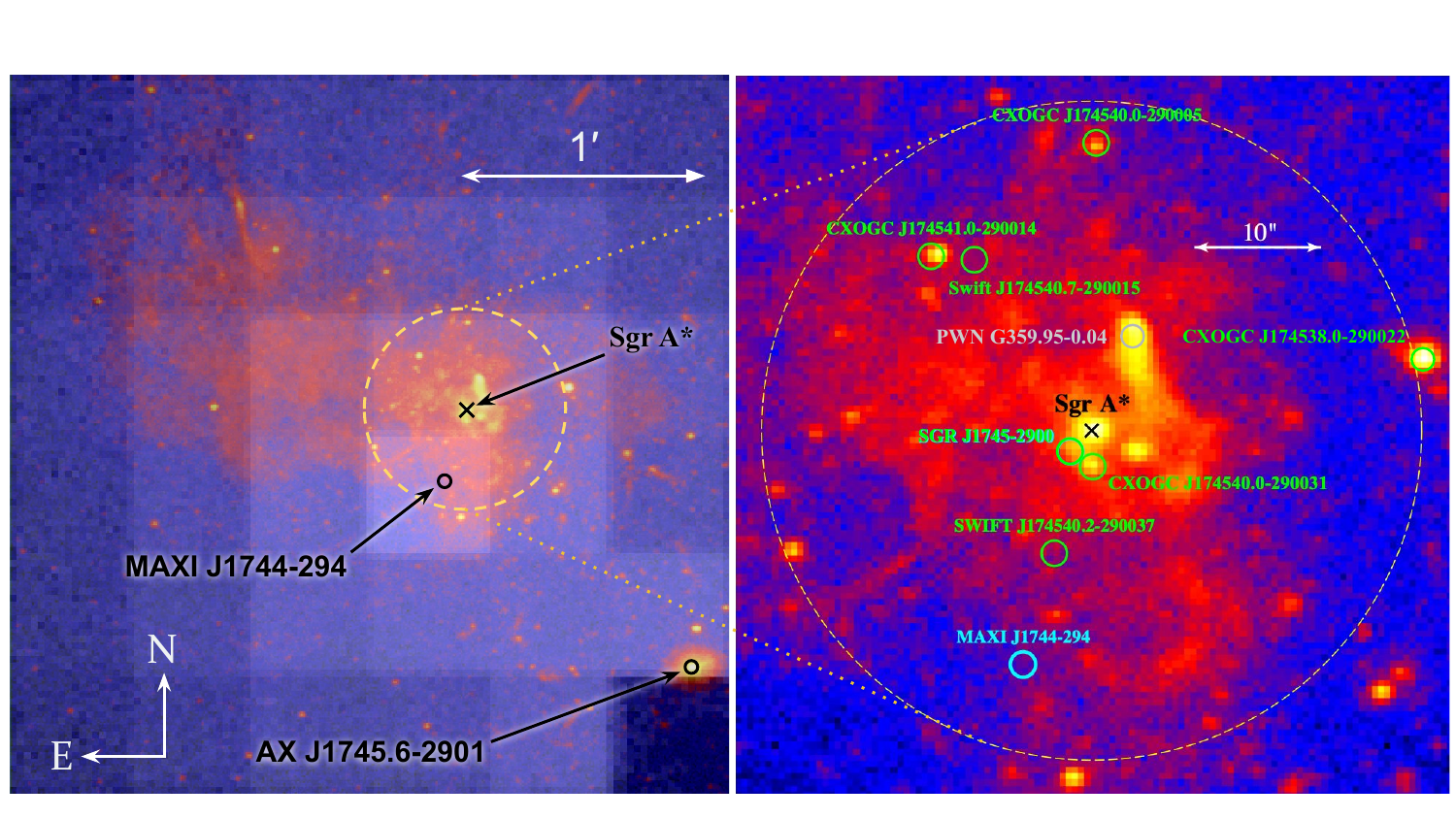}
    \caption{\textbf{(Left)} Comparison between a stacked [0.7-7] keV \chandra{} image of the galactic center and the \xrism{} Resolve 6x6 pixel field of view for the 2025 Director's Discretionary Time observation of \src{}. We highlight the position of Sgr A*, \src{} and \axj{}. \textbf{(Right)} Zoom of the central region of the \chandra{} image, highlighting the position of known X-ray transients in green, \src{} in cyan, and Sgr A* in black.}
    \label{fig:Resolve_Chandra_blend_DDT}
\end{figure*}

We showcase in Fig.~\ref{fig:Resolve_Chandra_blend_DDT} a comparison of the 6x6 pixel grid of \xrism{}'s microcalorimeter Resolve and a stacked \chandra{} observation of the Galactic center, along with a zoom of the region closest to Sgr A*. As can be seen in Fig.~\ref{fig:Resolve_Chandra_blend_DDT}, \src\ is located in a region that is home to copious X-ray emission from an array of both diffuse and point sources, including -- but not limited to -- flares from the SMBH Sgr A* \citep{Ponti2015, Ponti2017_Sgr_Flare, Haggard2019, Mossoux2020}; the supernova remnant (SNR) Sgr A East \citep{Nynka2013, Mori2024, XrismCol2025_GC_obs_diffuse}; the pulsar wind nebula (PWN) G359.95-0.04 \citep{Wang2006}; Fe K$\alpha$ fluorescence from the dense molecular clouds that form the circumnuclear disk (CND; \citealt{Stel2023}); and an untold number of faint unresolved point sources that contribute to the central hard X-ray emission (CHXE; \citealt{Hailey2016, Xu2019}).  In addition, the nearby dipping and eclipsing neutron star LMXB \axj \citep{Ponti2015c, Jin2018}, which is known to exhibit variable Fe emission and absorption features, was also in outburst at the time of the \xrism\ observation of \src. Given the various sources of strong, highly ionized X-ray lines in the FoV around the BH candidate and the poor half-power diameter of \xrism{}, robust modeling of the complex contamination they produce is a necessary precursor to any reliable study of the spectral properties of each source, and of the composition of the hot interstellar medium (ISM) along the line of sight. 
Fortunately, the vast majority of the FoV is covered by a previous \xrism\ observation of the Galactic center, performed in March 2024 during the Performance Verification (PV) phase of the instrument. From this observation, the underlying diffuse emission has been studied in detail in \cite{XrismCol2025_GC_obs_diffuse}, and the contribution of AX J1745.6-2901 -- which was 
also in outburst at the time -- in \cite{Tanaka2026_AXJ}, providing a useful template for the underlying background around \src.  

In this paper, we will describe how we utilize the previous observations to model the contributions from various sources of contamination and untangle those components from the \src\ emission. To this end, we detail the methodologies, data reduction, and background modeling applied to both \xrism\ observations, enabling the study of several science cases of interest in the \xrism\ observation of March 3, 2025. 
The spectral analysis of the \xrism\ spectrum of \src\ and its physical modeling are detailed in Parra et al. (submitted to ApJ), hereafter Paper II. The detailed study of the ISM features apparent in the observations is presented in Gatuzz et al. (submitted to A\&A, Paper III). 
In these three works, we use empirical models for the NS AX J1745.6-2901, in order to remove its contribution from that of \src. An in-depth analysis of the NS spectra from the same set of observations, its broad-band properties, and the high-resolution view of its wind features, along with photoionization modeling, will be presented in forthcoming papers (Matsunaga et al., in prep.). Finally, we refer to \cite{Mandel2026_ApJ} (hereafter M26) for a comprehensive study of the outburst evolution of MAXI J1744-294 beyond the March 3 observations.

We first describe the observations, region selection, and general approach to distinguishing the different sources in Section~\ref{sec:reg_choice}. Those choices condition our data reduction methodologies, themselves described in Section~\ref{sec:datared}. We provide details of the diffuse emission modeling from previous \xrism\ and \xmm\ observations in Section~\ref{sec:diffuse_mod}. We summarize our findings in Section \ref{sec:conclu}. Appendix~\ref{app:obs_compl} provides additional details on the observations, data reduction, and systematics. The full description of the best fit models applied to the diffuse emission is shown in Appendix~\ref{app:tables}.

\begin{figure*}[t!]
    \includegraphics[clip,trim=4.5cm 2cm 5.4cm 1.5cm,width=0.5\textwidth]{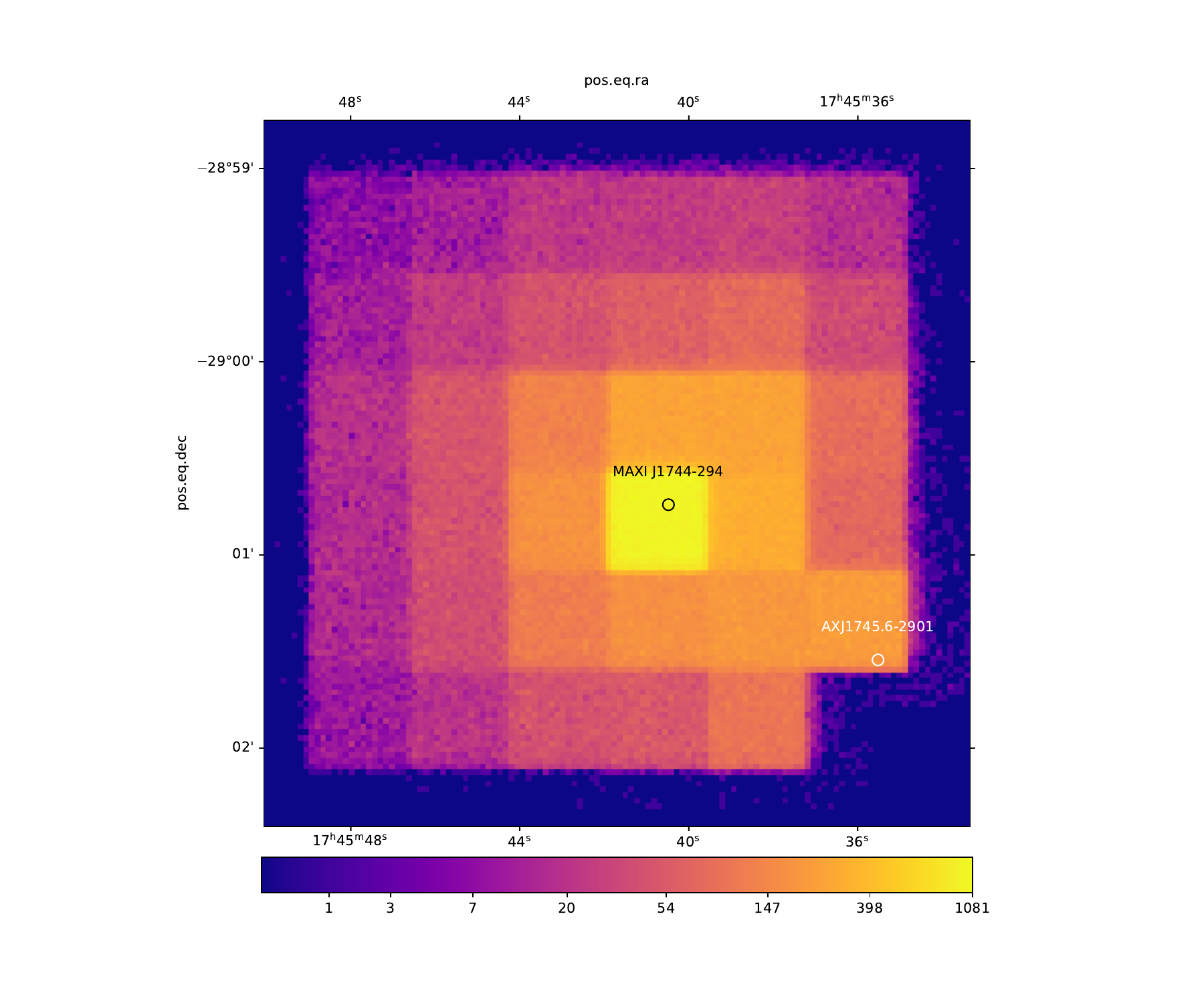}
    \includegraphics[clip,trim=4.5cm 2cm 5.3cm 1.5cm,width=0.5\textwidth]{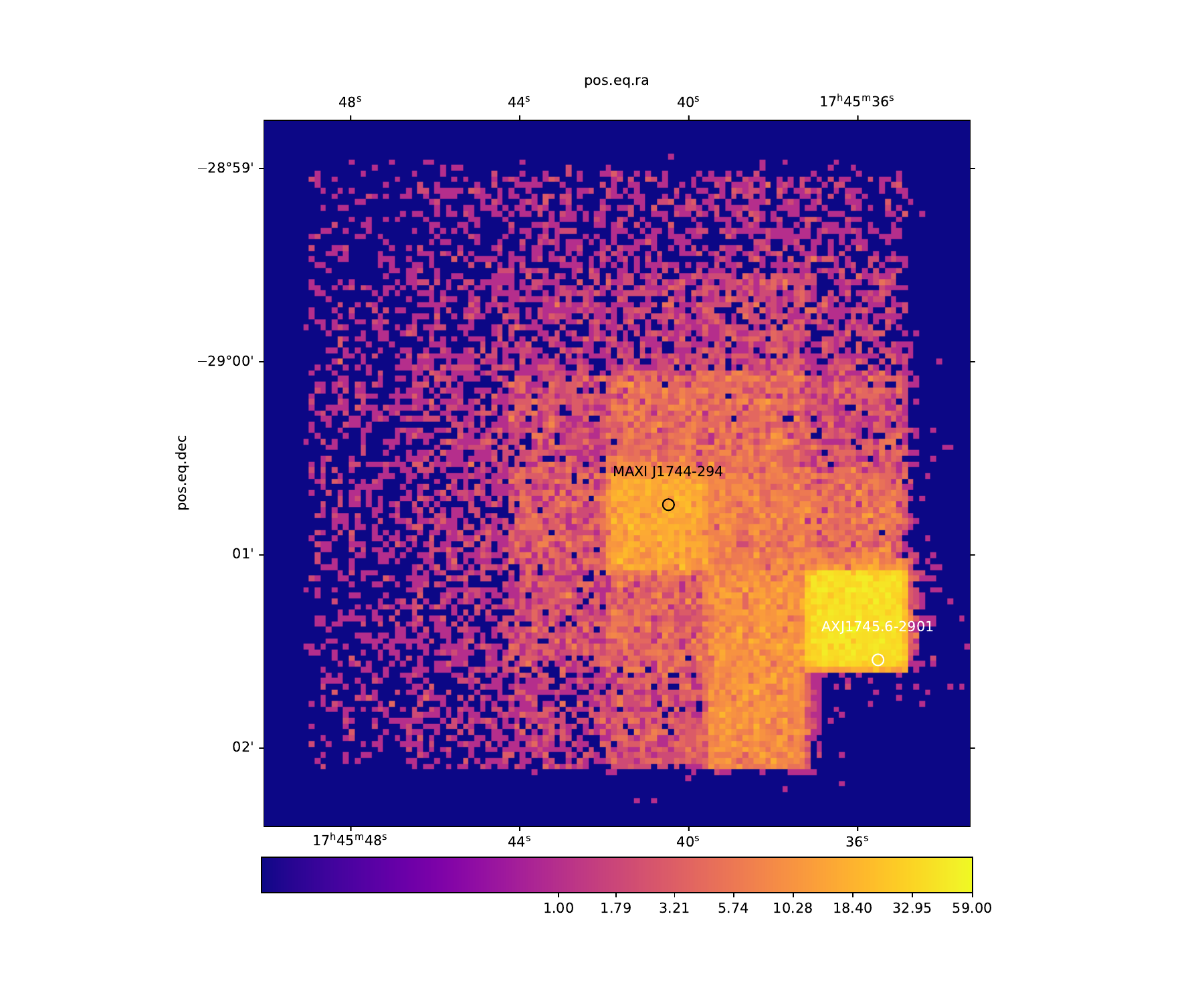}
    \includegraphics[clip,trim=4.5cm 2cm 5.3cm 1.5cm,width=0.5\textwidth]{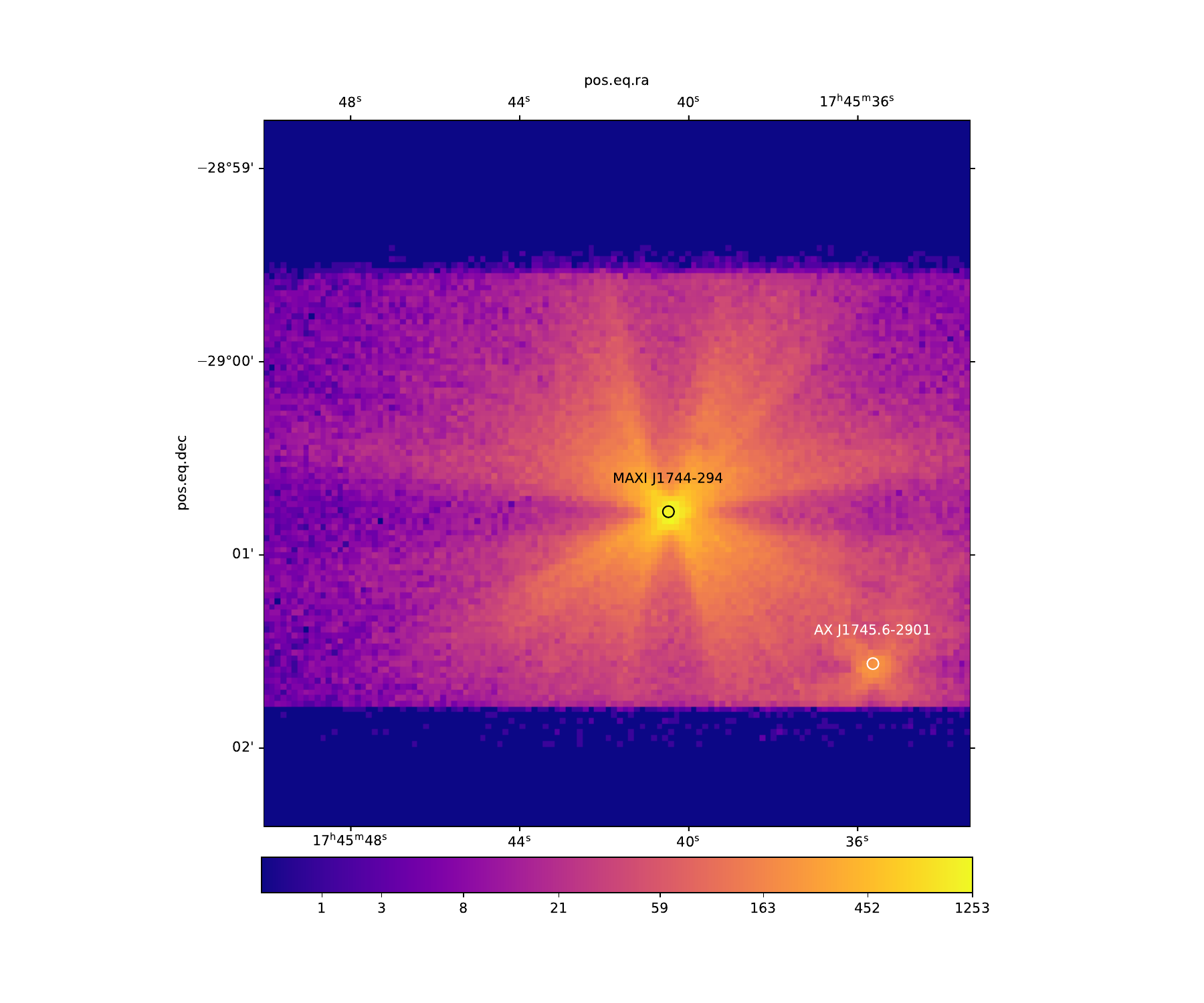}
    \includegraphics[clip,trim=4.5cm 2cm 5.3cm 1.5cm,width=0.5\textwidth]{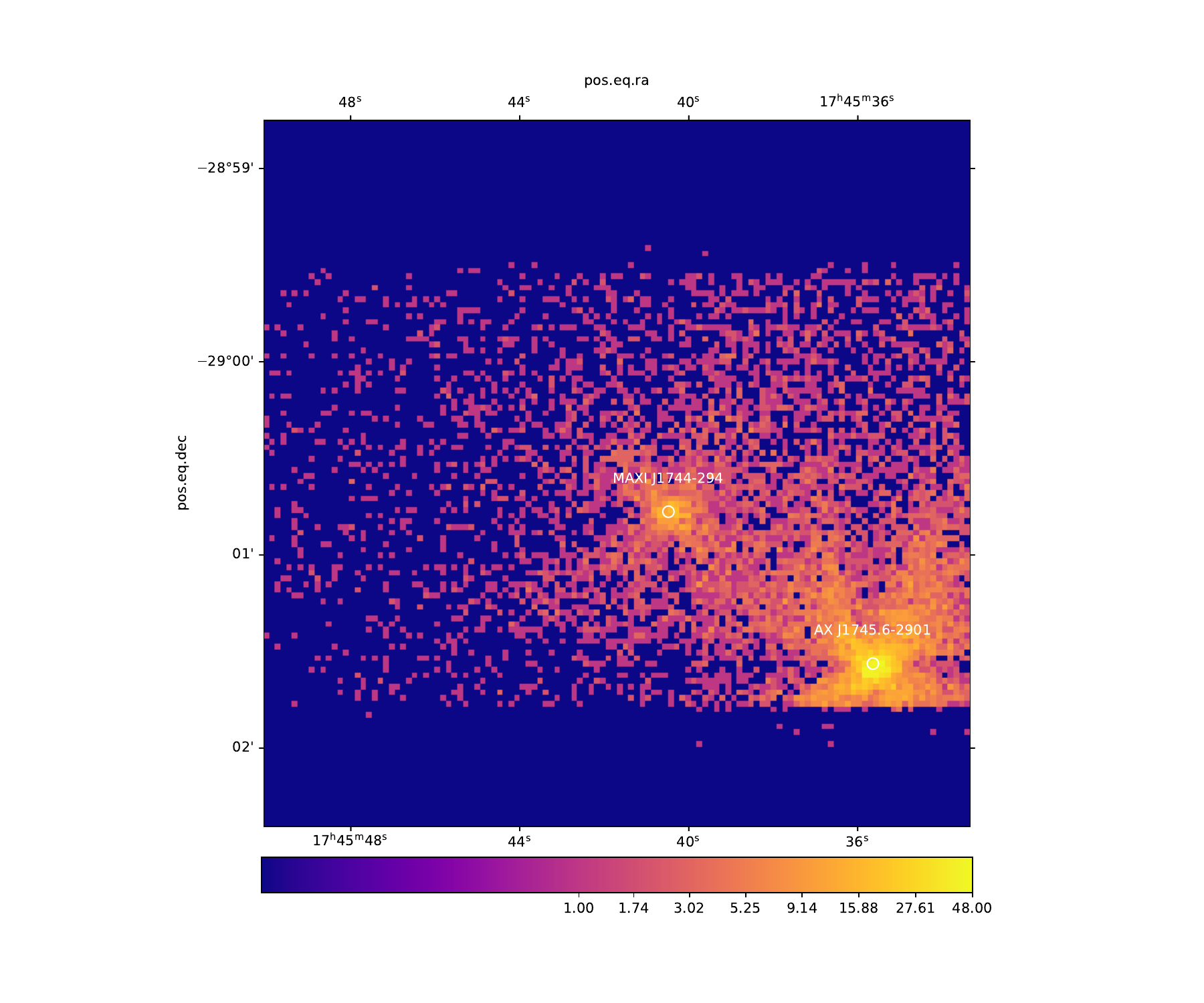}
    
    \caption{\xrism\ Sky images from the 2025 DDT Observation, in the [2-10] keV band (\textbf{top left}, Resolve), [0.3-10] keV band (\textbf{bottom left}, Xtend), and [7-10] keV band (\textbf{right}, Resolve and Xtend) computed from the reprocessed event files, after proximity and rise-time event screening, along with Flickering Pixel removal (see Section~\ref{subsubsec:xtend}) and with all Resolve event grades. The circles show the projected positions of MAXI J1744-294 and AX J1745.6-2901.
    }
    \label{fig:XRISM_FoV_2025}
\end{figure*}

\begin{figure*}[h!]
    \includegraphics[width=0.5\textwidth]{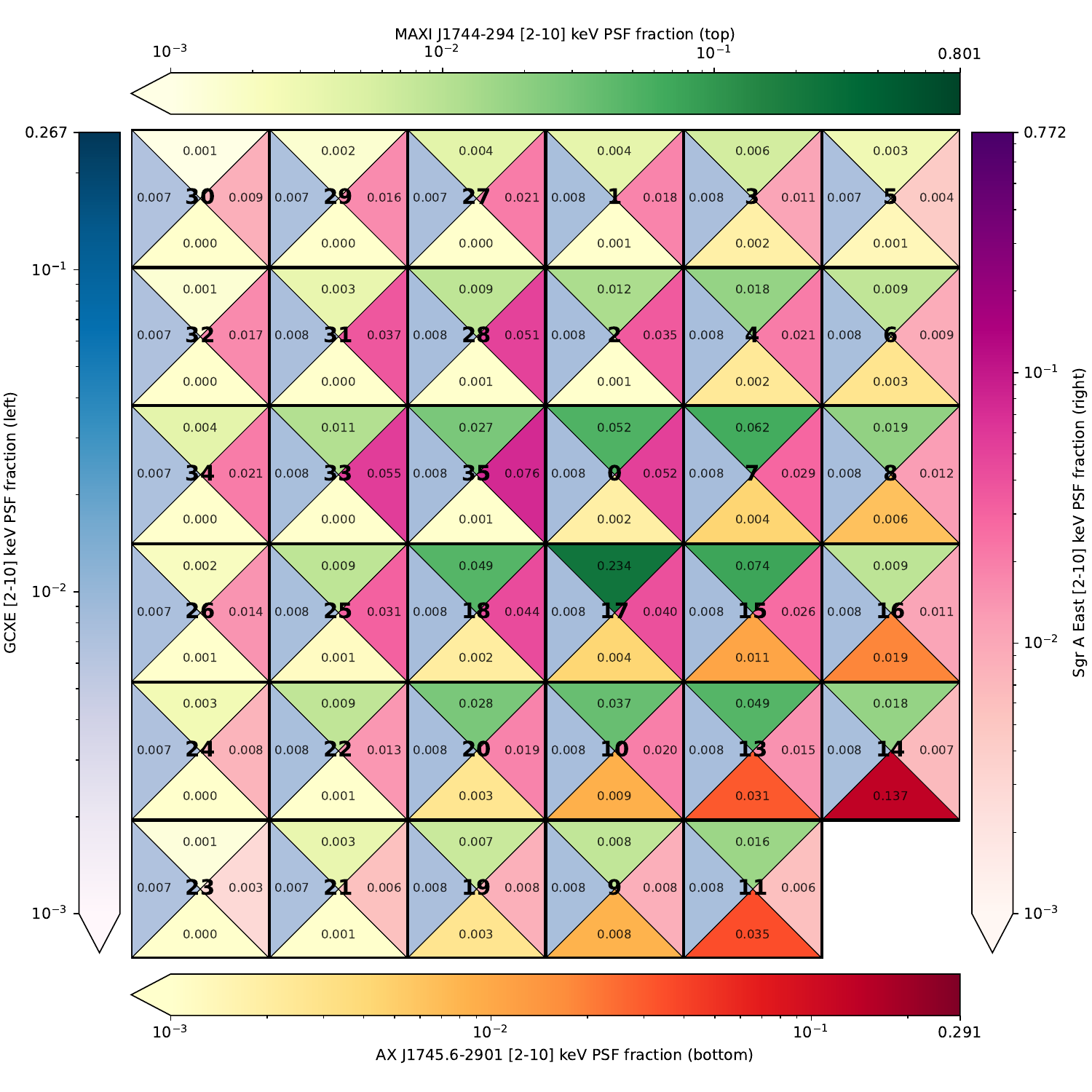}
    \includegraphics[width=0.5\textwidth]{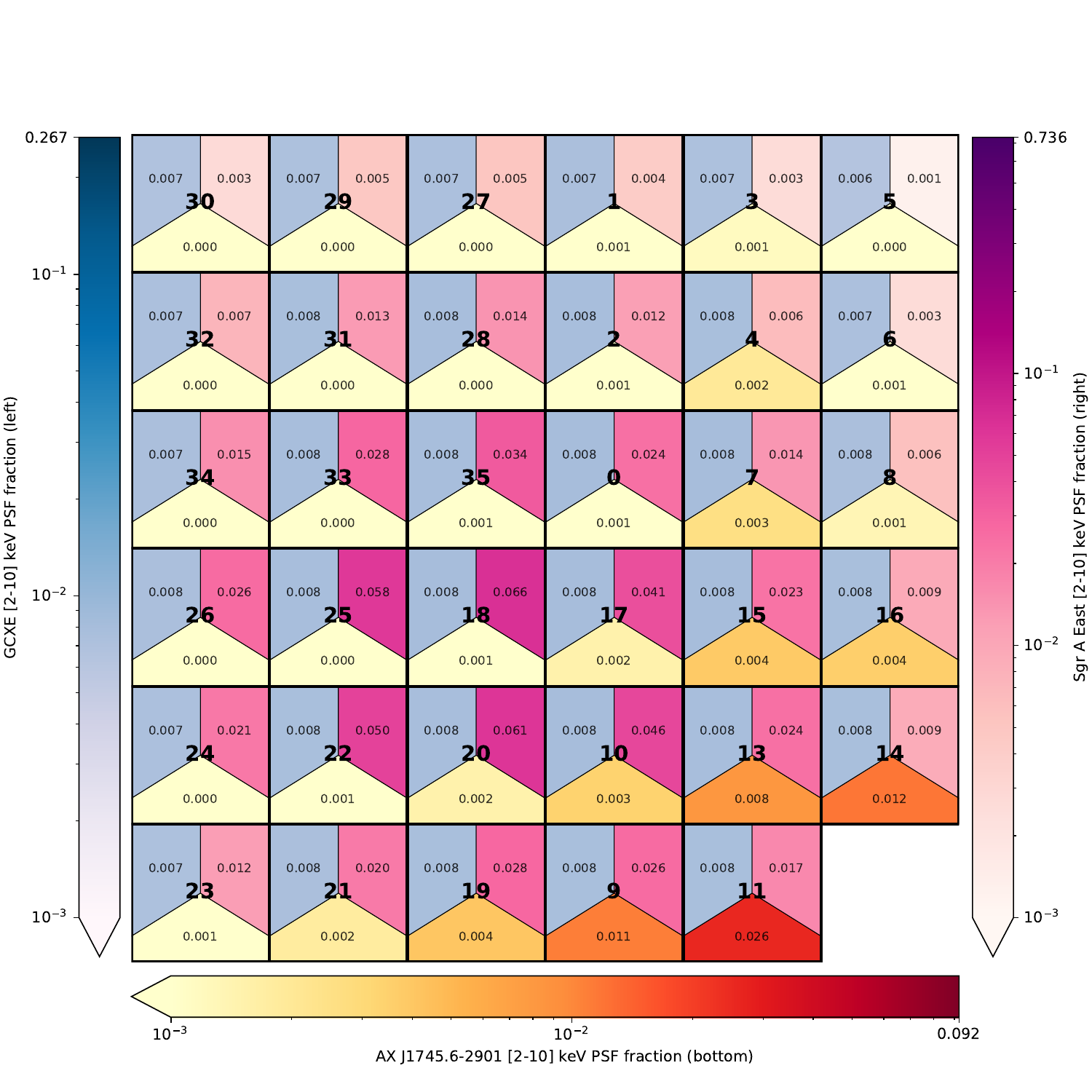}
    \caption{\textbf{(Left)} Distribution of PSF fractions in each pixel from the DDT observation, for MAXI J1744-294 (green, upper quadrants), AX J1745.6-2901 (red, bottom quadrants), the GCXE (blue, left quadrants), and Sgr A East (magenta, right quadrants). \textbf{(Right)} distribution of PSF fractions in each pixel from the PV observation, for AX J1745.6-2901 (red, bottom thirds), the GCXE (blue, left thirds), and Sgr A East (magenta, right thirds). The MAXI J1744-294 PSF fraction is not represented in the right panel as the source was not yet active in the PV observation. In both panels, the colormaps of each source are normalized with respect to their respective cumulated PSF fraction across the  Resolve FoV.}
    \label{fig:Resolve_PSF_frac}
\end{figure*}

\section{Observations and Region Selection}\label{sec:reg_choice}

A proper consideration of the point and diffuse sources in the Galactic center region is integral to revealing the intrinsic spectral features of MAXI J1744-294 (hereafter M1744) and AX J1745.6-2901 (hereafter AXJ). This is, however, made very challenging by the differences in instrumental Field of View (FoV), angular and energy resolution, and sensitivity of different instruments, along with the limited number of past (but recent) observations of this region.
 In this section, we thus present the observations of both \xrism{} instruments and \xmm{}, and our method to reliably describe in each of them the contribution of the different sources in the FoV. Their respective diffuse emission background contributions will be fitted in Section~\ref{sec:diffuse_mod}.
We note that a \nustar{} observation was also performed simultaneously to \xrism{}: while paramount to estimate the hard X-ray continuum of the source in our studies, its region selection and data reduction were already presented in detail in M26, and its low spectral resolution does not require a detailed background modeling.

Moreover, the presence of dust scattering haloes (DSH) around \src{} \citep{Mandel2026_ApJ} and \axj{} \citep{Jin2017_DSC} can significantly affect the spectral shape of each source, particularly for incomplete and/or off-axis regions. They can, however, be corrected with empirical models, which deconvolve the influence of the dust and recover the intrinsic spectral energy distribution (SED) of each source. For the DSH around \src{} and \axj{}, such models have been developed for \nustar{} and \xmm{} \citep{Jin2017_DSC,Mandel2026_ApJ}, but we have not yet adapted this approach to \xrism{}. Thus, for now, only with the former two instruments can we recover the "real" intrinsic SED of each source. \nustar{} and \xmm{}, which were already important to derive broadband SEDs across the entire X-ray band, thus become even more critical to compute physical models of the line features, as will be described for \src{} in Paper II. This also prompts us to separate the analysis with and without DSH correction. However, we stress that the influence on narrow emission lines in \xrism{} will be weak and mostly quantitative: the absorption due to dust only varies over energy ranges of at least a few hundreds eVs \citep{Jin2017_DSC}, and will thus not affect the shape, velocities, or equivalent widths of the narrow emission features found in our spectra. The scattered component may nonetheless affect the depth of strong absorption lines, biasing the intrinsic covering factor in physical models. In our case, such features are only found in \axj{} and at high energies --with less influence from the DSH--, and the scattered DSH component should thus be much weaker than the significant continuum contamination due to the different sources of diffuse emission and \src{} in the vicinity. We thus expect limited changes in \axj{}'s intrinsic line properties, especially considering their limited signal-to-noise ratio.

\subsection{\xrism}\label{subsec:xrism_reg}

Following the detection in outburst of M1744
, we requested a \xrism\ observation as part of the Director's Discretionary Time (DDT), which was performed on 2025-03-03 UT 12:04-2025-03-05 UT 00:05, with a net exposure of 71ks (Obsid 901002010, hereafter DDT observation). We refer to M26 for long-term monitoring light curves and a full description of the outburst evolution of M1744.
The microcalorimeter Resolve \citep{Ishisaki22_Resolve,Porter2024_Resolve} was operated in the OPEN filter configuration, and soft X-ray imager CCD Xtend \citep{Mori2022_Xtend,Noda2025_Xtend,Uchida2025_Xtend} in the 1/8 window+BURST mode due to the high source count rate, which remained stable in the soft state throughout the observation. We showcase the Resolve and Xtend FoVs of that observation in Fig.~\ref{fig:XRISM_FoV_2025}, using two different energy bands to highlight the competing contributions of M1744 and AXJ at low and high energies; while M1744's emission dominates below 6 keV, AXJ is stronger in the $7-10$ keV band. The flux of the sources is high enough to safely neglect the influence of the non-X-ray background (NXB) in both Resolve and Xtend spectra.

\begin{figure*}[t!]
    \includegraphics[clip,trim=4.5cm 2cm 5.4cm 1.5cm,width=0.5\textwidth]{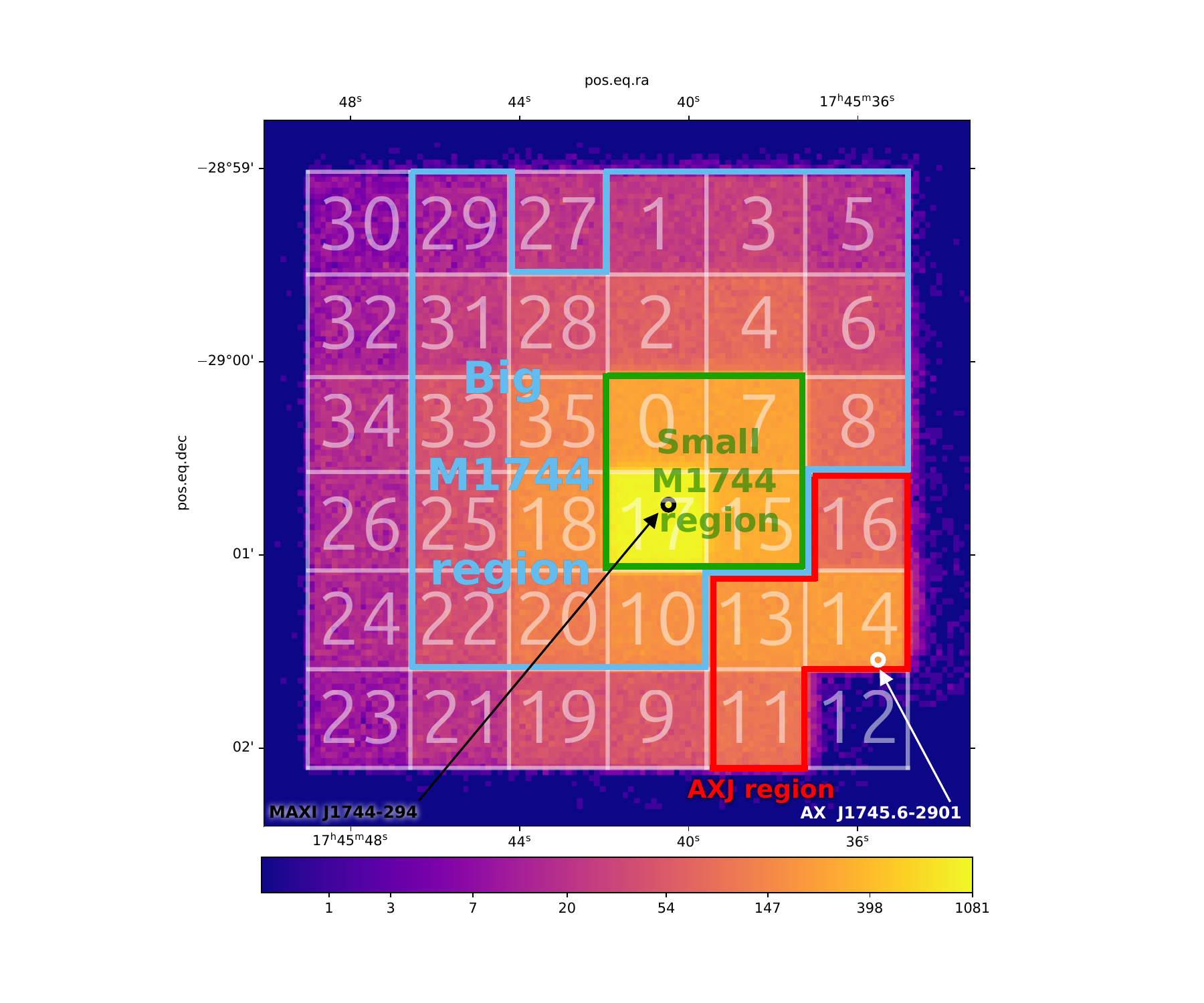}
    \includegraphics[clip,trim=4.5cm 2cm 5.4cm 1.5cm,width=0.5\textwidth]{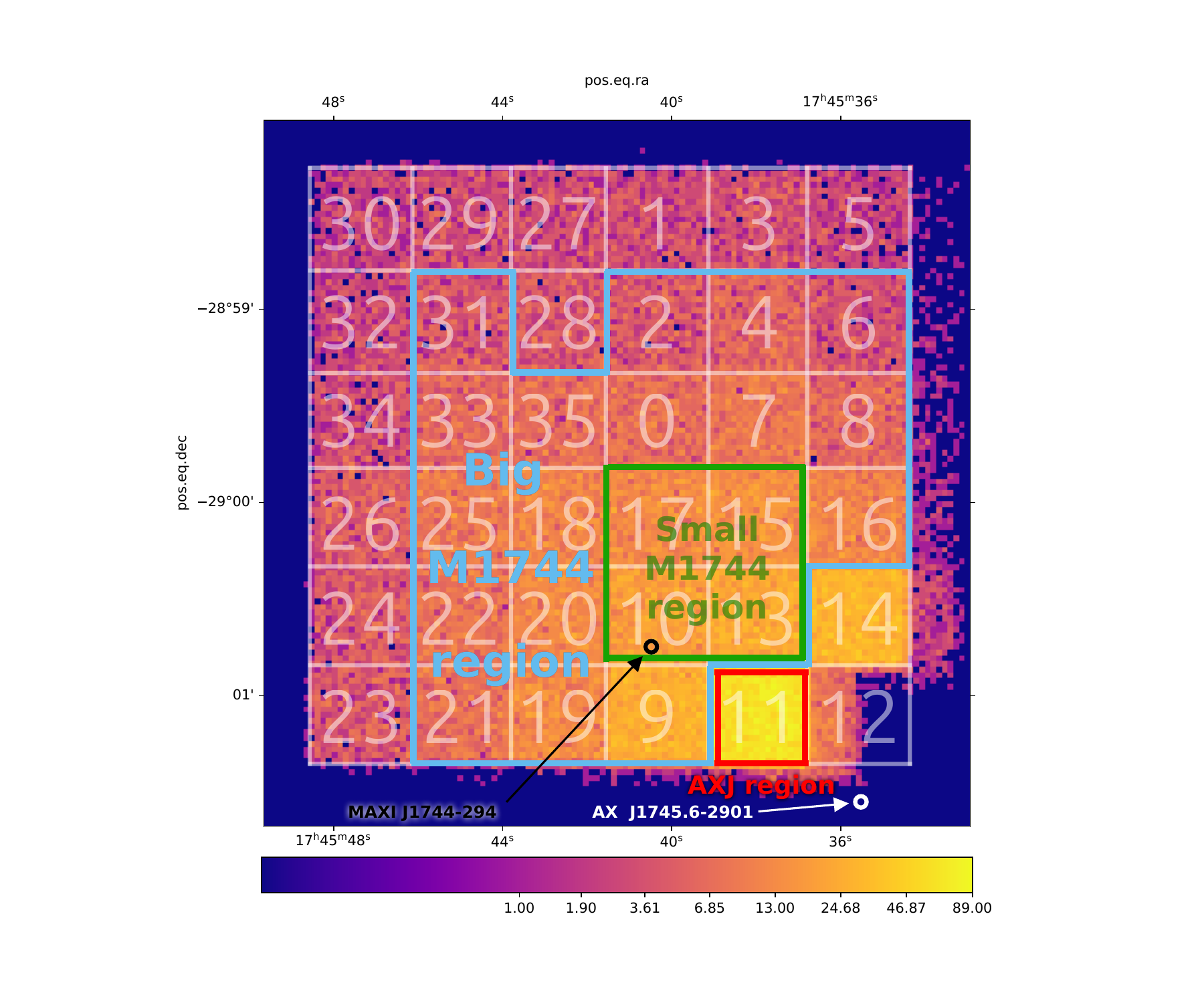}
    \caption{[2-10] keV band Resolve Sky images, computed similarly to Fig.~\ref{fig:XRISM_FoV_2025} for the 2025 DDT observation \textbf{(Left)}, and the 2024 PV phase observation used to estimate the diffuse emission \textbf{(Right)}. We highlight the two regions used for \src\ in cyan (big region) and green (small region), and the region used for \axj\ in red. Pixel numbers are overlaid for reference.}
    \label{fig:Resolve_regions_DDT_PV}
\end{figure*}

\subsubsection{Resolve}\label{subsub:reg_resolve}

The limited angular resolution of \xrism\ and the large pixel size compared to the PSF lead to significant spatial-spectral mixing between different sources. To make matters worse, two competing effects influence the fidelity of any chosen region. Selecting more pixels results in a higher PSF fraction (and thus a better signal-to-noise ratio), and lessens the influence of the dust scattering halo surrounding both M1744\ and AXJ\ \citep{Mandel2026_ApJ}. Selecting 
fewer pixels means less contamination from the diverse secondary sources, at the cost of a lower signal-to-noise ratio, and higher uncertainties on the PSF fraction and background matching. For M1744, we chose one "big" and one "small" source region to represent both approaches. We will systematically compare the analysis results yielded by the two regions.

To model the diffuse emission, we leverage the previous \xrism\ observation with the most similar FoV, ObsID 300044010, performed in the PV phase, before the outburst of M1744. It has been analyzed in \cite{XrismCol2025_GC_obs_diffuse} (for the diffuse emission) and \cite{Tanaka2026_AXJ} for AXJ. However, the FoV of that first \xrism\ Galactic center observation (hereafter "PV" observation) is shifted by a non-integer amount of pixels. Most importantly, the region covered by pixels 9, 11, 19, 21, and 23 (the southernmost row) in the DDT observation is beyond the FoV of the PV observation, making them unsuitable to model the diffuse emission in the region surrounding M1744.
We show in Fig.~\ref{fig:Resolve_PSF_frac} the different PSF fractions derived from effective area computations of the different sources in each individual pixel, for both the PV and DDT observations (see Section~\ref{sec:datared} for details of the computations). While these values rely on the inherently imperfect modeling of the \xrism\ PSF, they provide the best possible quantitative assessment of the weight of the sources in each pixel. These results, along with the sky images in different bands shown in Fig.~\ref{fig:XRISM_FoV_2025}-top, will serve as the basis for the different region choices. 

For the "big" M1744 region, we use a region containing 21 pixels, shown in blue in Fig.~\ref{fig:Resolve_regions_DDT_PV}-left. We start from the entire image, with the exception of the southernmost pixel row and 
pixel 27 (in the northernmost row, third from left), known to exhibit abnormal gain jumps and thus not suitable for scientific analysis. We then remove the leftmost (east) column, whose combined M1744 PSF contribution is only 1.1$\%$, and three of the four pixels closest to the position of the NS AXJ (13, 14, 16; 11 having been excluded already). These four pixels, shown in red in Fig.~\ref{fig:Resolve_regions_DDT_PV}-left, have the highest predicted NS PSF fraction in Fig.~\ref{fig:Resolve_PSF_frac}-left, and are the brightest pixels in the 7-10 keV band, where the AXJ\ emission dominates \citep{Mandel2026_ApJ}, as illustrated in   Fig.~\ref{fig:Resolve_regions_DDT_PV}-top right. These pixels will form the region used to model the contribution from AXJ. For the "small" M1744 region region, shown in green in Fig.~\ref{fig:Resolve_regions_DDT_PV}-left, we only consider the four brightest pixels in the array in the entire \xrism\ energy band (Fig.~\ref{fig:Resolve_regions_DDT_PV}-top left), as they have the highest predicted contribution from M1744 (Fig.~\ref{fig:Resolve_PSF_frac}-left).

To derive the "nearest" pixel positions in the PV phase observation, which will be used to estimate the diffuse emission, we shift the two main M1744 regions one row to the south. The resulting "background" regions are shown in Fig.~\ref{fig:Resolve_regions_DDT_PV}-right, with the same colors as used previously for the corresponding source regions. This translation is not perfect, as the FoV difference is close to 1.5 pixels 
north in addition to a small shift eastward, but these shifts will have a negligible influence on the "big" M1744 source region owing to its size.
In parallel, the contribution to the background from AXJ, which was the brightest point source in the PV observation and whose flux changed between the two observations, must be modeled in both M1744
regions. To represent the AXJ contribution, we opt to use only pixel 11 (shown in red in Fig.~\ref{fig:Resolve_regions_DDT_PV}-right), which has by far the highest PSF fraction for that source in the PV phase observation (Fig.~\ref{fig:Resolve_PSF_frac}-right). Indeed, as the position of AXJ was beyond the limits of the FoV for the PV observation (lower right in Fig.~\ref{sec:reg_choice}-right), all other pixels have a very low signal-to-noise ratio, and the emission lines from the diffuse  
components of Sgr A East and the Galactic center X-ray emission (GCXE) dominate the main ionized absorption lines seen between $6.5-7.0$ keV (the most important feature of the NS to model). 
As we do not have a "clean" observation to model the diffuse emission \textit{local to the NS region}, in both the DDT and PV phase observations, we assume the same diffuse emission sources found in the M1744  regions.

\begin{figure*}[t!]
    \includegraphics[clip,trim=4.5cm 2cm 5.4cm 1.5cm,width=0.5\textwidth]{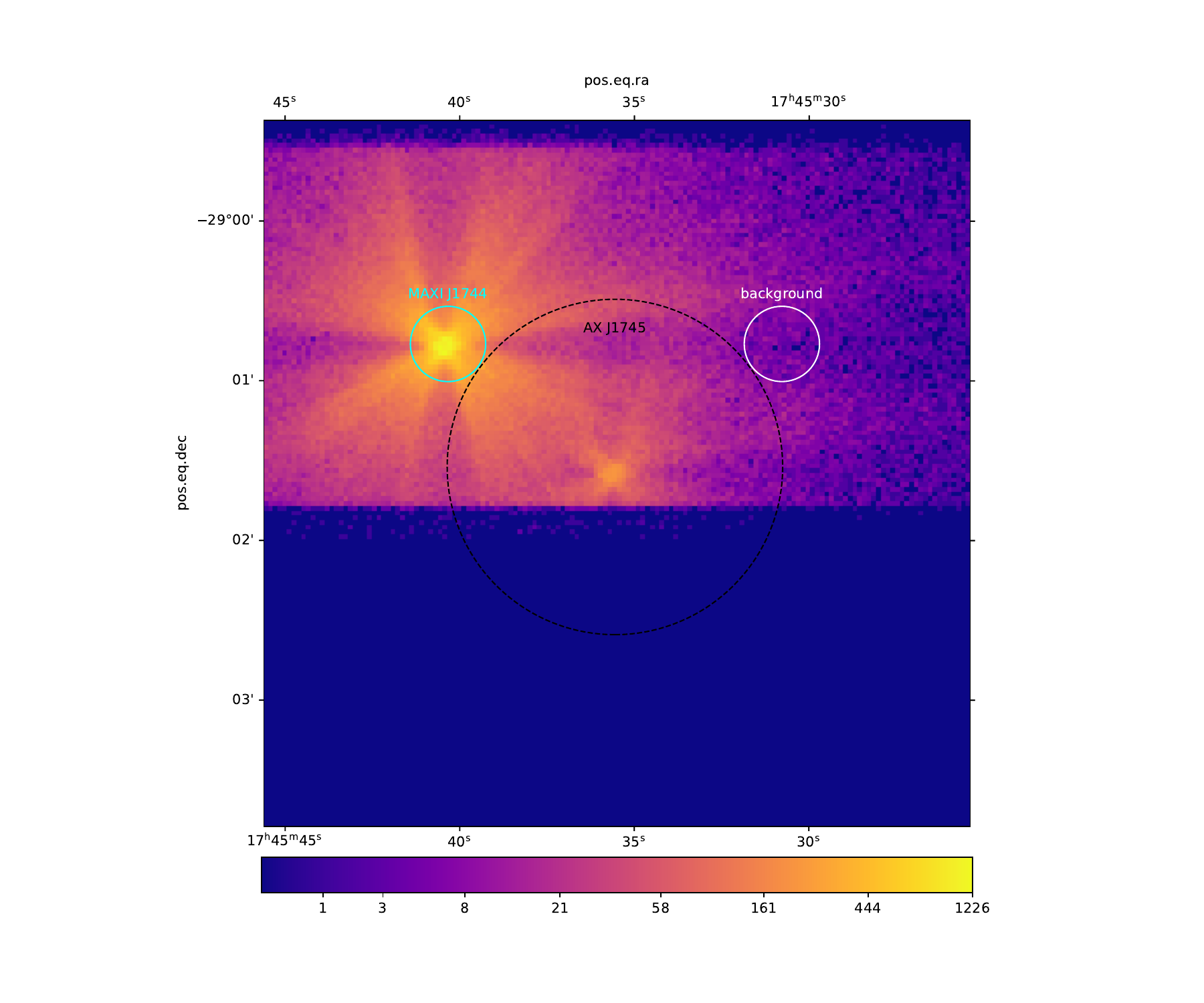}
    \includegraphics[clip,trim=4.5cm 2cm 5.4cm 1.5cm,width=0.5\textwidth]{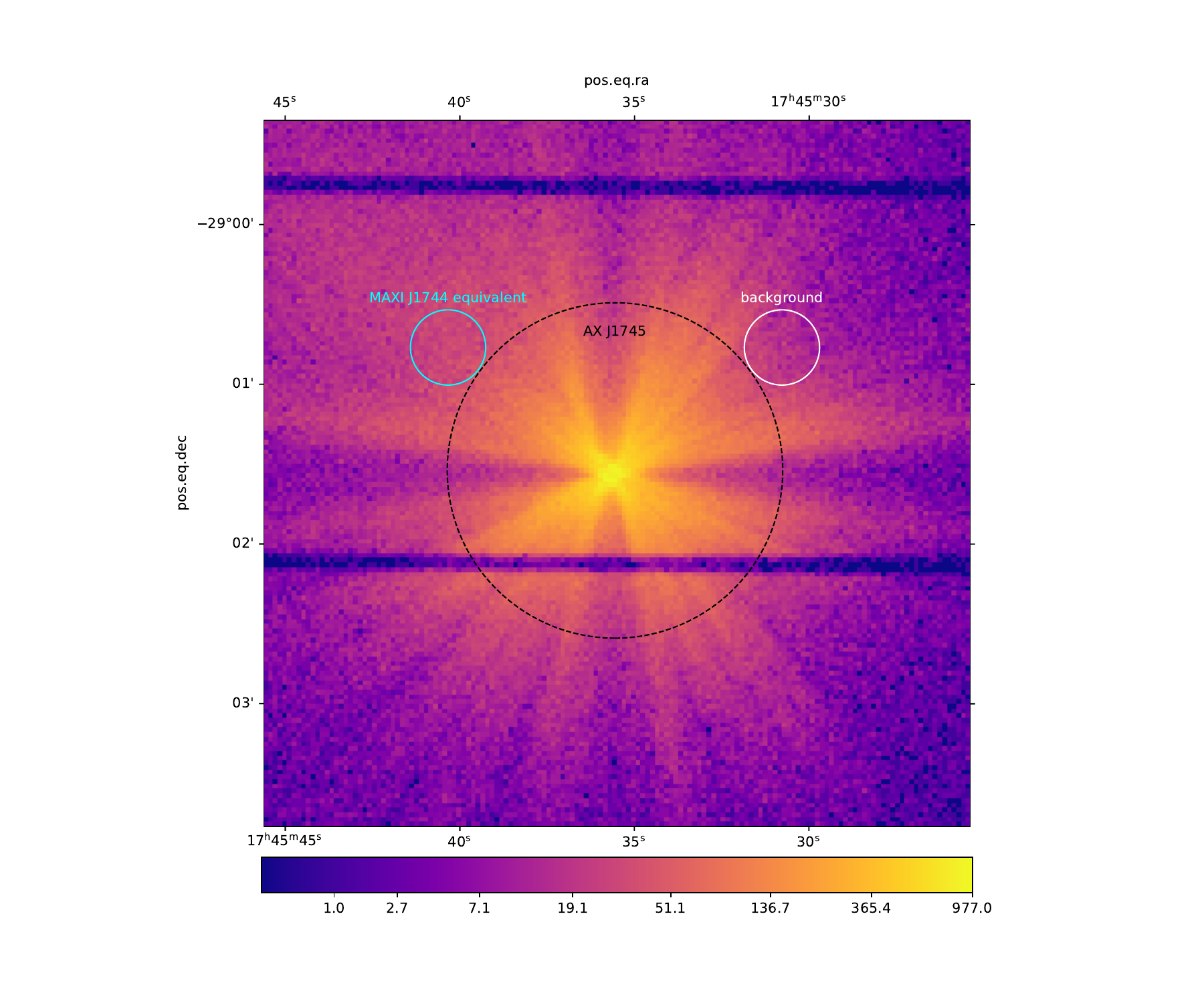}
    \caption{[2-10] keV band Xtend Sky images computed similarly to Fig.~\ref{fig:XRISM_FoV_2025} for the 2025 DDT \textbf{(left)} and the 2024 PV Phase \textbf{(right)} observations, highlighting the regions used to model MAXI J1744-294.}
    \label{fig:Xtend_regions_DDT_PV}
\end{figure*} 

\subsubsection{Xtend}\label{subsub:reg_xtend}

Xtend's imaging capabilities and wider FoV allow more flexible source extraction regions and more accurate backgrounds -- both of which reduce contamination from secondary sources. Unfortunately, it is also subject to PSF mixing, as Xtend and XRISM share the same PSF, as well as pile-up due to the nature of the instrument.
We thus computed the radial profile of the pile-up fraction in Xtend events by comparing the count rate to the readout time of the instrument across the FoV. We largely follow the methodology described for Suzaku in \cite{Yamada2012_pileup_Suzaku}, and compute the pile-up fraction following the formula of \cite{Davis2001_pileup}, applied to cell sizes of 5 pixels, centered between the pixel with the highest count rate for the relevant PSF and up to a radius of 60 pixels. For the DDT observation, operated in burst mode, we confirm a negligible pile-up fraction ($\lesssim 1\%$) for the brightest source in the FoV (M1744), as shown in the top panels of Fig. A.\ref{fig:Xtend_pileup} in App.~\ref{app:Xtend_pileup}. 
In the PV phase observation, which was operated in Full-Window mode, we derived a pileup maximum of 6\% at the PSF core of the brightest point source in the FoV, AXJ. This confirms that "background regions" taken from the PSF wings will not suffer from 
pile-up.

For M1744, we thus use an eight-pixel ($\sim14\arcsec$) region centered on the position of the source, as shown in cyan in Fig.~\ref{fig:Xtend_regions_DDT_PV}-left. As the background needs to account for AXJ, Sgr A East, and the GCXE, we use a background region of identical size at the western end of the NS PSF (shown in white in Fig.~\ref{fig:Xtend_regions_DDT_PV}-left), to limit our reliance on PSF modeling. This region will include the NS,  GCXE (assumed to be uniform on these angular scales), and any uniform NXB contribution. In addition, the contribution from Sgr A East is modeled from a "background" region in the PV phase observation, whose coordinates are identical to the source region of the DDT (Fig.~\ref{fig:Xtend_regions_DDT_PV}-right, cyan). To isolate the Sgr A East component, that region is once again background subtracted using a background from the (same) symmetrical position 
 relative to the NS PSF, shown in white in Fig.~\ref{fig:Xtend_regions_DDT_PV}-right. As these NS background choices rely on the symmetry of the \xrism{} PSF, we quantify its influence in Appendix~\ref{app:Xtend_PSF_asym}, and confirm that it has a negligible effect on our analysis. 

\subsection{\xmm}\label{subsec:reg_xmm}

A simultaneous \xmm\ observation was performed starting on 2025-03-03 UT 21:20, with a net exposure of 9 ks. Although the instrument was operated in small window mode, it was still affected by significant pile-up.
Our approach for \xmm's region selection is similar to that applied to other CCD instruments, although the contamination from AXJ\ cannot be directly estimated, as it was outside the FoV. In order to minimize pile-up for the M1744 source region, we selected an annulus of inner/outer radii $12\arcsec/25\arcsec$, centered on the source position.  We refer to M26 for further details of the observation and region selection. 
To model the diffuse background contribution, we used identical regions from the latest  Galactic center observations in the \xmm\ archive, uncontaminated by an X-ray transient: ObsIDs 0893811101 (2022-03-18) and 0893811101 (2022-03-26). In both cases, as the observations were performed in full-window mode, we subtracted "empty" background regions near the edge of the FoV to account for contamination from the GCXE.

\section{Data reduction} \label{sec:datared}

We refer to M26 for a description of the \nustar{} and \xmm{} data reduction procedures, which largely follow the standard analysis guides. We highlight that in the archival \xmm\ observations used to model the diffuse emission, we filtered data contaminated by soft proton flares, as well as periods with flares appearing to come from Sgr A*.

\subsection{\xrism}
\label{xrismdata}

The \xrism\ data was reduced using Heasoft 6.36, following the \xrism\ data reduction threads\footnote{\href{https://heasarc.gsfc.nasa.gov/docs/xrism/analysis/}{https://heasarc.gsfc.nasa.gov/docs/xrism/analysis/}} and~{\it Quick-Start Guide}\footnote{\href{https://heasarc.gsfc.nasa.gov/docs/xrism/analysis/quickstart/index.html}{https://heasarc.gsfc.nasa.gov/docs/xrism/analysis/quickstart/index.html}} (v3.1). We restricted ourselves to \xrism\ CALDB 11 (20250315), as \xrism\ CALDB 12 (20250915) is known to have an erroneous implementation of newly introduced optical axis tilt adjustments, as described in the \xrism~{\it Quick-Start Guide} (v3.2). For standard point-source observations at the aim-point, the effects of the errors in Resolve data are not significant, but for the complex field that we are dealing with, the effects are large, especially because the source AXJ is situated at the edge of one of the pixels at the edge of the Resolve detector (the effects are quantified in Appendix~\ref{app:caldb_influence}). For Xtend, it so happens that errors in the units of some of the CalDB quantities conspired to guarantee that the results from CalDB 11 and CalDB 12 are indistinguishable.

The first step of the data reduction, common to both instruments, was to reprocess the data with the aforementioned versions of Heasoft and CALDB using \texttt{xapipeline}. This notably ensures that the Middle-resolution primary events (see Section \ref{subsec:resolve3} below) are corrected with the new tasks and calibration added to the \xrism\ pipeline in its recent versions.

\subsubsection{Resolve}\label{subsec:resolve3}

We provide details on the absolute energy calibration of Resolve and our manual update of the gain correction in Appendix~\ref{app:gain_calb_rsl}. 
One of the main specificities of the instrument is the link between the spectral resolution of detected photons and the count rate. In the processed data, events are separated into 5 individual grades \citep{Ishisaki2018_Hitomi}, representing different scenarios and time intervals between successive events, which can diminish the precision of the energy reconstruction procedure. Higher count rates lead to shorter intervals, and the so-called "branching ratios" of different grades compared to the entire event list thus become progressively dominated by Middle-resolution ($\gtrsim5$ eV at 6 keV) and Low resolution ($\gtrsim 20$ eV at 6 keV) events.
After applying the standard proximity and rise-time event cuts (RTS)
, we computed the branching ratios of each pixel in the 2-12 keV band using the task \texttt{rslbratios}, along with the theoretical values at the count rate of each pixel. These branching ratio values are summarized in the left panel of Fig. A.\ref{fig:resolve_branch} in App.~\ref{app:Resolve_branch}.
As expected for a bright source, the central pixels exhibit a non-negligible fraction of Middle- and Low-resolution events. For this analysis, we thus extracted spectral products from both High-resolution primary (Hp) and Middle-resolution primary (Mp) events, following improvements in the calibrations of Mp events introduced in Heasoft 6.35 and CALDB v20250315.  The spectra were extracted with \texttt{xselect}, using the regions described in Section~\ref{subsub:reg_resolve}.

As the Response Matrix File (RMF) is generated from only part of the event grades, it will be normalized according to the branching ratios of the selected grades. However, most \xrism\ observations are subject to uncertainty regarding the origin of excess Low-resolution secondary (Ls) events. In our observation, a direct comparison of the evolution of the Ls branching ratios with energy indicates an effective area uncertainty of up to 3.7$\%$ in the 2-10 keV band. Moreover, even if the false Ls rate is not known, as can be seen in Fig. A.\ref{fig:resolve_branch} of App.~\ref{app:Resolve_branch}, the 2-12 keV energy-averaged branching ratios of Ls events is sufficiently compatible with theoretical expectations for the brightest pixels, with branching ratio discrepancies of less than 1\% relative to the sum of all event grades. Meanwhile, the faintest pixels have sufficiently high count rates to be at most marginally affected by potentially "false" Ls events, with Ls branching ratios either compatible with theoretical expectations or below $\sim 10^{-2}$. Finally, the 3 faintest pixels (30, 23, 32), which also have the highest Ls branching ratios, are already excluded due to our choice of spatial regions. 
We thus consider this source of uncertainty to be negligible in our observation, and generated the response file from the 2-12 keV event file with all event grades included and Ls events considered as "real". We apply the same methodology for the PV phase data, which is fainter but retains a proportion of Hp events of $\sim98\%$, with no significant contamination from unexpected Ls events.

As a sanity check, we further investigated the energy dependence of the branching ratios, and show in Fig. A.\ref{fig:resolve_branch_Edep_DDT} of App.~\ref{app:Resolve_branch} the energy-dependent count rates (top) and branching ratios (bottom) of each event grade in the DDT and PV phase observations, integrated over 20 eV bins, for all pixels in the array. In the DDT observation (left panels), the count rate increase around 6.7 keV in the Hp events, due to strong emission lines, does not significantly affect the Hp branching ratios. However, as a precaution, since the count rates vary widely for different pixels, and the spurious contamination of Ls events is strongly energy dependent, we computed the branching ratios of individual pixels in the 6.6-6.73 keV band, where the Fe lines completely dominate the continuum. We show the ratio of the fraction of Hp+Mp events in this band compared to the 2-12 keV band in the right panel of Fig. A.\ref{fig:resolve_branch} in App.~\ref{app:Resolve_branch}. As we see only minimal differences (max $\sim3\%$ for Pixel 17), we do not create an additional RMF with different normalizations specifically to fit the energy regions with the Fe lines. In the PV phase observation (right panels of Fig. A.\ref{fig:resolve_branch_Edep_DDT}), which is more than 10 times fainter, the branching ratios of Hp events remain virtually indistinguishable from 1 even in the 6.6-6.73 keV band.

Our method for computing the response files followed the introduction of a new procedure, suited to bright sources, in Heasoft 6.36. We used the "previous" RMF + ARF creation method for the (faint) PV observation, and applied both methods to the DDT observation as an independent check. We first generated the RMF from \texttt{rslmkrmf} using the "extra-large" (X) option to consider all effects, splitting the Individual electron loss continuum (ELC) and core matrices into separate tables, which were then merged in a single "combined" response file. We then generated the Ancillary Response File (ARF) from \texttt{xaarfgen} \citep{Yaqoob2018_spectral_effarea_Hitomi_imaging}. For 
M1744, we assume a point-like source at its \chandra\ position  
\citep{Mandel2025d}. For the AXJ region, we assume a point-like source at the \chandra\ position of the NS. For the GCXE contribution and the empirical modeling of all diffuse emission, we assume a "flat" circular ARF contribution with a radius of 3 arcminutes. Finally, for the Sgr A East contribution, we apply a flux map distribution using a smoothed, stacked \chandra\ image of Sgr A East in the 6.6-6.8 keV band, similarly to \cite{XrismCol2025_GC_obs_diffuse}. These choices result from our complementary approaches to modeling the different diffuse emissions, which we detail further in Section~\ref{sub:diffuse_mod_Resolve}. 

For the DDT observation, we also used the new task \texttt{rslmkrsp}, which runs \texttt{rslmkrmf} and \texttt{xaarfgen} together to derive more precise effective area computations when the event grade fraction varies spatially with the pixels. This is notably the case for the central pixels around M1744 (see App.~\ref{app:Resolve_branch} and Fig. A.\ref{fig:resolve_branch}). The task outputs a spectral response file (RSP) that combines RMF and ARF, and was run with the same parameters used for our previous RMF and ARF computations. In addition, \texttt{rslmkrsp} requires an energy range in which to compute the grade fractions. We chose the $2-12$ keV band, having confirmed that the variations in branching ratios around the Fe lines were negligible. For reference, we show in Fig. A.~\ref{fig:Resolve_arf_rsp_compa_CALDB11_DDT} of App.~\ref{app:caldb_influence} the comparison and ratios between effective areas computed with Heasoft 6.36 (\texttt{rslmkrsp}) and Heasoft 6.35.1 (\texttt{xaarfgen}) for each source. As expected, the difference is highest ($\sim15-20\%$) for the M1744 source in its own regions, but remains non-negligible for all sources, especially for regions with higher count rate and lower Hp branching ratios. Since this task is more accurate than the simple RMF + ARF creation for bright sources, we use RSP responses with all spectral products in the DDT observation. We keep all spectra ungrouped to work with Cash statistics, and restrict our spectral analysis to the $2-10$ keV range. Below that range, the Resolve effective area is negligible with the Gate Valve in its closed configuration, and above 10 keV, the contribution from M1744 is negligible compared to the other background sources.

To complement the Resolve images and select suitable regions for the different sources, we computed the PSF fractions for each source in each pixel in the DDT and PV phase observations (using \xrism\ CALDB 11). 
We first generated ray-tracing files for all four (three) sources in the DDT (PV phase) observations
, using the entire pixel array as the detector region, and following the methodology described above for ARF computations. For the DDT data, we generated the grade-fraction-corrected RMF, exposure maps, and ray-tracing files independently for each source using \texttt{rslmkrsp}. In the PV phase data, we generated the RMF and exposure map with \texttt{rslmkrmf} and \texttt{xaexpmap}, and a raytracing file for each source with \texttt{xaarfgen}. To minimize systematic uncertainties and accurately represent the PSF fraction in regions far from the positions of the different sources, we set the \texttt{numphoton} parameter to $3\times10^6$ in all computations. We then computed 35 individual instances of \texttt{xaxmaarfgen}, with the regions of each individual pixel, the previously computed raytracing file, the exposure map, and the RMF. The number of photons per energy channel in each pixel was then retrieved from the log files of the \texttt{xaxmaarfgen} runs and normalized to the total number of photons generated in the raytracing files across energies. This process was repeated for all four (three) sources in the DDT 
(PV phase) observations. Finally, to represent the contribution of each source, we generated figures with each pixel color-coded according to the contribution of its different sources, with the maximum values of the colormaps normalized to the total PSF fractions of each source in the array. We show in Fig.~\ref{fig:Resolve_PSF_frac} the resulting figures,  computed for the broad energy band relevant to our analysis, namely $2-10$ keV. Comparing different values from more precise energy ranges, notably at 2 keV and 7 keV, confirmed that the PSF fractions are virtually independent of energy, aside from statistical fluctuations on the order of $1\%$ between different pixels. We note that these values are raw outputs from a single run of {\tt xraytrace} for each source and thus do not include the simulation's systematics, nor those due to the global uncertainties in the modeling of the Resolve PSF. These results should be interpreted qualitatively only.

Two additional elements not yet taken into account in the current \xrism\ pipeline, and which could warrant additional corrections, are the energy-scale shift and the decrease in energy resolution of Resolve for bright sources. \cite{mizumoto2025_Resolve_bright_effect} recently analyzed \xrism\ observations of GX 13+1 and the Crab, and identified an absolute offset of up to $\sim-2$ eV at 6 keV for $\sim22$ cts/s/pix, and a linear increase of the full-width at half-maximum (FWHM) of the spectral response at 6 keV with the count rate of the direct neighbors of a given pixel, from $\sim4.5$ eV at low count rates, to $\sim6.3$ eV for a first neighbor total of 17.5 cts/s/pix. As shown in App.~\ref{app:Resolve_branch} and Fig. A.\ref{fig:resolve_branch}, in our observation, the brightest pixel has a count rate of 4.3 counts/s in the entire \xrism\ band, and all other pixels are below $\sim1.3$ counts/s. Based on the data analyzed by \cite{mizumoto2025_Resolve_bright_effect}, we may expect an energy shift of at most $\pm0.5$ eV, which translates to 22 km/s for the Fe XXV K$\alpha$ complex at 6.7 keV; the FWHM at 6 keV should increase by the same amount. However, to first order, the energy shift scales linearly with the average energy of the spectrum
. For our spectrum, the mean photon energy is $E_{avg}=4.15$ keV, which is lower than the $5-5.3$ keV averages for the Crab and GX 13+1 spectra used in \cite{mizumoto2025_Resolve_bright_effect}. We therefore expect the real energy shift to be closer to $\sim18$ km/s. All of these additional errors remain much smaller than the statistical uncertainties we derive in the spectral analysis, and do not qualitatively affect our results. 

\subsubsection{Xtend}\label{subsubsec:xtend}

After reprocessing the Xtend event files, we searched and removed flickering pixels in the DDT observation. 
We computed an image of the Xtend CCD in Burst mode, then measured the brightest pixel in a 15\arcsec\ region around the \chandra{} measured position of the brightest source in the FoV, M1744. This pixel should be the brightest pixel originating from real X-ray photons in the observation, and thus all pixels with count rates above this limit must be removed. We thus used the \texttt{searchflixpix} task to apply a flat-top filter and create a clean event file, using 110\% of the brightest pixel as the threshold value, and a cell size of zero, which translates to simple pixel-to-pixel comparisons. We then created a new image from the cleaned event file, inspected it visually, and confirmed that in the cleaned event file, the part of the FoV around the M1744, AXJ, and background regions was free of flickering pixels. Further confirmation of the removal of the flickering pixels with respect to the low-energy anomalies in the Xtend spectrum of M1744 is presented in Paper II. In the PV phase observation, visual inspection of the full-window FoV confirmed that the CCD chip including all regions of interest was not affected by flickering pixels. All Xtend images shown in this paper are generated after flickering-pixel removal.

Lightcurves and time-averaged spectral products were derived from the cleaned event file, using xselect and the regions defined in Section~\ref{subsub:reg_xtend}. We generated RMFs and ARFs using \texttt{xtdrmf} and \texttt{xaarfgen}. We used the same angular distributions for each source as for Resolve, with M1744 as a point source, and Sgr A East as an extended source modeled from a smoothed, stacked \chandra\ image of the SNR in the 6.6-6.8 keV band. Considering the limited signal-to-noise of our sources above 10 keV, we restricted the use of our spectral products to the 0.3-10 keV band. We grouped the spectra using the "opt" binning in \texttt{ftgrouppha}, following the method described in \cite{Kaastra2016_binning_opt}. 

\section{Diffuse emission modeling in the PV Phase}\label{sec:diffuse_mod}

Here, we describe the steps of our diffuse emission modeling with both \xrism\ instruments and \xmm. We recall that for \nustar, as mentioned in M26,  its single observation mode and limited spectral resolution allowed to directly combine the individual background spectra, derived with similar criteria to our Xtend analysis, using the \texttt{MATHPHA} tool.
Throughout this work, for spectral modeling, we use Xspec v12.15.1 \citep{Arnaud1996_xspec}, along with AtomDB 3.1.3. Unless stated otherwise, uncertainties are computed at the 90$\%$ confidence level. We note that while the Resolve spectra are left ungrouped in our analysis, our figures will be visually rebinned for readability. We stress that with this rebinning, the significance of the features in the residual plots will be systematically underestimated compared to the actual computations in our fits, which use the full spectral resolution of the spectrum. Furthermore, to highlight the narrow line complexes that may otherwise blend in the broadband residual plots, we systematically display the Resolve model components to a higher resolution than the data itself, using a 3$\sigma$ significance level, which refers to the model components displayed for spectra rebinned at this significance. 

\subsection{Resolve}\label{sub:diffuse_mod_Resolve}

The Galactic center region is an extremely crowded field and includes a very high number of point sources (see e.g. M26), along with several sources of diffuse emission, the most predominant of them being the SNR Sgr A East. The GCXE (reviewed in \cite{Koyama2018}), which can be modeled with hot plasma components and may correspond to the stacked contribution of an unresolved population of cataclysmic variables, is another source of contamination. Given \xrism's limited angular resolution, we used two complementary methods to model diffuse emission using PV-phase observations. First, we performed a fully empirical modeling of the spectral features, which avoids the shortcomings of physical models, but can lead to overfitting spurious fluctuations in the spectrum. With this method, we do not distinguish the origin of the components, and thus treat them as a single source with a flat (i.e. uniform) contribution. Albeit imperfect, this approach is not unreasonable given the very low angular resolution of the \xrism\ PSF and the large or off-axis regions considered in our analysis. Secondly, we attempted physical modeling using overionization models, limiting ourselves to the two main contributing sources: Sgr A East and the GCXE.

 In this section, we apply each methodology to the source region that best complements it. On one hand, the empirical, single-source, uniform modeling is more accurate for the "big" M1744 region, where the PSF contributions will be heavily averaged, and a much higher fraction of the diffuse emission will be considered. On the other hand, the physical, multi-source modeling is more accurate for the "small" M1744 region, where the non-uniform contribution of Sgr A East is stronger and can be modeled with a higher degree of precision. These two approaches are complementary: the spectral features that will remain consistent between the two are unlikely to be imputable to the shortcomings of either methodology. The interpretation of the models describing the diffuse emission (e.g., their origin and composition), regardless of the approach, is beyond the scope of this work. We refer to \cite{XrismCol2025_GC_obs_diffuse} for the ionized Fe K-shell lines, and Uchiyama et al. (in prep.) for the rest of the transitions. Furthermore, we do not impose any physical constraints on the normalization ratios of the neutral Fe K lines used in the fit. A more detailed exploration of their parameter space to contextualize the residual features observed in the spectrum of M1744 is presented in Paper II.

 No matter the methodology used to describe the diffuse emission, the AXJ contributions are modeled jointly using the point-source effective areas generated for the M1744 and AXJ regions. A physical approach to the  AXJ absorption features in the PV phase has already been provided by \cite{Tanaka2026_AXJ}. However, given the abnormal line profiles observed in the NS spectrum during the DDT observation (which cannot be represented by a standard single-zone photoionization model), and thanks to a higher signal-to-noise ratio, we opt to use a fully empirical model to describe the spectrum of AXJ. For brevity, we do not detail here the results of modeling AXJ's PV phase spectrum using each methodology. This new characterization of the AXJ line profiles with Resolve will be presented in Matsunaga et al. (in prep.).

\subsubsection{Big M1744 region - empirical modeling}\label{subsub:diffuse_bigpix}

Our empirical diffuse emission fit was performed on two spectra simultaneously: the "big" M1744 region shown in cyan in Fig.~\ref{fig:Resolve_regions_DDT_PV}-right, and the single NS pixel used to estimate the AXJ spectrum, shown in red in the same image. Both sources contribute to both spectra with their respective on-axis and off-axis ARFs. 
\begin{figure*}[t!]
\centering
    \includegraphics[clip,trim=0.5cm 0cm 0.3cm 0.2cm,width=1.0\textwidth]{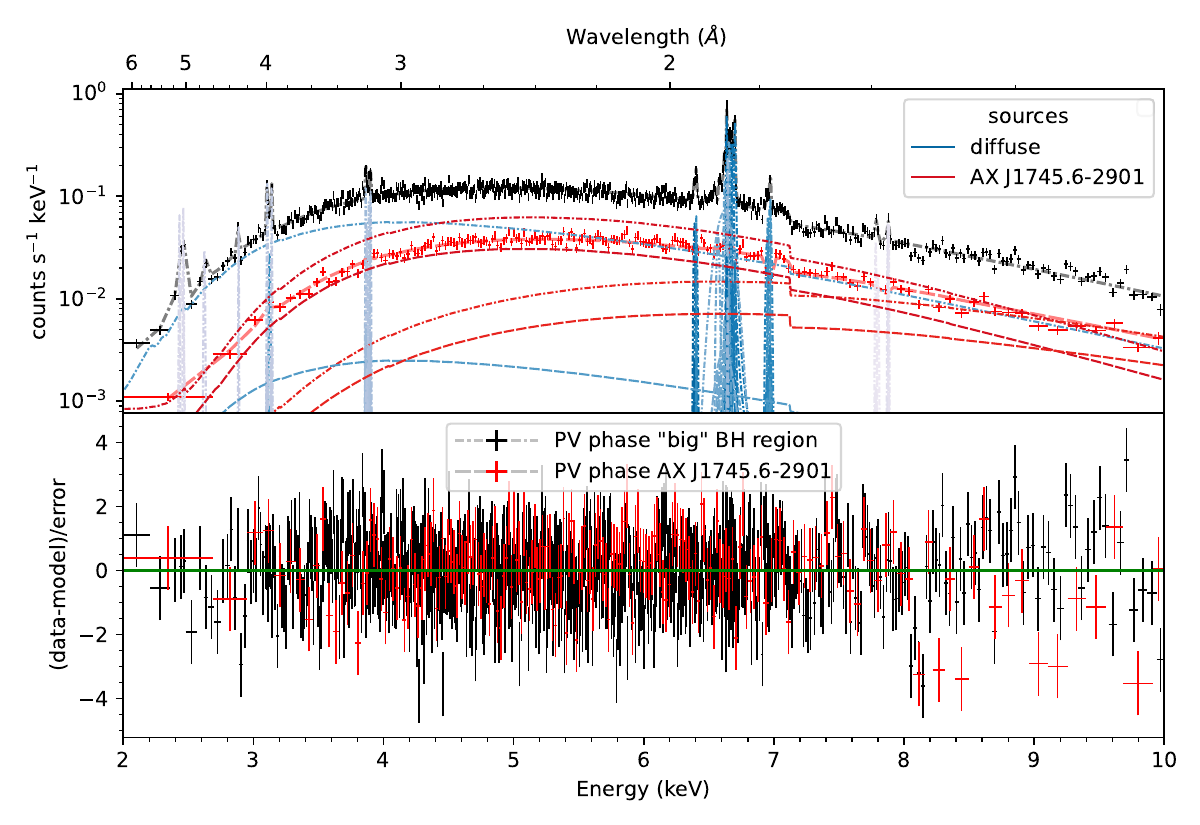}
    \vspace{-2em}
    \caption{$2-10$ keV Resolve spectra and residuals of the "big" MAXI J1744-294 region and single pixel AX J1745.6-2901 region in the PV phase observation, after the empirical diffuse emission modeling described in Section~\ref{subsub:diffuse_bigpix}. Both spectra were rebinned at a 10$\sigma$ significance level for visibility, and individual components at 3$\sigma$. }
    \label{fig:PV_bigpix_resid_2-10}
\end{figure*}

For the diffuse emission, we found that a single absorbed powerlaw model (\texttt{TBabs(powerlaw)} in XSPEC format) was sufficient to reproduce the spectral shape of the continuum. For AXJ, which was in the soft state during that observation \citep{Tanaka2026_AXJ}, the continuum was modeled with an absorbed combination of a multicolor thermal disk and blackbody component (\texttt{TBabs(diskbb+bbody)} in XSPEC format). The AXJ absorption component was fixed at a column density of $\mathrm{N}_H=3\times10^{23}$ cm$^{-2}$ following the values derived in \cite{Ponti2018_AXJ1745.6-2901}. Although we confirmed that thawing this component did not lead to any significant change in either the fit statistic or parameter value, our measurement may be region-dependent due to the dust scattering halo, and we explored an alternative value when modeling the "small" M1744 region. In addition, due to limited spectral constraints in the vicinity of the PV phase \xrism\ observation, we fixed the (poorly constrained) temperature of the blackbody component to 3 keV. This matches, at first order, the values derived for the soft-state observations in \cite{Ponti2018_AXJ1745.6-2901}, and the value obtained in 2025 in the spectra around the DDT period.

After an initial common fit to both sources, with all noticeable line complexes ignored, we froze the continuum and added emission/absorption lines to the diffuse emission/AXJ spectrum until no strong line residuals remained in either. To help the fitting process, each added line/line complex was initially fitted in a narrow energy range around its rest energy. Simple \texttt{gaussians} were found sufficient to reproduce the line shapes, except for the four main transitions of the \Fexxv{} K$\alpha$ complex in the diffuse emission, for which \texttt{Lorentzians} provided a much better fit ($\Delta C>100$). The individual parameters of the different components of each complex (velocity shift, widths) were linked, unless thawing them led to a significant (99\% at F-test) improvement in the fit.  
For the \Fexxv{} K$\alpha$ complex, which results from a complex combination of several ionized plasmas at different temperatures, and includes a significant contribution from many satellite lines, we found that a small number of emission lines reproduced the spectral shape very accurately if combined with two absorption lines at 6.565 and 6.707 keV, respectively. We stress that these two lines currently have no physical interpretation, and a satisfactory fit was reached in both the "small" M1744 region and \cite{XrismCol2025_GC_obs_diffuse} with pure physical emission models. Zoomed residuals of the diffuse emission spectrum over the 6.3-7.2 keV band are shown in the left panel of Fig. B.\ref{fig:PV_diffuse_resid_zoom} in App.~\ref{app:diffuse_Resolve}, and confirm a very good fit in this energy range, which hosts the main Fe transitions. 

After fitting all the lines independently, we froze the entire emission/absorption component set to refit the continuum, this time over the full energy range of the spectrum, before refitting the different line complexes for consistency. 
The final residuals of the two spectra over the 2-10 keV band are shown in Fig.~\ref{fig:PV_bigpix_resid_2-10}, for a global C-statistic/d.o.f. of 33214/31850; The only remaining feature is an absorption line between 8.05 and 8.2 keV in the diffuse emission spectrum. This feature is unexpectedly stronger than the corresponding absorption line in the AXJ spectrum, as expected, but not significant enough to warrant adding an additional line component. As we do not expect intrinsic absorption components in the diffuse emission, we choose not to include this feature in our model.
As a sanity check, after deriving the best fit models for both sources, we tested the presence of significant residuals with blind searches for narrow line features in both the diffuse emission and AXJ spectra individually, following the methodology described in \cite{Parra2024_winds_global_BHLMXBs}. For this, we computed 2D $\Delta$C-stat maps resulting from the addition of a narrow (5 eV width) Gaussian line, whose energy was incremented by steps of 5 eV, and normalization between $10^{-2}$ and 30 times the local continuum flux in each energy step, for both positive and negative normalizations. Here, we focused on the 6.3-7.1 keV range, where the absorption and emission components of the diffuse emission and AXJ are most prominent. We show the results of the blind searches for both spectra in the left panels of Fig. B.\ref{fig:blind_search_PV_Resolve} in App.~\ref{app:diffuse_Resolve}. No significant residuals remained for the entire band, confirming that our model reproduced the spectra well over this energy range.
The full list of parameters for the empirical components (including 34 individual lines) used to represent the diffuse emission spectrum is provided in Table B.~\ref{tab:comp_param_diffuse_bigpix_PV} of App.~\ref{app:diffuse_Resolve}. The detailed empirical modeling of the NS components will be described in future works.  

\subsubsection{Small M1744 region - physical modeling}\label{subsub:diffuse_smallpix}

\begin{figure*}[t!]
\centering
    \includegraphics[clip,trim=0.5cm 0cm 0.3cm 0.2cm,width=1.0\textwidth]{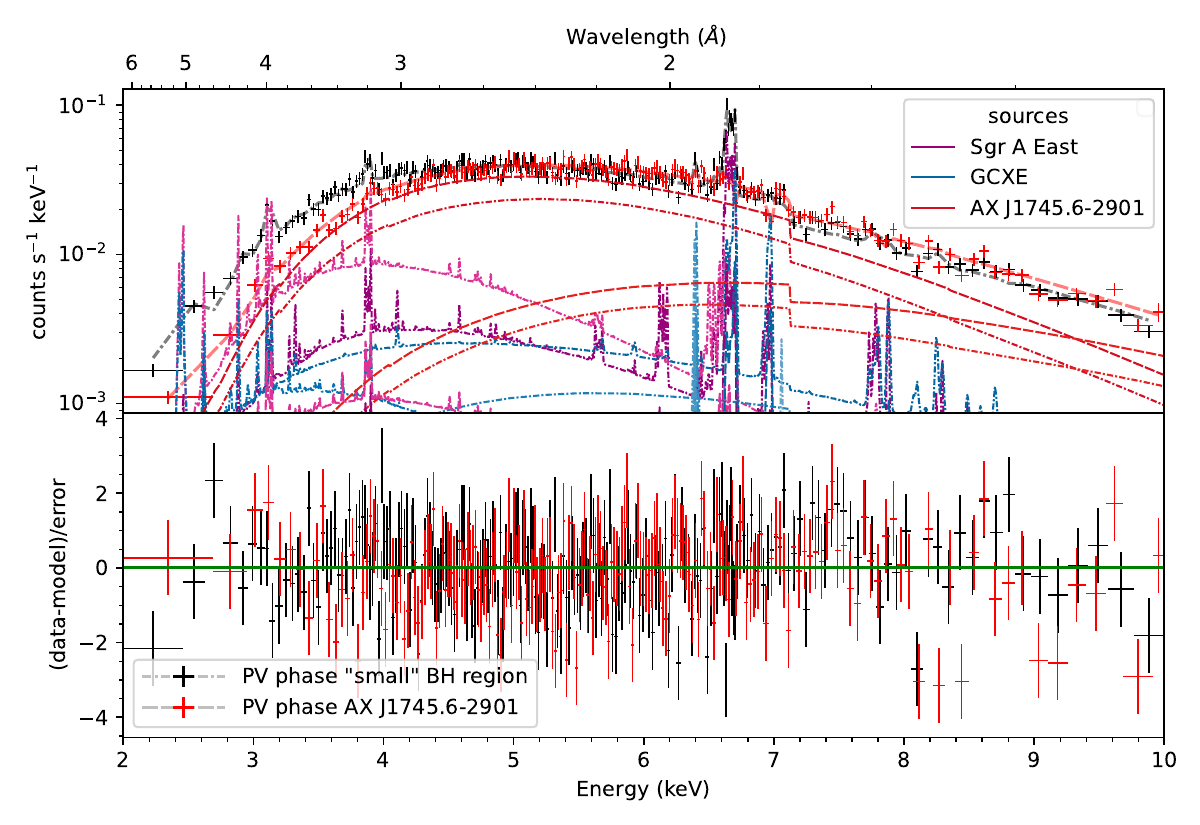}
    \vspace{-2.5em}
    \caption{$2-10$ keV Resolve spectra and residuals from the "small" MAXI J1744-294 region and single pixel AX J1745.6-2901 region in the PV phase observation, after the physical diffuse emission modeling described in Section~\ref{subsub:diffuse_smallpix}. Both spectra were rebinned at a 10$\sigma$ significance level for visibility, and individual model components at 3$\sigma$.}
    \label{fig:PV_smallpix_resid_2-10}
\end{figure*}

Our physical diffuse emission fit was performed on two spectra simultaneously: the "small" M1744 region shown in green in Fig.~\ref{fig:Resolve_regions_DDT_PV}-right, and the single AXJ pixel shown in red in the same image. We adapted the methodology presented in \cite{XrismCol2025_GC_obs_diffuse}, which has since been refined for a study of other line features (Uchiyama et al. in prep), to our objectives.
In \cite{XrismCol2025_GC_obs_diffuse}, three sources contributed to each spectrum: Sgr A East, the GCXE, and AXJ, each with its own ARF distribution, derived with the same methods as in our study. The NS model was imported from the photoionization analysis of \cite{Tanaka2026_AXJ}, and the GCXE model was imported from an independent fit of a third \xrism\ observation of a neighboring Galactic center region (ObsID 300045010), with no contribution from other point sources. The model included an absorbed power-law continuum, along with two recombining CIE plasma components (\texttt{bvvrnei} in XSPEC format) at high and low temperatures to represent the main ionized lines. The highly ionized Fe lines were modeled empirically using Lorentzians by removing the emissivities of iron from the atomic table. The neutral lines from Fe and Ni were also described using Lorentzians. Although the same spectral model was reused to describe the GCXE contribution in the PV phase FoV, the GCXE surface brightness (normalization) was readjusted using a common fit of Sgr A East and the GCXE to the northernmost 2 pixel rows (12 pixels) in the PV phase observation. Sgr A East itself was modeled in a similar way (\texttt{TBabs(powerlaw+bvvrnei+bvvrnei)} + several Lorentzians), and its main source region used the southwest 4x4 pixel square in the PV phase observation. 

\begin{figure*}[t!]
\centering
    \includegraphics[clip,trim=0.5cm 0cm 0.3cm 0.2cm,width=1.0\textwidth]{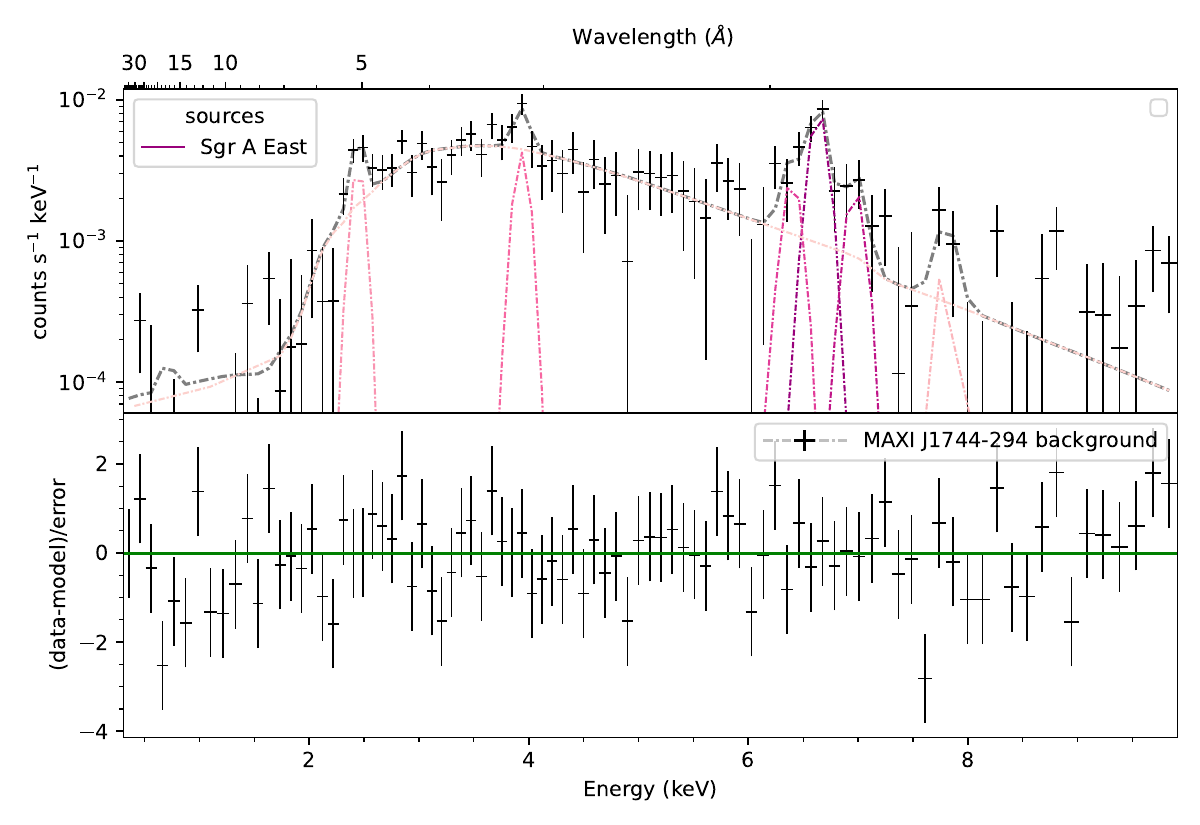}
    \vspace{-2.5em}
    \caption{$0.3-10$ keV Xtend spectrum and residuals of the background-subtracted \src\ region in the PV phase observation, following the empirical diffuse emission modeling described in Section~\ref{sub:diffuse_xtend}. The spectrum is rebinned using the optimized scheme of \cite{Kaastra2016_binning_opt}.}
    \label{fig:xtend_diffuse_resid_03-10}
\end{figure*}

Our own approach includes a few modifications to the methodology described above. First, we modeled AXJ self-consistently with diffuse emission and replaced its static background with contributions from off-axis ARFs of Sgr A East and the GCXE. Secondly, our diffuse emission regions differed from those modeled above to tailor our description to the study of M1744. Thirdly, the first diffuse emission study focused on energies above 5 keV, while we now model the entire 2-10 keV band. Finally, we now include the emissivities of iron within the \texttt{bvvrnei} of each model, for a fully physical description.
For these reasons, the only source we directly imported and froze in our fit is the GCXE. For the AXJ model, we used the same absorbed continuum and significant absorption lines as in the "big" M1744 region fit. Sgr A East was refitted from the best-fit model described above, with the main spectral parameters left free to vary to account for potential differences in spectral brightness and shape across pixels and to correctly fit the low-energy lines.

With this methodology, one of the main differences compared to the previous empirical modeling of the "big" M1744 region is that thawing the AXJ absorption component yields a column density of $\mathrm{N}_H=2.47\times10^{23}$ cm$^{-2}$, with an improvement of $\Delta C>200$ compared to the initial value of 3$\times10^{23}$ cm$^{-2}$. This may trace the effect of the dust scattering halo, although it generally hardens the spectra taken from incomplete angular regions \citep{Jin2017_DSC}, or hint at an additional diffuse soft X-ray component, which would not have been necessary in previous physical fits of the diffuse emission, excluding the soft X-ray band of \xrism. We stress that the assumed value for N$_H$ only affects the continuum modeling of AXJ and the diffuse emission, and will thus have a negligible impact on the analysis of the line features in each point source. The final residuals of the two spectra over the 2-10 keV band are shown in Fig.~\ref{fig:PV_smallpix_resid_2-10}, for a global C-statistic/d.o.f. of 32379/31970; Similarly to the empirical modeling of the "big" M1744 region, weak residuals remain at high energies, hinting at additional absorption lines expected from higher order transitions in the AXJ spectrum. Zoomed-in residuals of the diffuse emission spectrum over the 6.3-7.2 keV band are shown in the right panel of Fig. A.\ref{fig:PV_diffuse_resid_zoom} in App.~\ref{fig:PV_diffuse_resid_zoom}, and confirm a satisfactory spectral fitting in this energy band.

Once again, we assessed the presence of significant residual features with blind searches. We used the same procedure as in the previous analysis (Section \ref{subsub:diffuse_bigpix}) and applied it to the diffuse emission and AXJ spectra independently. The results, which we show in App.~\ref{app:diffuse_Resolve} in the right panels of Fig. B.\ref{fig:blind_search_PV_Resolve} for the critical 6.3-7.1 keV band, confirm that no additional component should be introduced to the fit in this energy range. The full description of the two diffuse emission models and their parameters is provided in Table B.\ref{tab:comp_param_diffuse_smallpix_PV} of App.~\ref{app:diffuse_Resolve}. The detailed empirical modeling of the AXJ components will be described in future works.

\subsection{Xtend}\label{sub:diffuse_xtend}

For Xtend, our empirical Sgr A East diffuse emission fit was performed on a single background-subtracted spectrum, using the regions derived in Section~\ref{subsub:reg_xtend}. The model includes a single diffuse source, whose ARF was computed using the \chandra\ flux map of the Sgr A East SNR.  

The diffuse emission was well modeled with a single absorbed \texttt{powerlaw} continuum and six \texttt{gaussian} emission lines for the different (unresolved) complexes of the most abundant elements, following the main lines already seen in the Resolve spectrum (see Section~\ref{subsub:diffuse_smallpix}). 
As all lines remain unresolved at CCD resolution, we fixed their widths to 0 and their rest energy to the averaged values from Resolve, except for the \Fexxv{} K$\alpha$ complex, due to the high number of lines blending at its energy.
The final residuals of the two spectra in the 0.3-10 keV band are shown in Fig.~\ref{fig:xtend_diffuse_resid_03-10}, for a final C-statistic/d.o.f. of 86/79; the shape of the continuum is only weakly constrained, particularly below 2 keV and above 7 keV due to insufficient statistics at lower and higher energies.  

\begin{figure*}[t!]
\centering
    \includegraphics[clip,trim=0.5cm 0cm 0.3cm 0.2cm,width=1.0\textwidth]{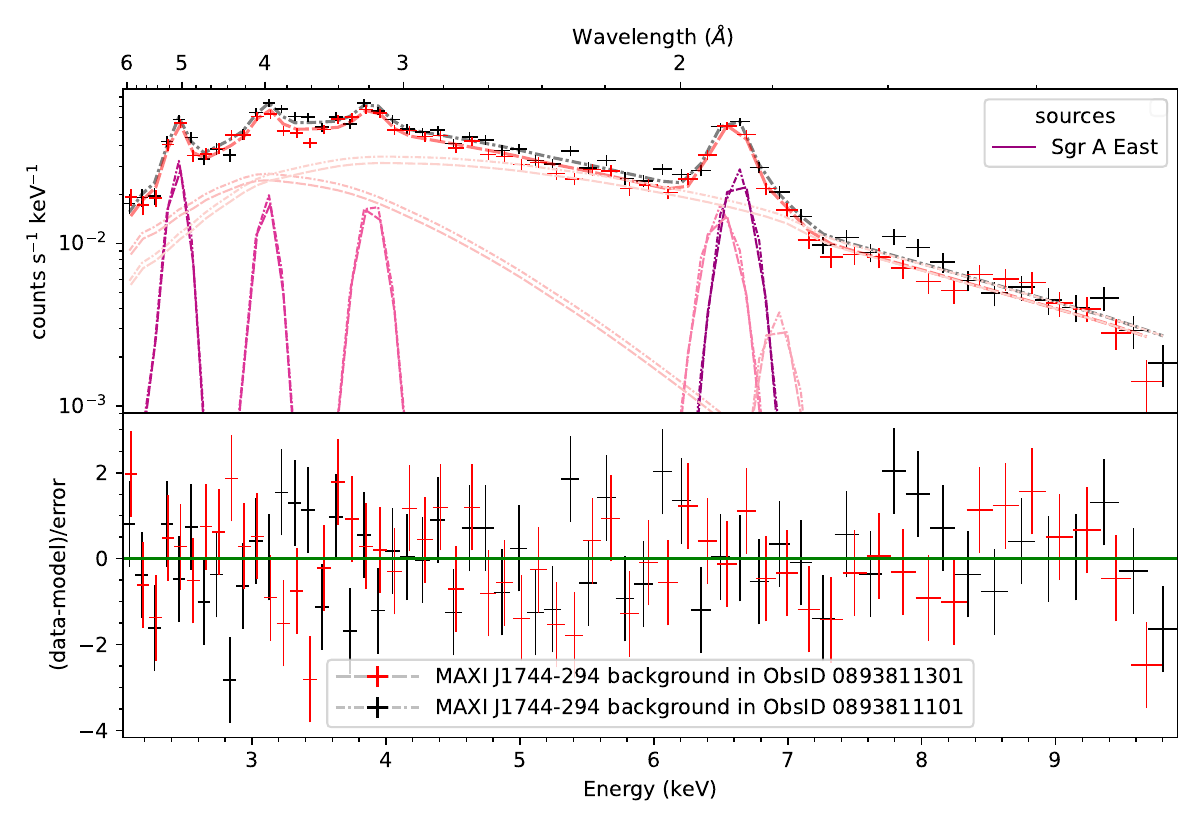}
    \vspace{-2.5em}
    \caption{$0.3-10$ keV \xmm\ spectra and residuals of the background-subtracted MAXI J1744-294 
    region
    , following the empirical diffuse emission modeling described in Section~\ref{sub:diffuse_xmm}. The spectra are rebinned using the optimized scheme of \cite{Kaastra2016_binning_opt}.
    }
    \label{fig:pn_diffuse_resid_03-10}
\end{figure*}

Considering the limited spectral resolution of Xtend and the fact that our model was adapted from the well-constrained Resolve line features, we do not perform blind searches for this spectrum. The full description of the empirical diffuse emission model and its parameters is provided in Table B.\ref{tab:comp_param_diffuse_xtend_PV} of App.~\ref{app:diffuse_CCD}.

\subsection{\xmm}\label{sub:diffuse_xmm}

For \xmm, our empirical Sgr A East background modeling was performed on 2 background-subtracted spectra from archival data, using the region selection detailed in Section~\ref{subsec:reg_xmm}. Aside from a constant (cross-normalization) factor, the two spectra were modeled using the same diffuse source model, and in both cases, an ARF computed using the \chandra\ flux map of the Sgr A East SNR.  Owing to the differences in signal-to-noise ratio, source regions, and angular resolution compared to Xtend, we used a different model to describe the diffuse emission. The main difference was the requirement of a two-component absorbed continuum model, including both a \texttt{powerlaw} and \texttt{diskbb}. Furthermore, while a new emission line was clearly required around 3.1 keV, there was no \Nixxvii{} K$\alpha$ / \Fexxv{} K$\beta$ line around 7.8keV, and we thus ended up with a total of 6 Gaussians at 2.45, 3.1, 3.9, 6.3, 6.6, and 6.9 keV, respectively. Unlike with Xtend, we did not impose strong restrictions on their energies, owing to the known energy calibration limitations of \xmm\ (see e.g. \citealt{Parra2024_winds_global_BHLMXBs}). 

The model provided a very good fit to the data above 2 keV, but very high residuals remained below that threshold, likely due to the complex interstellar absorption in this region. We thus restricted the fit to the $2-10$ keV band in order to avoid these residuals biasing our model estimates. We derived a final best fit with a C-statistic/d.o.f. of 123/89, with negligible residuals, which did not warrant a blind search due to the low resolution of the instrument. The full description of the empirical diffuse emission model and its parameter values is provided in Table B.\ref{tab:comp_param_diffuse_XMM_PV} of App.~\ref{app:diffuse_CCD}.

Due to the limited spectral resolution of \xmm, and to facilitate the systematic fitting of a great number of observations, for this instrument only, we created a fake background spectrum using the  \texttt{fakeit} command in XSPEC, assuming the average of the constant factor of both archival spectra, appropriate RMFs and ARFs computed from the 2025 \xmm\ observation, and Poisson noise for an exposure of 100 ks. We followed the same procedure described in M26. This "fake" background was then used as a standard static background file for the spectral analysis in M26 and in Paper II.

\section{Discussion and conclusions} \label{sec:conclu}

Now that the diffuse emission models have been derived for the relevant instruments, and having presented the 2025 observations and their data reduction, the next step is to focus on the spectral analysis of the sources in the field of view -- namely, \src\ and \axj. In Parra et al. (submitted to ApJ, Paper II), the direct follow-up to this study, we present a detailed analysis of the intrinsic line features of \src{}. The high spectral resolution of \xrism{}, along with the broadband coverage of \xmm{} and \nustar{}, uncovers a set of intrinsic highly ionized Fe emission lines, along with several narrow emission features at atypical energies above 6.7 keV -- interpretable as highly ionized Fe lines with outflow velocities of several thousand km s$^{-1}$ -- and a weak neutral Fe K$\alpha$ line feature. As static and blueshifted emission layers are both highly unusual in non-obscured soft state BH-LMXBs, we model and discuss different and equally viable scenarios, such as photoionized plasma layers from a wind seen at low inclination, and collisional plasma from relativistic ejections in a jet, comparable to the microquasar SS 433. In addition,  detailing the contributions of the different sources involved in our simultaneous fitting of \src{} and \axj{} in Paper II 
allows us to correct the analysis and interpretation of \cite{spreadingmisinformationontheinternet}, who used the entire \xrism{} Resolve field of view simultaneously, and thus obtained a spectrum that blends \src{}, \axj{}, and multiple sources of diffuse emission -- obscuring the properties of all.

For now, this paper provides the technical foundation for a series of studies focusing on the high-resolution spectral features seen in a \xrism\ observation of the Galactic center region, conducted in March 2025 as Director's Discretionary Time. This observation, along with its simultaneous \xmm\ and \nustar\ coverage, fits within a large multi-wavelength program which targeted the 2025 outburst of the black hole X-ray binary candidate \src/Swift J174540.2-290037 \citep{Mandel2026_ApJ}, one of the closest compact objects to the supermassive black hole Sgr A*, with an angular separation of 18\arcsec.

The aim of our study was to disentangle the contributions of the two bright point sources in the field of view, \src\ and the neutron star AX J1745.6-2901, from the blend of unresolved background sources and diffuse emission. In Section~\ref{sec:reg_choice}, we combined complementary region selection methodologies, which maximize the signal-to-noise ratio of each point source ("big" region) or minimize the proportion of spatial-spectral mixing ("small" region), using optimal archival observations for each instrument. Considering the high degree of complexity of the observations, in Section~\ref{sec:datared}, we carefully reduced the data of both \xrism\ instruments using the latest tools available. The \xmm\ and \nustar\ observations, which complement the continuum modeling and can be corrected with respect to the influence of a dust scattering halo, are presented with details in \cite{Mandel2026_ApJ}, along with their data reduction. In addition, we performed precise simulations of the PSF fractions of the different sources in the \xrism-Resolve field of view, using the spatial distributions estimated from high-angular-resolution instruments such as \chandra, and comprehensive computations of the systematics introduced by the latest changes in the \xrism\ data analysis tasks and calibration. Then, in Section~\ref{sec:diffuse_mod}, we modeled the diffuse emission for our chosen background regions in archival \xrism\ and \xmm\ observations, simultaneously with AX J1745.6-2901 in the Resolve data. To ensure the reliability of our results, we combined an empirical line-by-line description of the diffuse emission components and physical modeling using overionized plasma models, expanding on the methodology previously used to study the SNR Sgr A East -- one of the main sources of diffuse emission overlapping \src\ -- in \cite{XrismCol2025_GC_obs_diffuse}. These results provide the baseline for subsequent individual studies of the high-resolution spectral features of \src\ (Parra et al. submitted to ApJ, Paper II), AX J1745.6-2901 (Matsunaga et al., in prep.), and the interstellar medium along the line of sight (Gatuzz et al. submitted to A\&A, Paper III). These works will complement the comprehensive study of the \src\ outburst presented in \cite{Mandel2026_ApJ}. 

Our effort to study the Galactic center perfectly highlights the challenges of the \xrism\ analysis of crowded fields of view, for which complementing high spectral resolution and high angular resolution instruments is essential to study multiple individual science cases in a restricted angular region. This benchmark of \xrism's abilities to disentangle precise, narrow spectral features from a highly complex blend of overlapping emitters and absorbers will prove very helpful for future observations of the Galactic center. Finally, our complementary studies showcase some of the many unexpected scientific outputs that can result from exploratory \xrism\ observations, and thus support further dedication of \xrism{} Director's Discretionary Time to unexpected and observationally challenging targets.

\section*{Data availability}
All reduced datasets, spectral models tables, and combined spectral fits will be made available upon reasonable request to the authors.

\begin{acknowledgments}

We thank the XRISM operation team for accepting our DDT proposal and conducting the observation, along with the XRISM Science Data Center, help desk, and calibration teams for their continued assistance. MP acknowledges support from the JSPS Postdoctoral Fellowship for Research in Japan, grant number P24712, as well as the JSPS Grants-in-Aid for Scientific Research-KAKENHI, grant number J24KF0244. Support for SM, KM and the Columbia University team was provided by \nustar\ AO-10 (80NSSC26K0286), \nustar\ AO-11 (80NSSC26K0154), \chandra\ AO-26 (SAO GO5-26016X) and \xmm\ AO-23  (80NSSC25K0651) programs. 
SM acknowledges support by the National Science Foundation Graduate Research Fellowship under Grant No. DGE 2036197 and the Columbia University Provost Fellows Program.  
Part of this work was financially supported by Grants-in-Aid for Scientific Research 19K14762, 23K03459, 24H01812 (MS) from the Ministry of Education, Culture, Sports, Science and Technology (MEXT) of Japan. 
          This research has made use of software provided by the High Energy Astrophysics Science Archive Research 
Center (HEASARC), which is a service of the Astrophysics Science Division at NASA/GSFC. 
          This research has made use of the
    NuSTAR Data Analysis Software (NuSTARDAS) jointly developed by the ASI Space Science
    Data Center (SSDC, Italy) and the California Institute of Technology (Caltech, USA). TY acknowledges support by NASA under award number 80GSFC24M0006.
\end{acknowledgments}




%
\facilities{\xrism{} (Resolve and Xtend), \xmm{}(EPIC), \nustar{}}


\bibliography{ref,bibliocompl}{}
\bibliographystyle{aasjournalv7}

\appendix
\vspace{-1em}

\section{Supplementary information on the observations}\label{app:obs_compl}
\subsection{Resolve}

\subsubsection{Branching ratios}\label{app:Resolve_branch}
\begin{figure*}[h!]
    \centering
    \vspace{-1em}
    \includegraphics[clip,trim=0cm 0cm 0cm 1cm,width=0.49\textwidth]{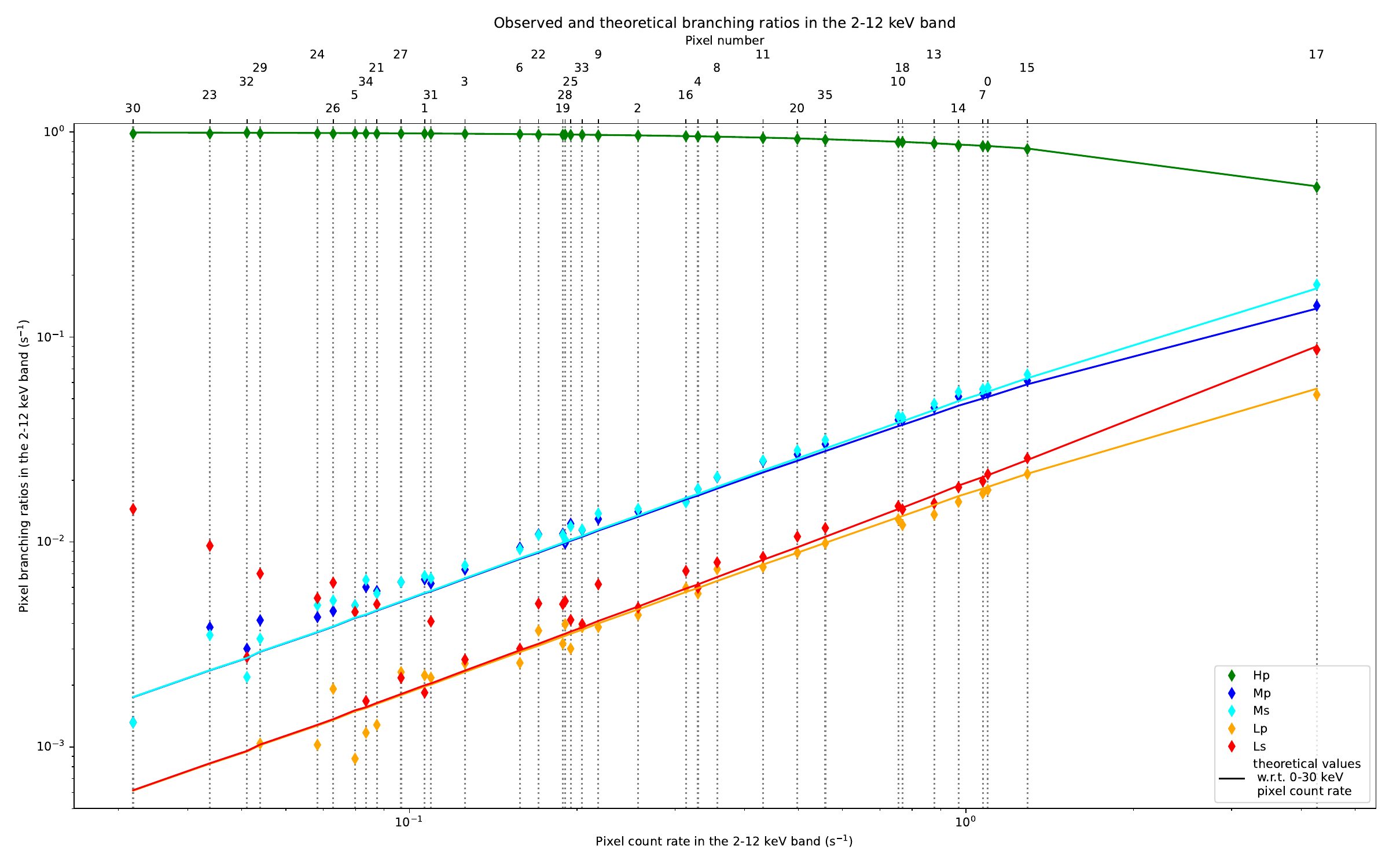}
    \includegraphics[clip,trim=0cm 0cm 0cm 0.95cm,width=0.49\textwidth]{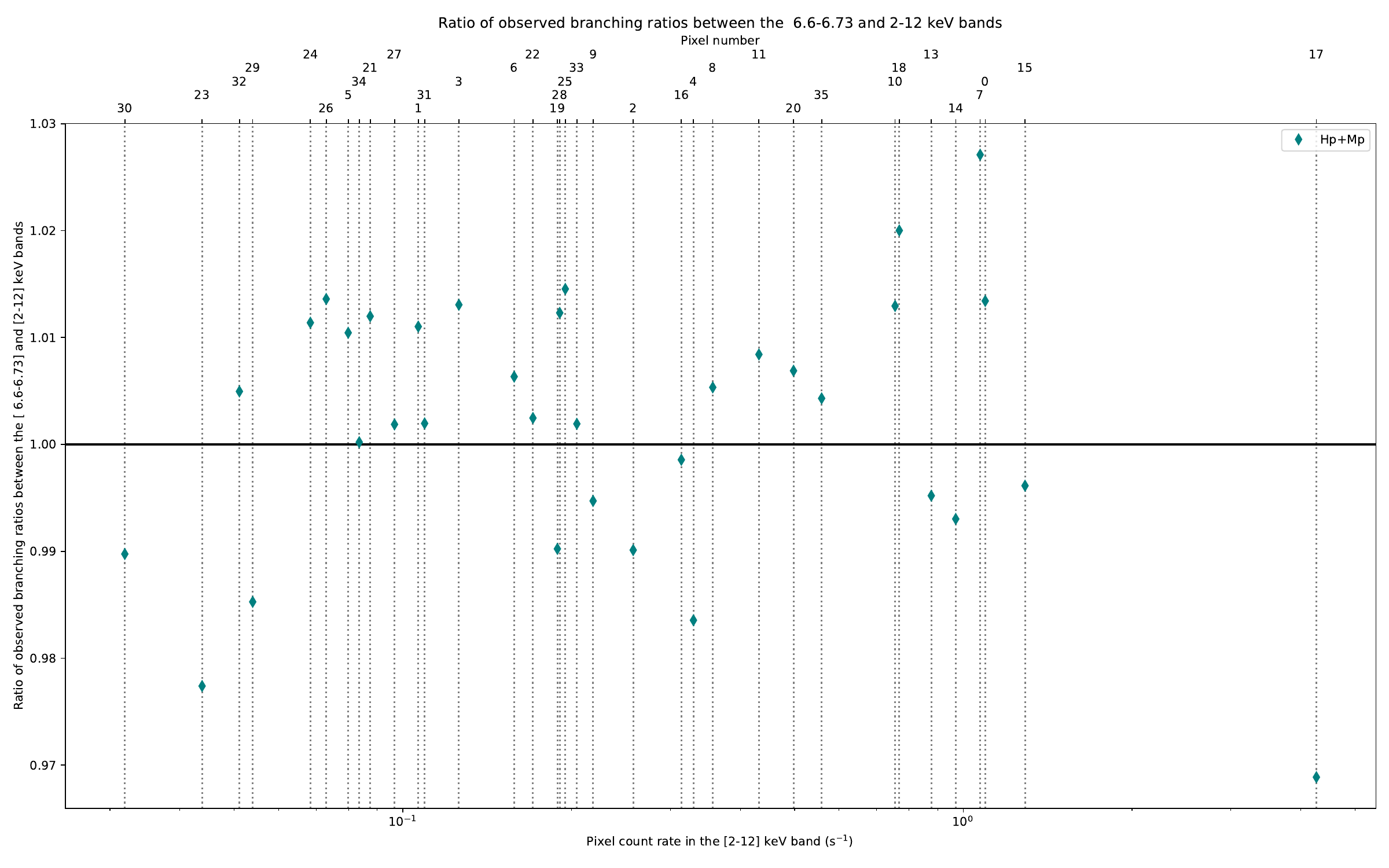}
    \vspace{-1em}
    \caption{\textbf{Left:} Branching ratios of each Resolve pixel in the $2-12$ keV band for the ToO observation after proximity and rise-time event cut (RTS) filtering, identified by count rate (bottom horizontal axis) and pixel number (top horizontal axis). Diamonds indicate real values and lines theoretical predictions at the corresponding count rate.
    \textbf{Right:} Ratio of the Hp+Mp grade branching ratios between the $6.6-6.73$ keV and $2-12$ keV bands for the DDT observation.} 
    \label{fig:resolve_branch}
\end{figure*}

\begin{figure*}[h!]
    \centering
    \includegraphics[clip,trim=0cm 0cm 0cm 0cm,width=0.49\textwidth]{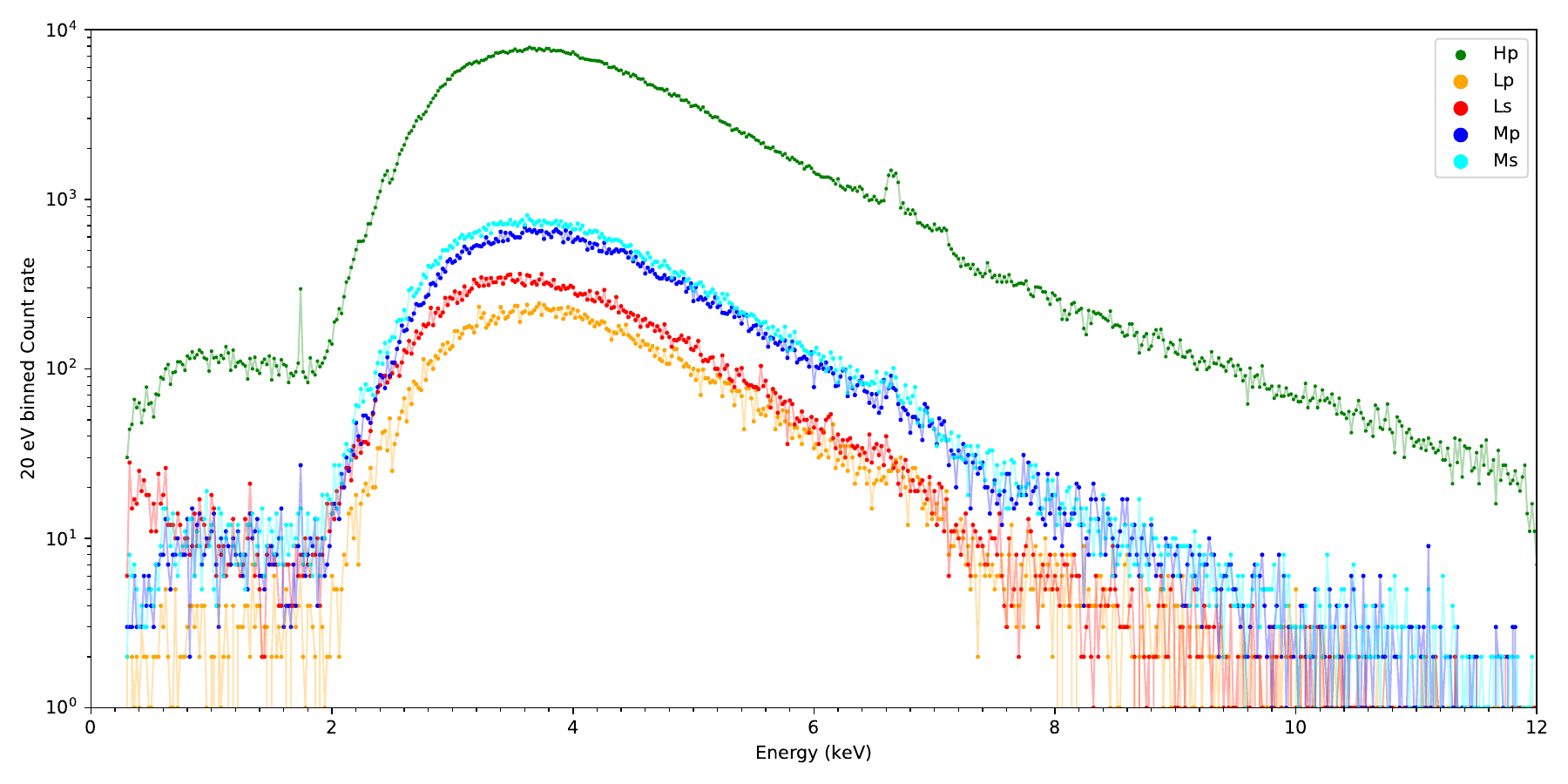}
    \includegraphics[clip,trim=0cm 0cm 0cm 0cm,width=0.49\textwidth]{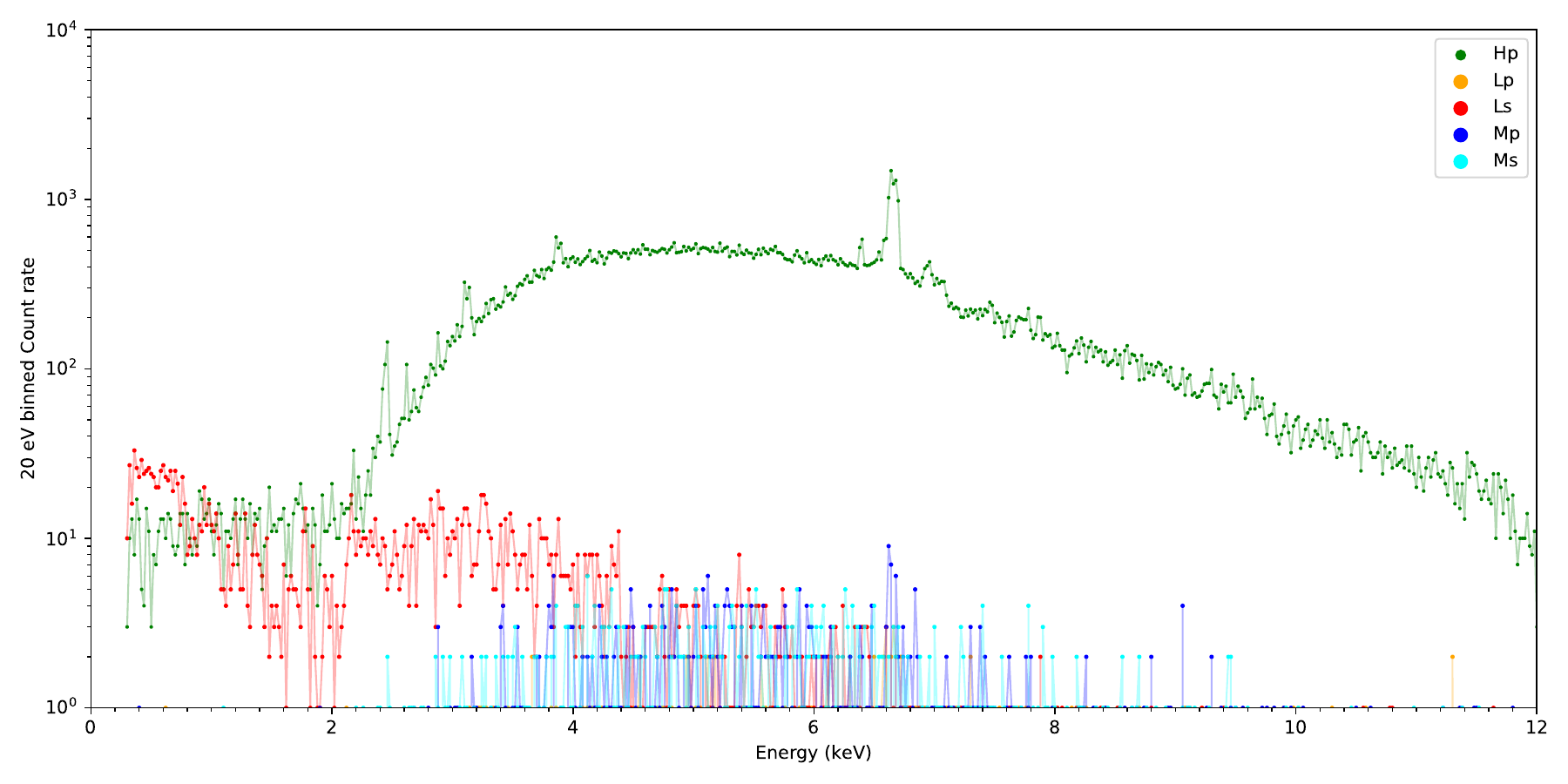}
    \includegraphics[clip,trim=0cm 0cm 0cm 0cm,width=0.49\textwidth]{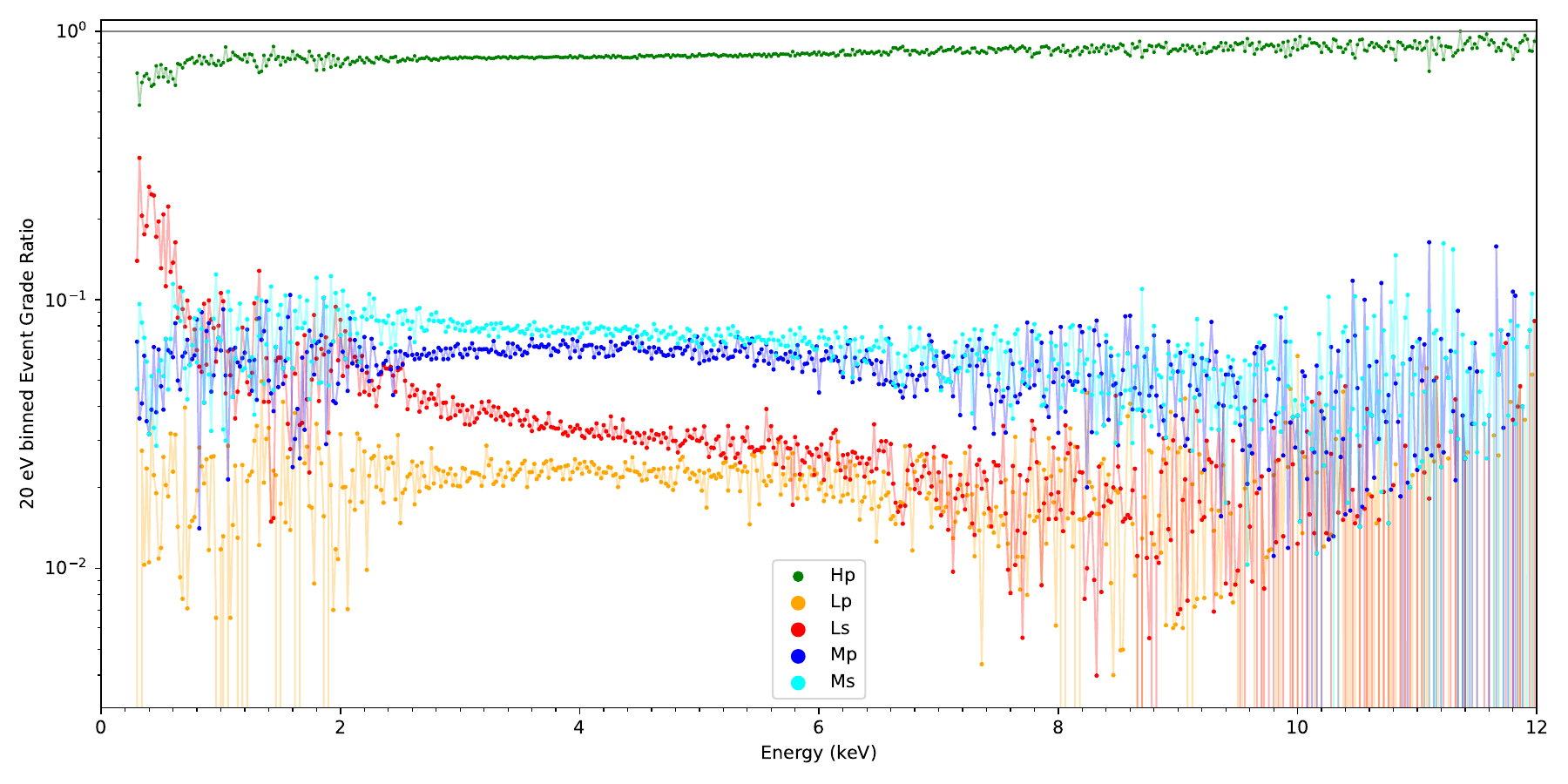}
    \includegraphics[clip,trim=0cm 0cm 0cm 0cm,width=0.49\textwidth]{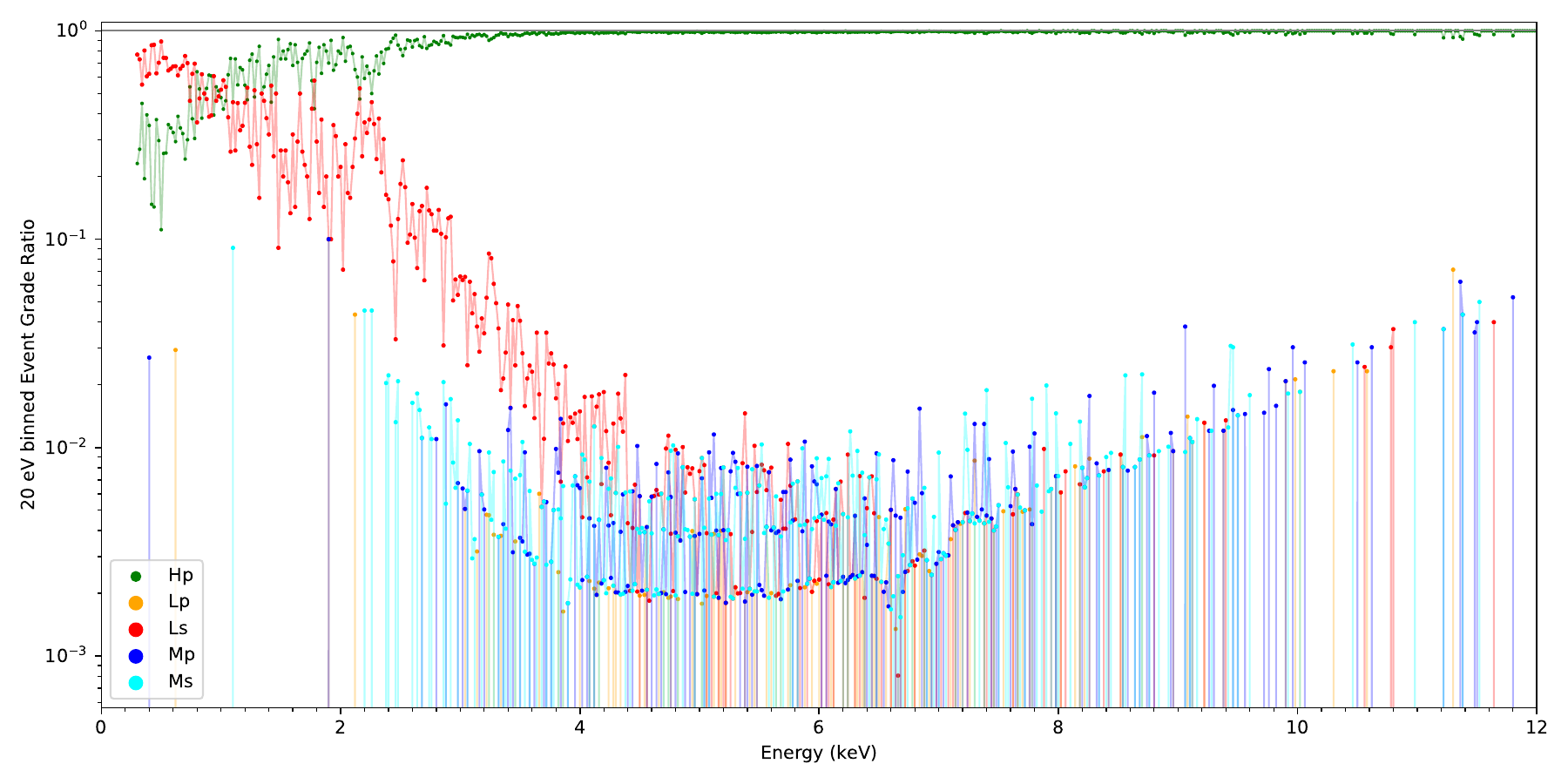}

    \caption{Distribution of Counts \textbf{(top)} and Branching ratios \textbf{(bottom)} with energy using a 20 eV binning, for the entire pixel array, in the 2025 DDT observation dominated by MAXI J1744-294 \textbf{(left)}, and in the 2024 PV phase observation used to compute the background \textbf{(right)}.} 
    \label{fig:resolve_branch_Edep_DDT}
\end{figure*}

\subsubsection{Gain calibration}\label{app:gain_calb_rsl}

The energy scale of the Resolve detector is measured independently for each pixel by fitting a $^{55}$Fe Mn K$\alpha$ line from a calibration source that illuminates the entire detector. This fit is corrected by measuring the evolution of each pixel's temperature at regular intervals throughout the observation. For the PV observation, the results of this calibration are detailed in \cite{XrismCol2025_GC_obs_diffuse}. In the DDT observation, from the modeling of the Mn K$\alpha$ line, we derived a FWHM detector resolution of 4.58$\pm$0.03 eV, and an absolute energy shift of 0.031$\pm0.013$ eV, accurate within $\pm1$ eV over the $2-8$ keV range \citep{porterInflightPerformanceXRISM2024,eckartEnergyGainScale2024}. Data from Pixel 27 were excluded due to gain jumps and a much higher temperature. The detailed energy scale report from the automatic pipeline is available on the HEASARC database\footnote{\href{https://heasarc.gsfc.nasa.gov/FTP/xrism/postlaunch/gainreports/9/901002010\_resolve\_energy\_scale\_report.pdf}{https://heasarc.gsfc.nasa.gov/FTP/xrism/postlaunch/gainreports/9/901002010\_resolve\_energy\_scale\_report.pdf}}.
Typically, the energy scale is calibrated using the gain history file during the standard pipeline processing. However, in this observation, because the calibration source was irradiated while the target was within the field of view, some pixels did not receive a sufficient number of Mn K$\alpha$ photons at the Hp grade. As a result, the default gain history file contained missing values for pixel 17, the brightest pixel in the entire array. To address this, we recalculated the gain correction by relaxing the minimum photon count requirement for Mn K$\alpha$ line fitting and regenerated the cleaned event file using the updated gain information.

\subsubsection{Effect of the CALDB version on the effective areas}\label{app:caldb_influence}

\begin{figure*}[h!]
\centering
    \includegraphics[clip,width=0.31\textwidth]{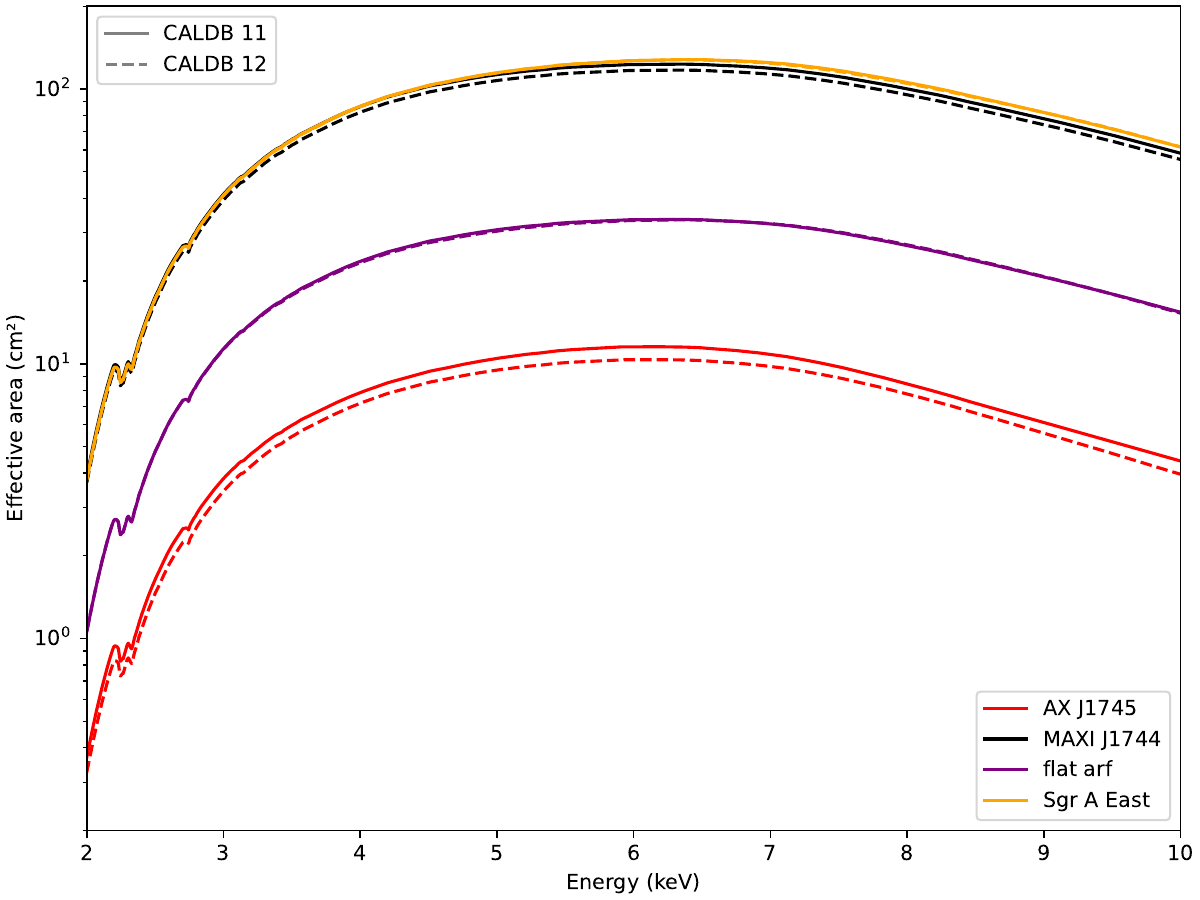}
    \includegraphics[clip,width=0.31\textwidth]{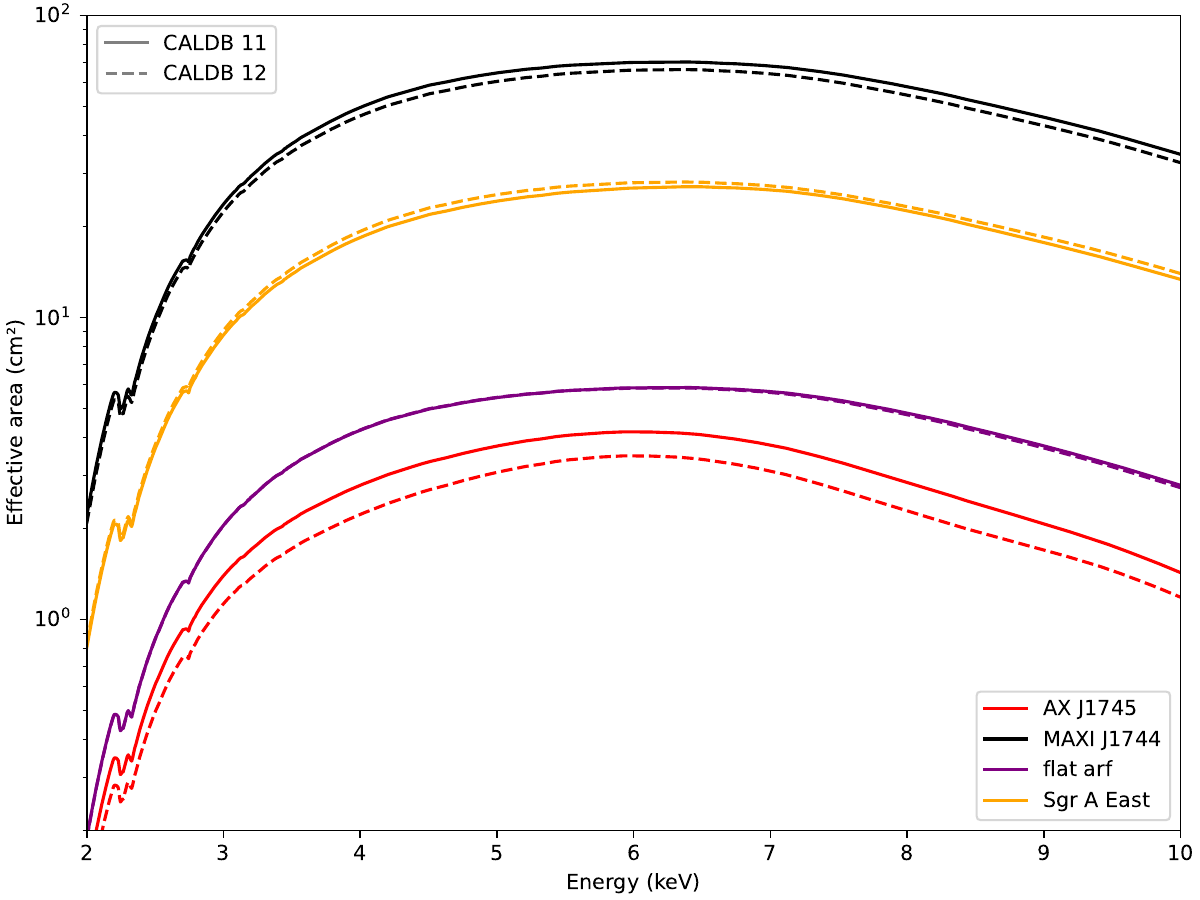}
    \includegraphics[clip,width=0.31\textwidth]{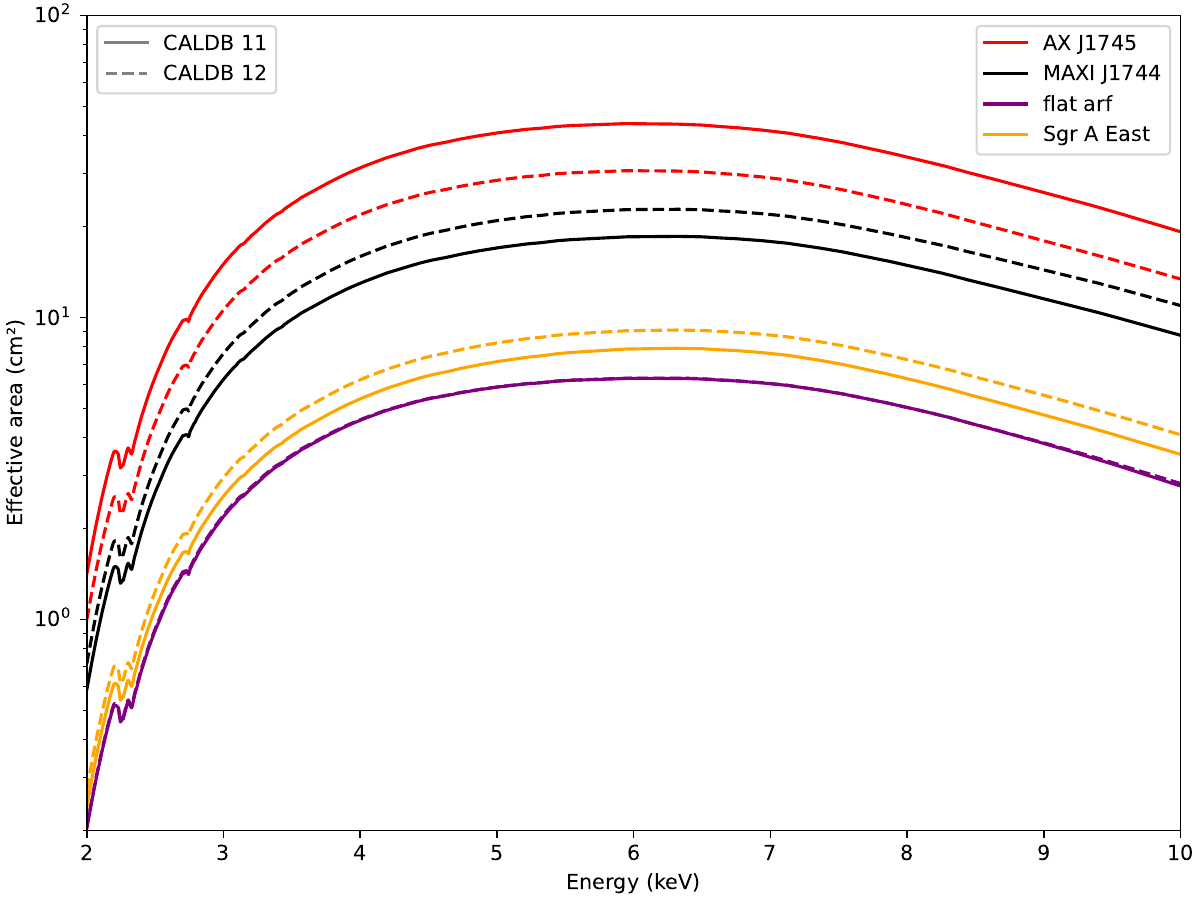}
    \includegraphics[clip,width=0.31\textwidth]{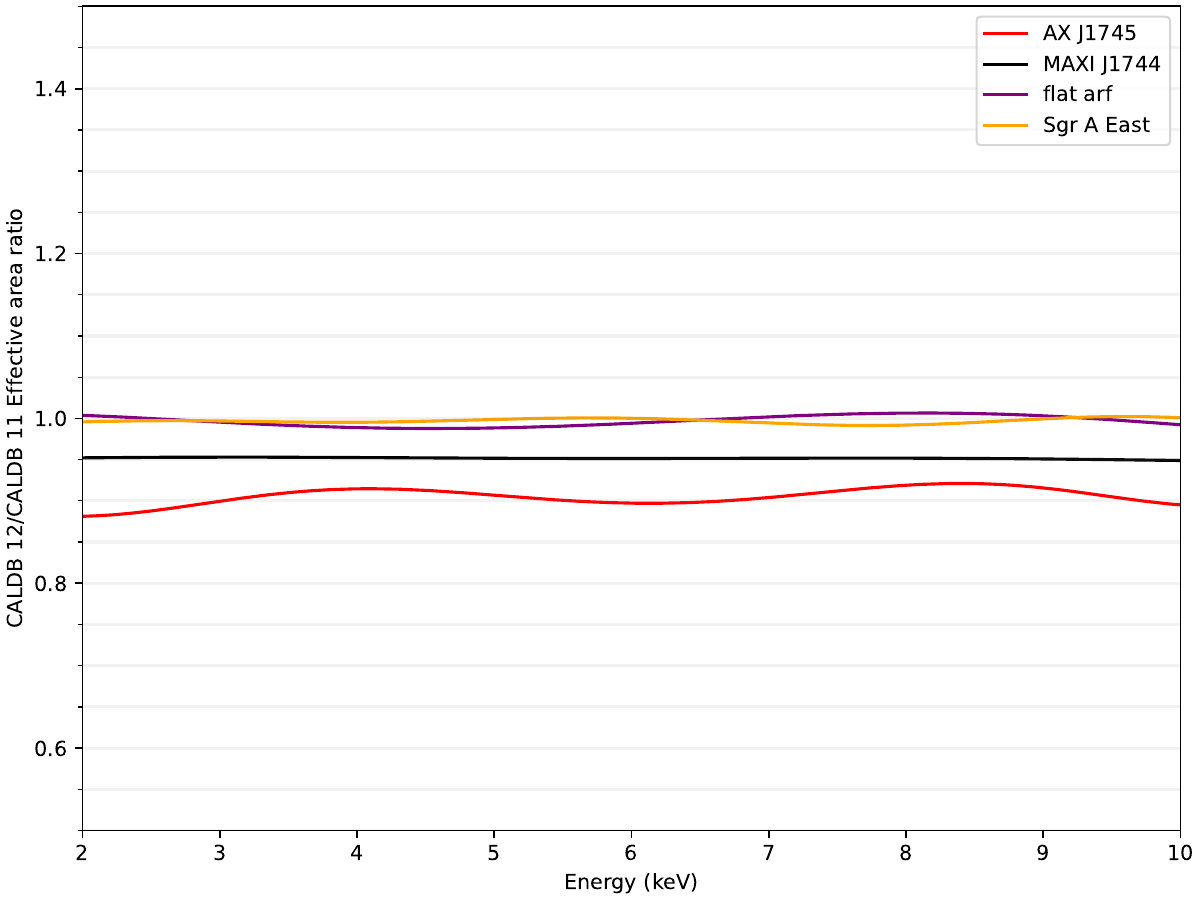}
    \includegraphics[clip,width=0.31\textwidth]{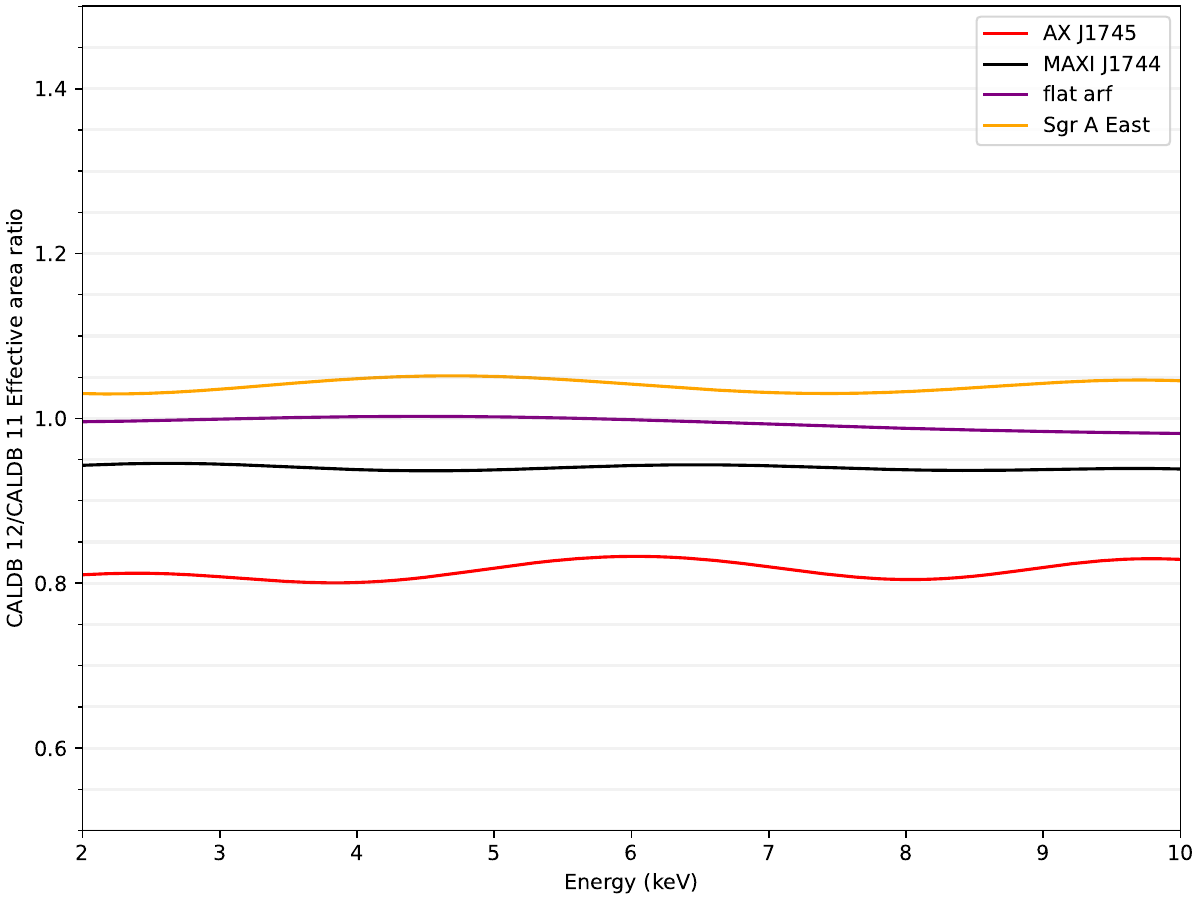}
    \includegraphics[clip,width=0.31\textwidth]{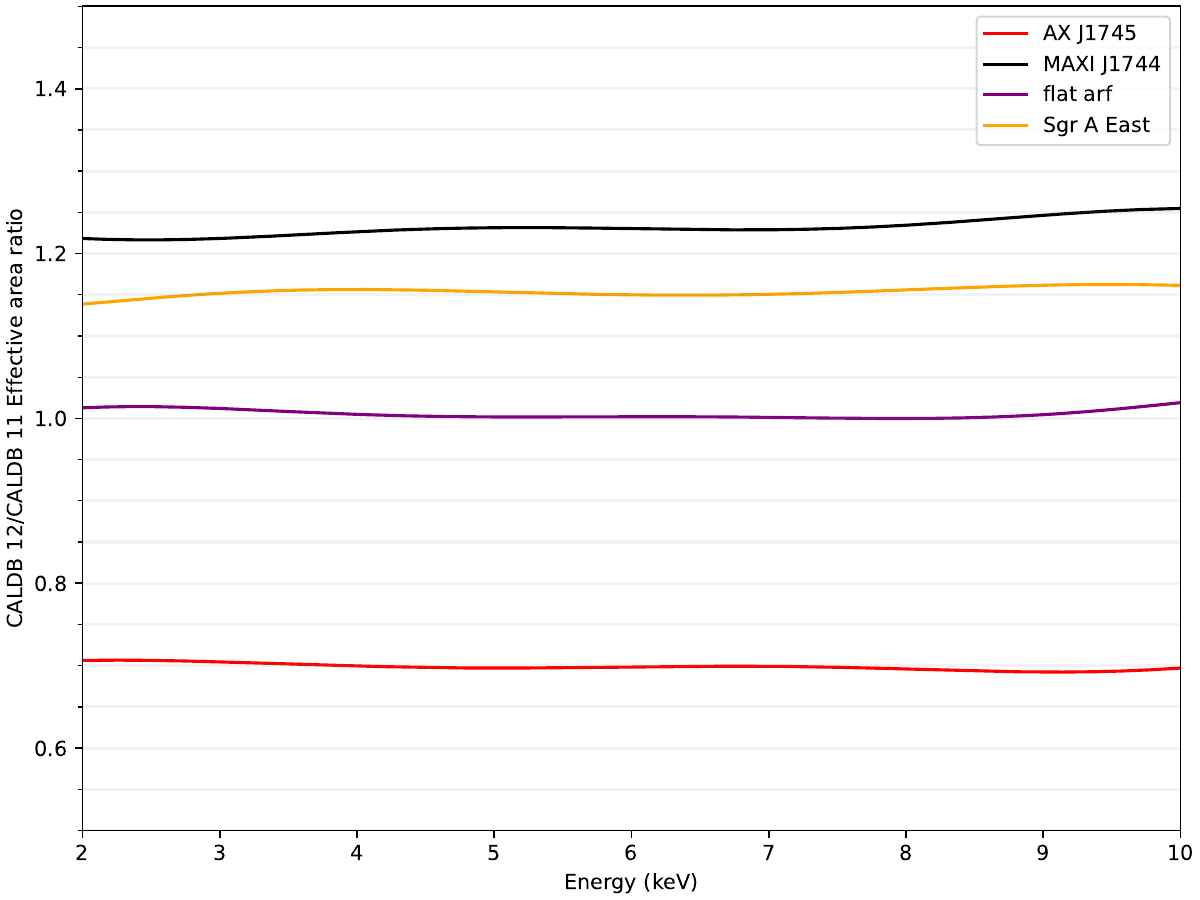}
    \caption{Comparisons \textbf{(top)} and ratios \textbf{(bottom)} of effective areas computed with CALDB 11 and 12 for the different sources in the "big" MAXI J1744-294 region \textbf{(left)}, "small" MAXI J1744-294 region \textbf{(center)}, and AX J1745.6-2901 region \textbf{(right)} in the DDT observation. All effective areas were computed with \texttt{rslmkrsp}.}
    \label{fig:Resolve_arf_compa_CALDB_DDT}
\end{figure*}

\begin{figure*}[h!]
\centering
    \includegraphics[clip,width=0.31\textwidth]{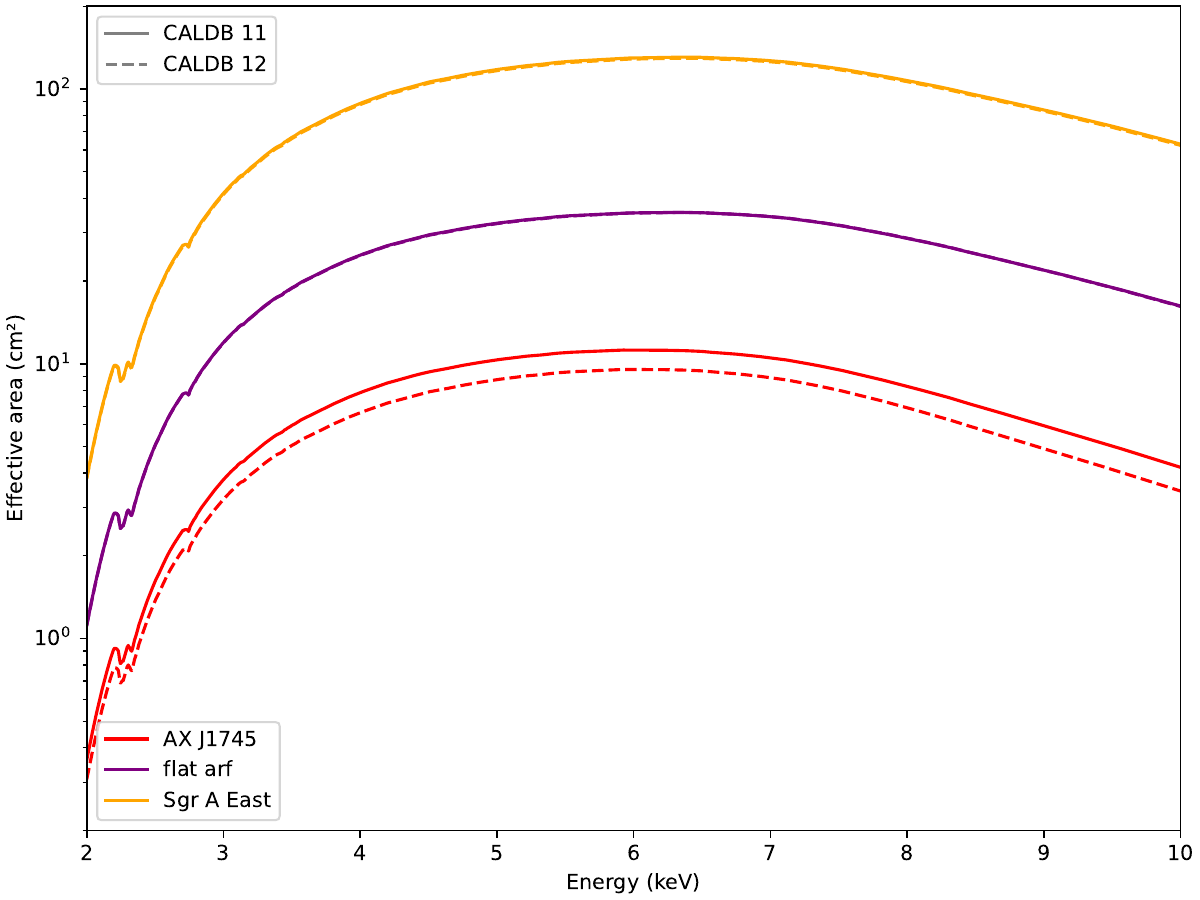}
    \includegraphics[clip,width=0.31\textwidth]{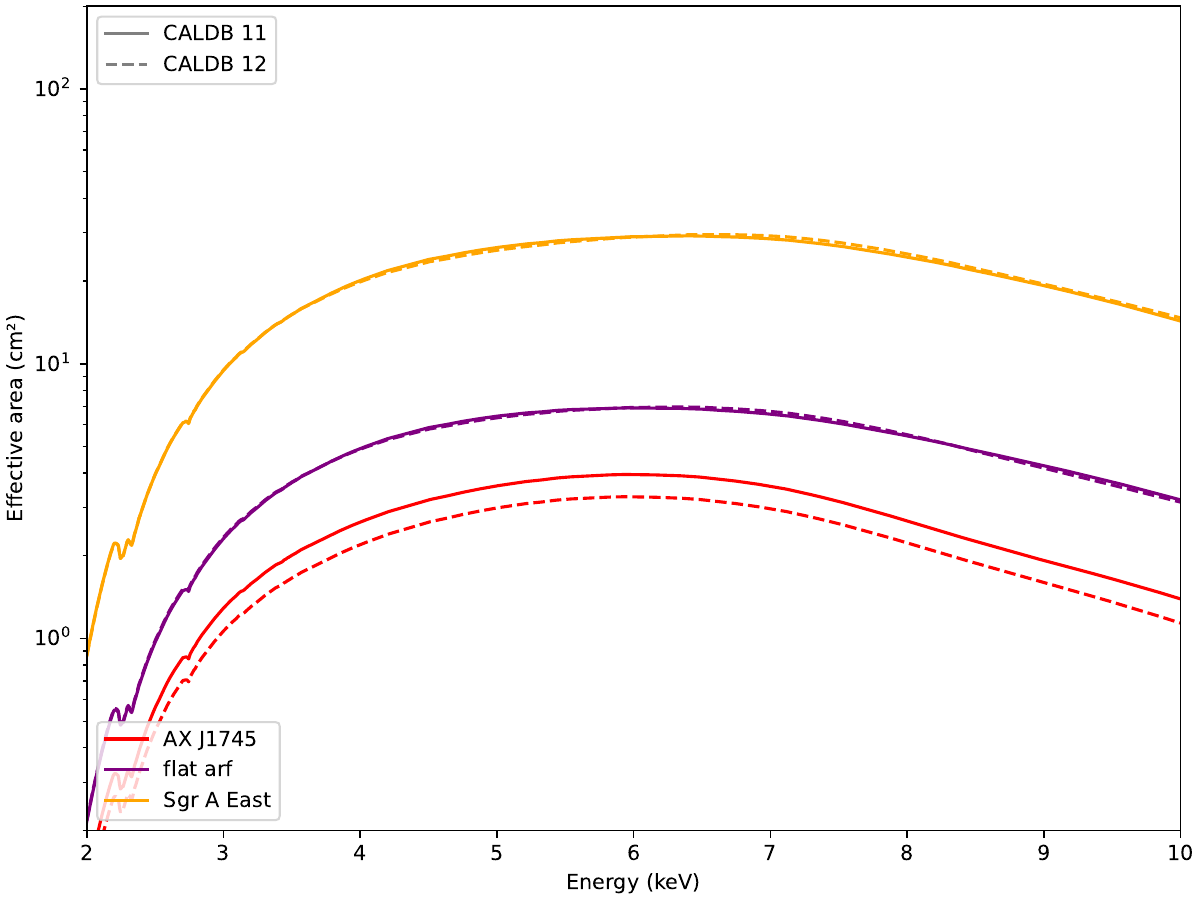}
    \includegraphics[clip,width=0.31\textwidth]{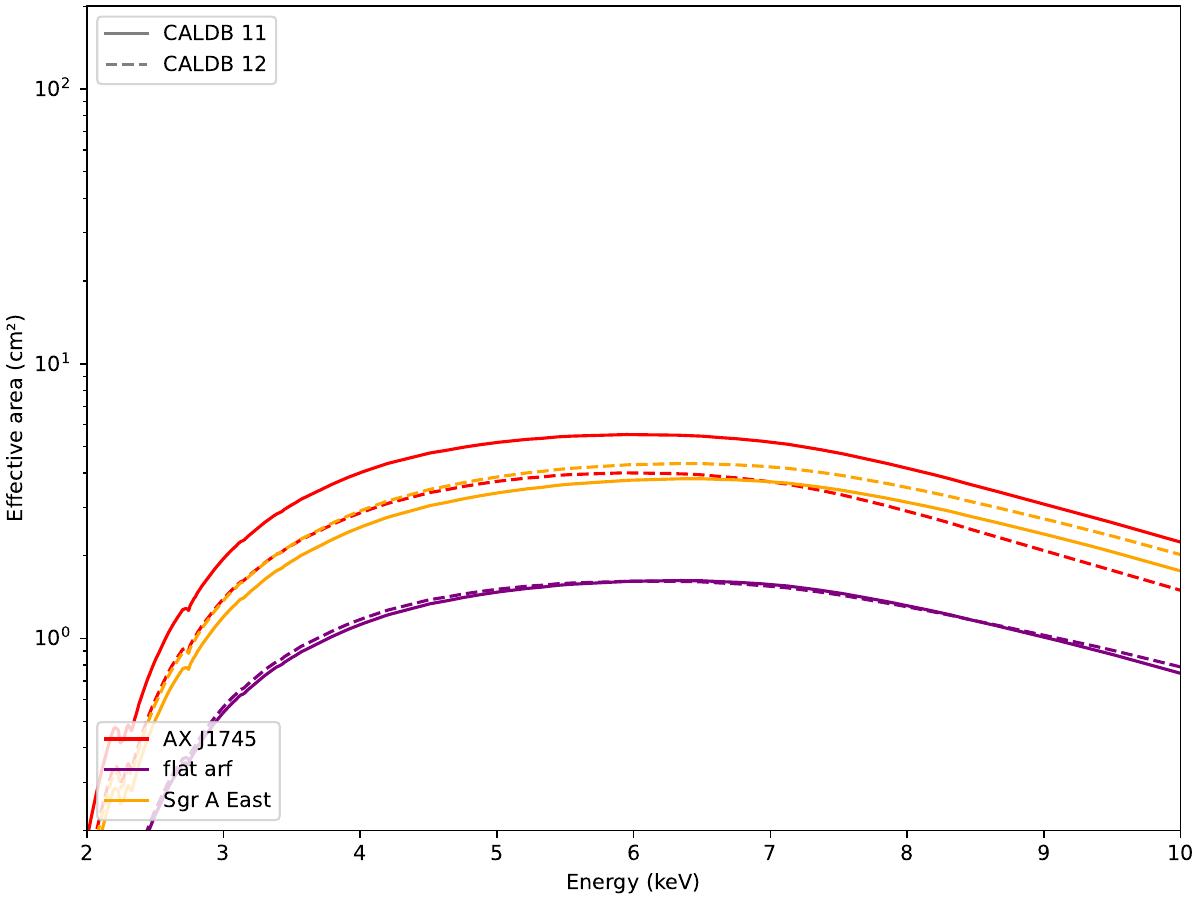}
    \includegraphics[clip,width=0.31\textwidth]{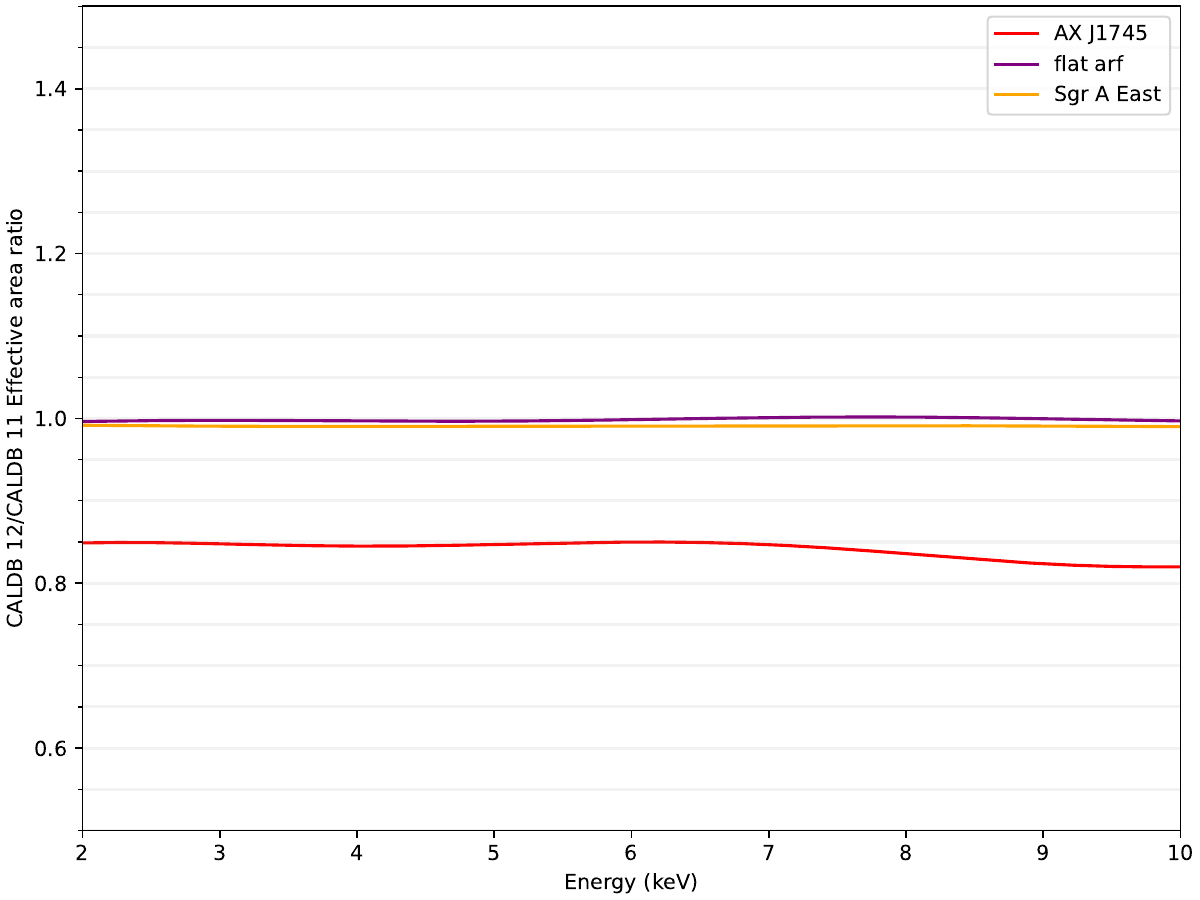}
    \includegraphics[clip,width=0.31\textwidth]{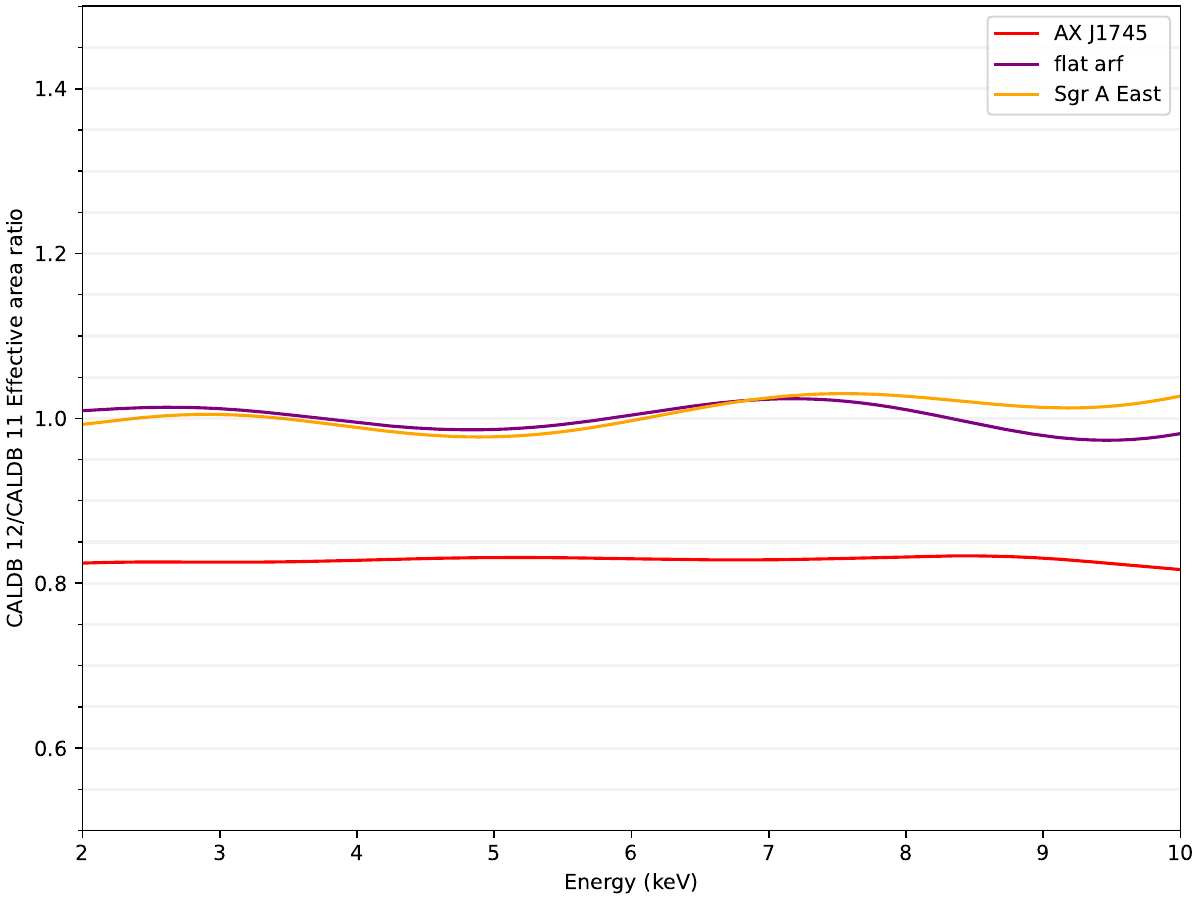}
    \includegraphics[clip,width=0.31\textwidth]{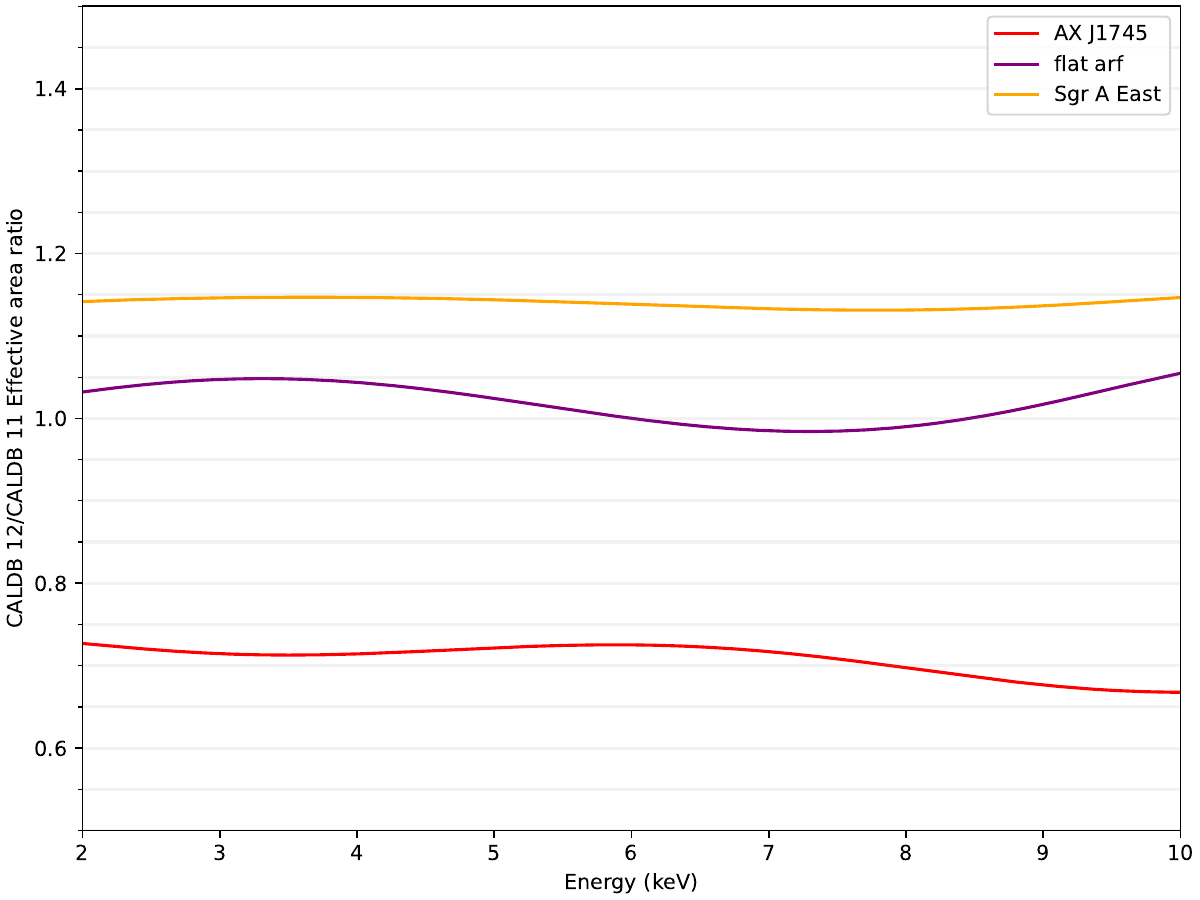}
    \caption{Comparisons \textbf{(top)} and ratios \textbf{(bottom)} of effective areas computed with CALDB 11 and 12 for the different sources in the "big" MAXI J1744-294 region \textbf{(left)}, "small" MAXI J1744-294 region \textbf{(center)}, and AX J1745.6-2901 region \textbf{(right)} in the PV observation. All effective areas were computed with \texttt{xaarfgen}.}
    \label{fig:Resolve_arf_compa_CALDB_PV}
\end{figure*}

It was stated in \S\ref{xrismdata} that CalDB 12 has errors that result in large differences in the Resolve effective area compared to CalDB 11, and we quantify this here. Fig. A.\ref{fig:Resolve_arf_compa_CALDB_DDT} (for the DDT observation) and A.\ref{fig:Resolve_arf_compa_CALDB_PV} (for the PV observation)
show the comparisons and ratios of effective areas computed using CALDB 11 and CALDB 12 for each source. In both observations, 
the difference for on-axis sources (M1744, flat ARF, Sgr A East) and central pixels ("small" M1744 region, "big" M1744 region) remains limited to $\lesssim 5\%$. However, they become important ($\gtrsim15\%$) for all sources except the flat ARF in the off-axis NS regions. AXJ, particularly when at the very edge of the FoV, is subject to a $\sim30\%$ difference in effective area across CALDB versions in both observations.

\clearpage

\subsubsection{Effect of the response creation methodology on the effective areas}

\begin{figure*}[h!]
\centering
    \includegraphics[clip,width=0.31\textwidth]{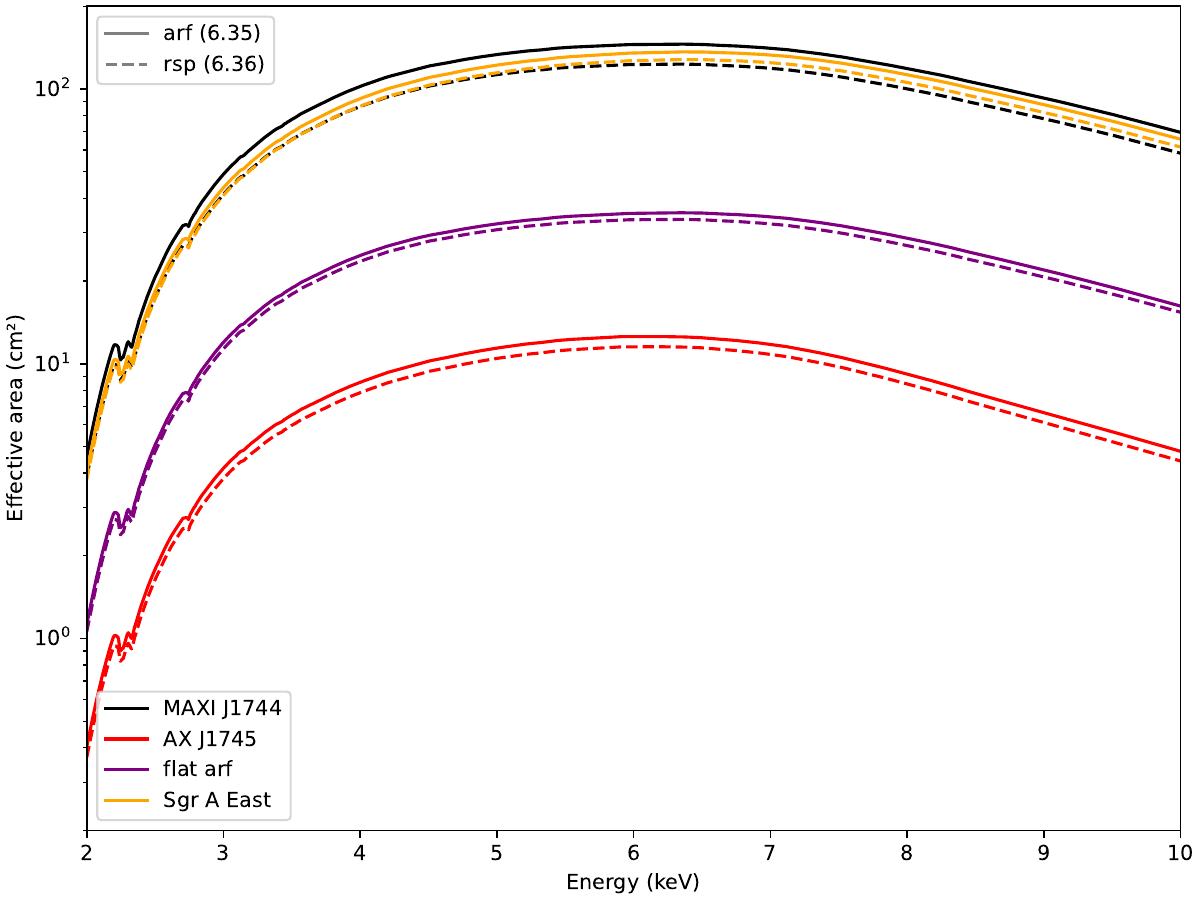}
    \includegraphics[clip,width=0.31\textwidth]{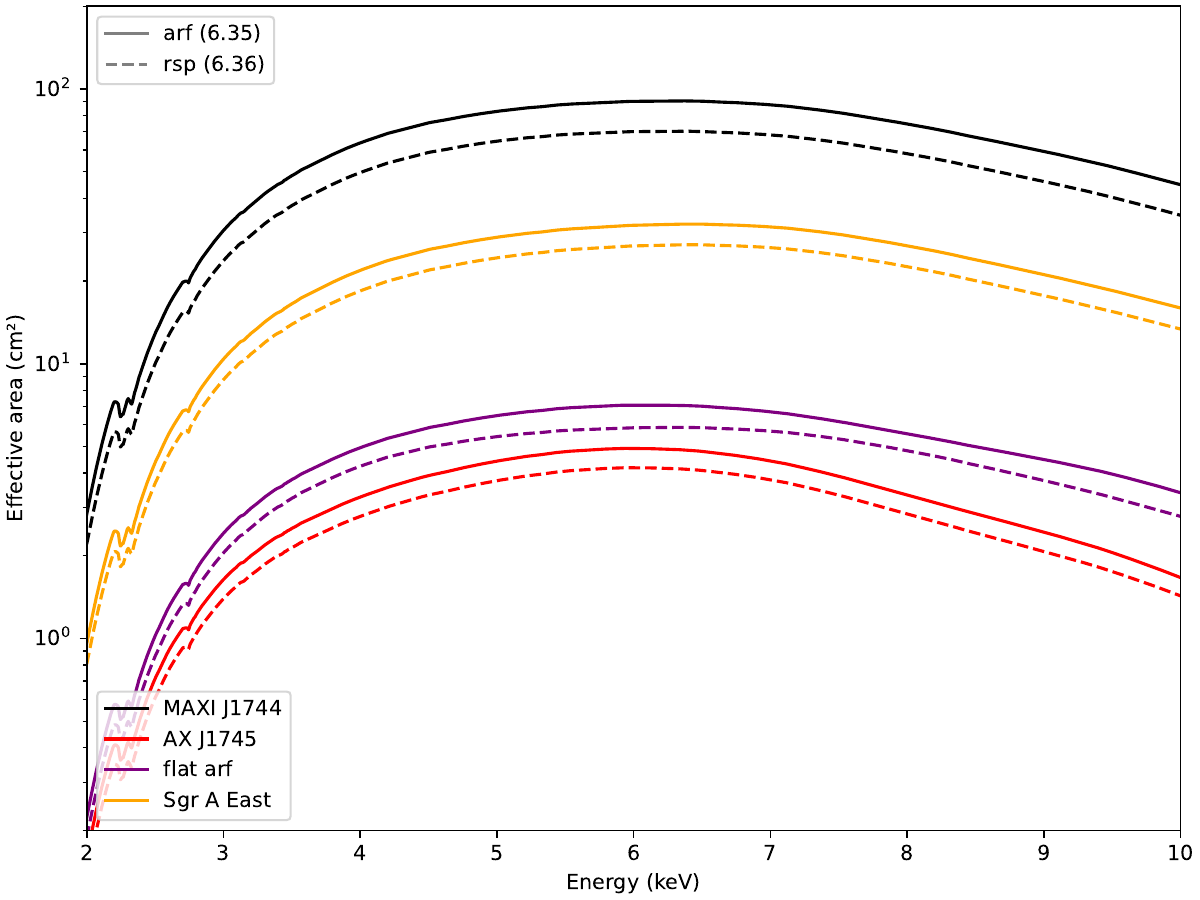}
    \includegraphics[clip,width=0.31\textwidth]{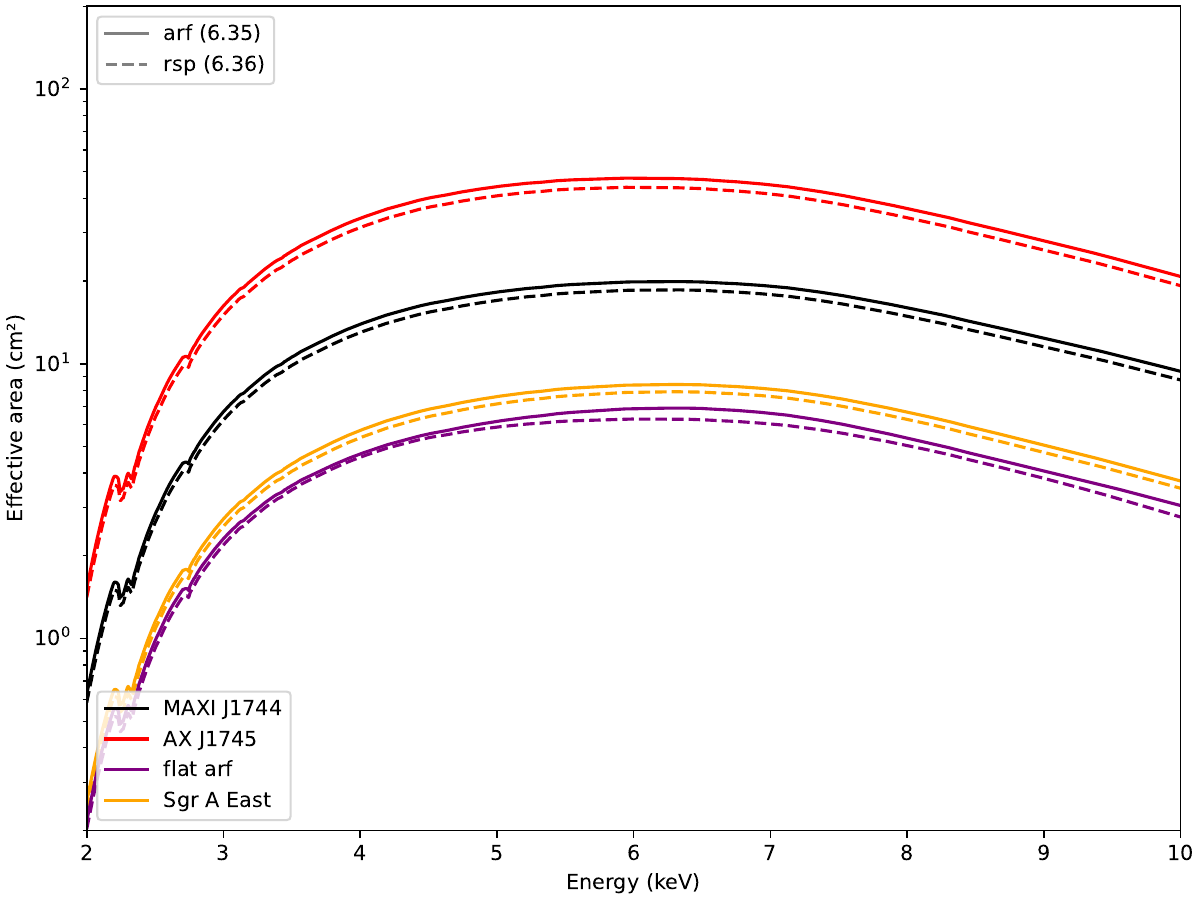}
    \includegraphics[clip,width=0.31\textwidth]{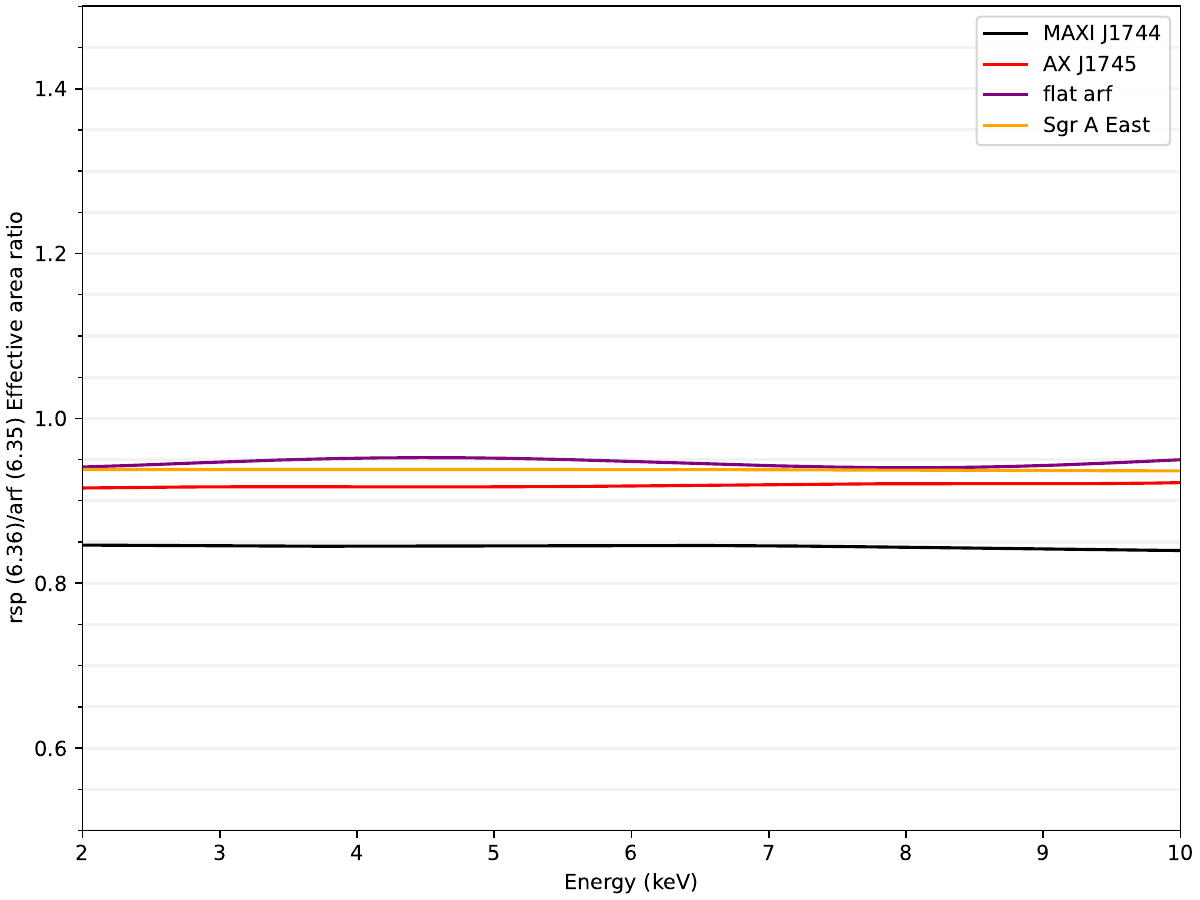}
    \includegraphics[clip,width=0.31\textwidth]{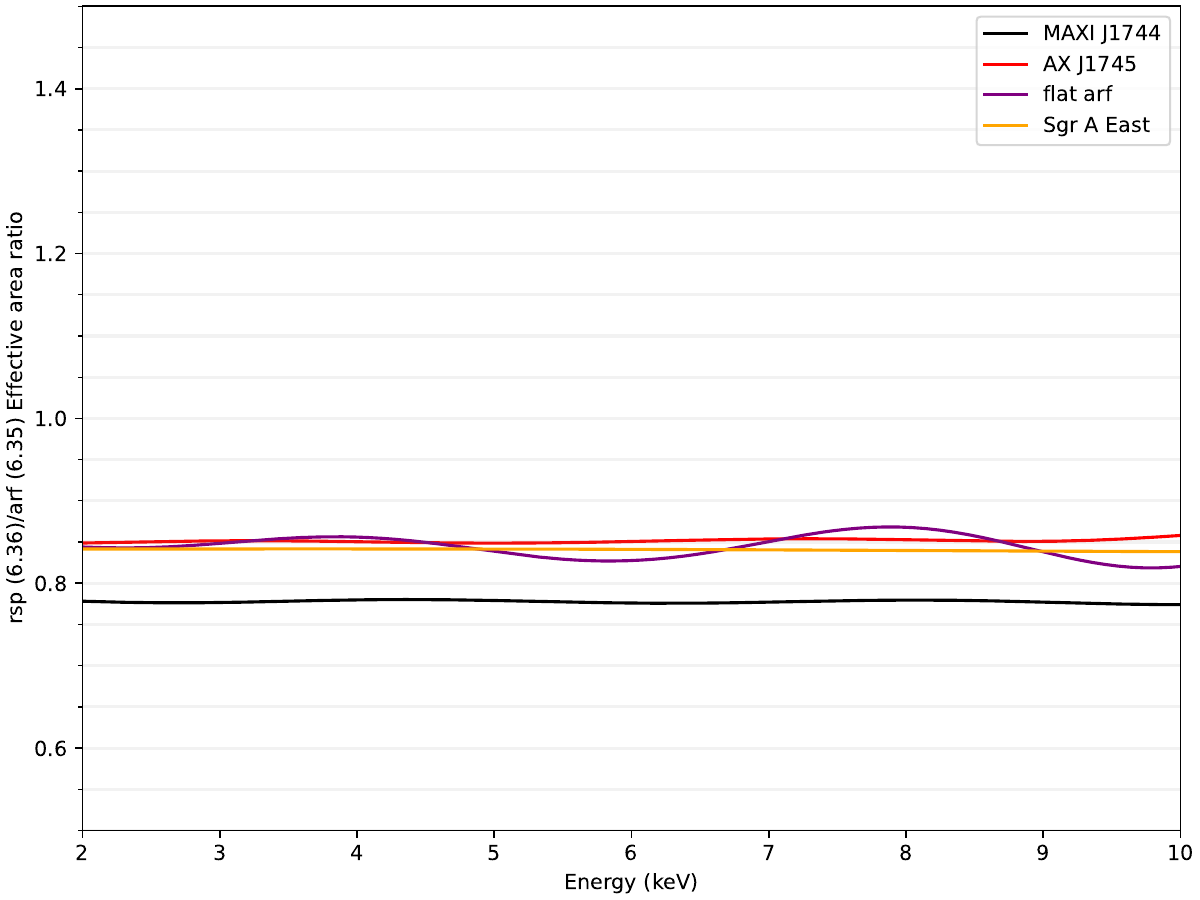}
    \includegraphics[clip,width=0.31\textwidth]{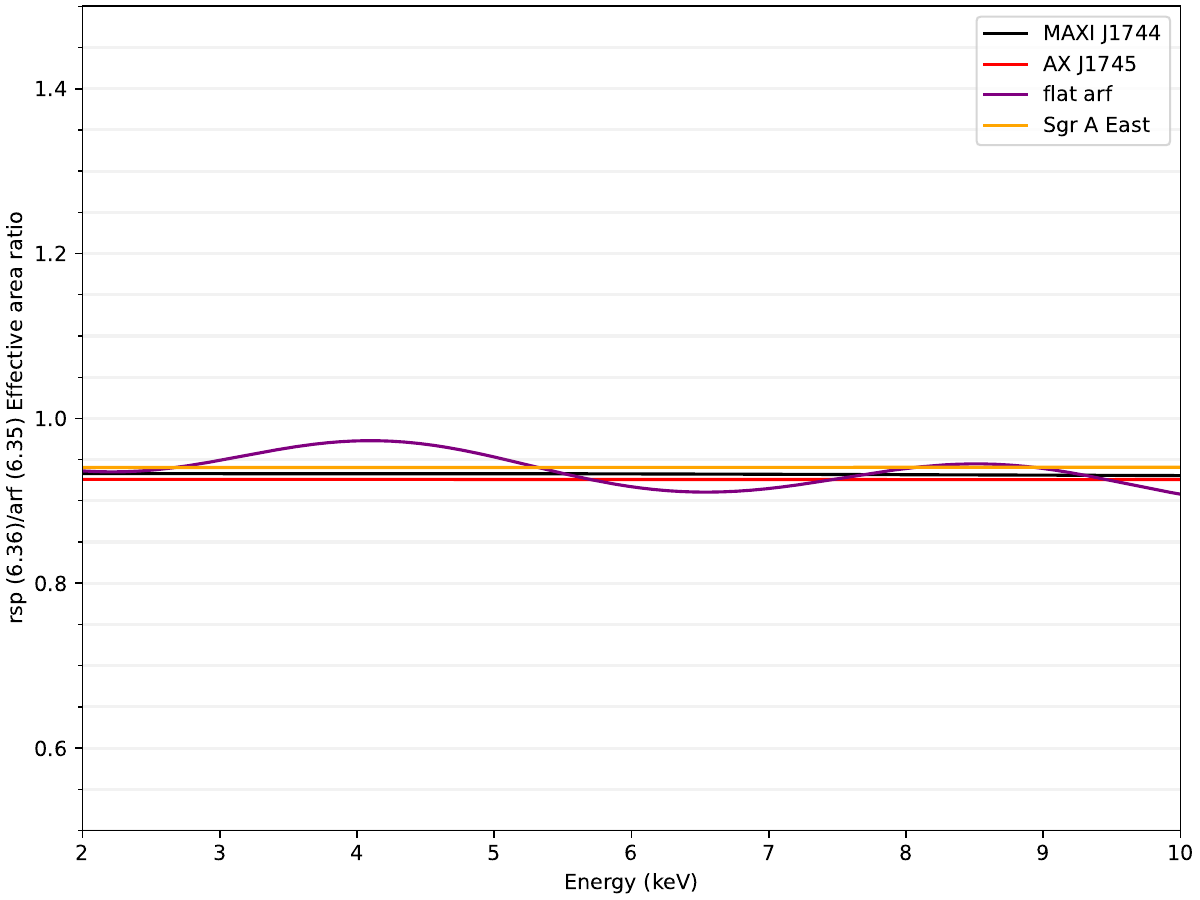}
    \caption{Comparisons \textbf{(top)} and ratios \textbf{(bottom)} of effective areas computed with \texttt{rslmkrsp} and \texttt{xaarfgen} for the different sources in the "big" MAXI J1744-294 region \textbf{(left)}, "small" MAXI J1744-294 region \textbf{(center)}, and NS region \textbf{(right)} in the DDT observation. All products were computed with \xrism\ CALDB 11.}
    \label{fig:Resolve_arf_rsp_compa_CALDB11_DDT}
\end{figure*}

\subsection{Xtend}

\subsubsection{Influence of the PSF asymmetry}\label{app:Xtend_PSF_asym}

 Ground calibration has shown that the \xrism{} PSF has azimuthal asymmetry, both in the "core" and in the "wings": in Fig.~\ref{fig:Xtend_regions_DDT_PV}-right, the eastern (left) part of the PSF appears globally brighter. However, this may also be due to the presence of Sgr A East. To quantify this difference, we compared the count rates of the M1744\ 
 region and its background in the PV phase, restricting ourselves to the $7-10$ keV band, where the CIE-overionization components of the SNR are expected to be much smaller than the high-temperature ($\sim1.7-1.8$ keV) \texttt{diskbb} from AXJ. In that band, the M1744 region has a total count rate of 8.0$\times10^{-3}$ counts/s, while our chosen background has a total of 6.7$\times10^{-3}$ counts/s. The NS contribution in both of our M1744\ regions may thus be underestimated by $\lesssim16\%$. However, this does not directly translate to a 16\% error for the AXJ contribution in the M1744 DDT spectrum, as the "additional" AXJ contribution in our PV phase M1744 region is modeled as part of the Sgr A East emission; the discrepancy comes from the changes in AXJ's spectral shape between the two observations. As detailed in Matsunaga et al. (in prep.), AXJ was in the same spectral state in 2024 and 2025, although it was 40$\%$ brighter in 2025 in the $7-10$ keV band. The relative error for the NS background modeling in the DDT observation should thus be on the order of $\lesssim6\%$. This should be contextualized with respect to the count rate of M1744 in the DDT data; similarly to the PV phase, we can derive a $7-10$ keV count rate in the M1744 region of $\times10^{-1}$ counts/s, compared to a background count rate of 9.4$\times10^{-3}$ counts/s. A $\lesssim6\%$ error on the NS estimation will thus remain below 1$\%$ of the M1744 spectrum, and can be safely considered as negligible.

\clearpage
\subsubsection{pile-up}\label{app:Xtend_pileup}

\begin{figure*}[h!]
    \includegraphics[clip,trim=0cm 0cm 0cm 0.6cm,width=0.44\textwidth]{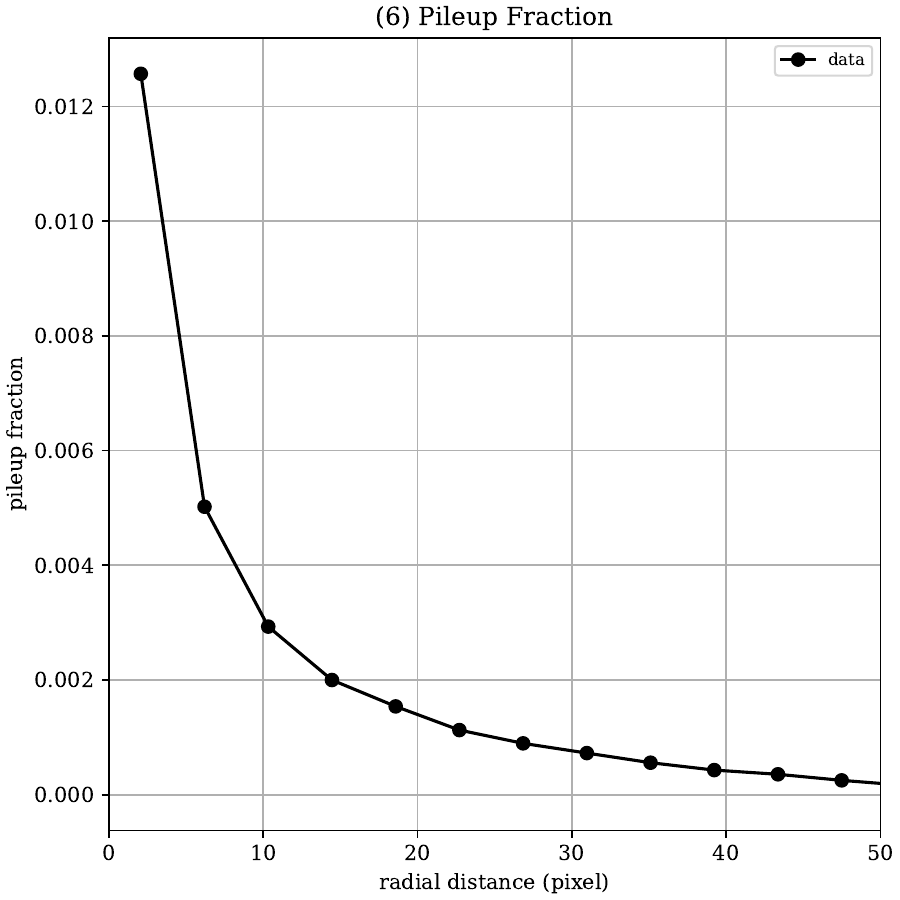}
    \includegraphics[clip,trim=0cm 1.4cm 0cm 2.35cm,width=0.56\textwidth]{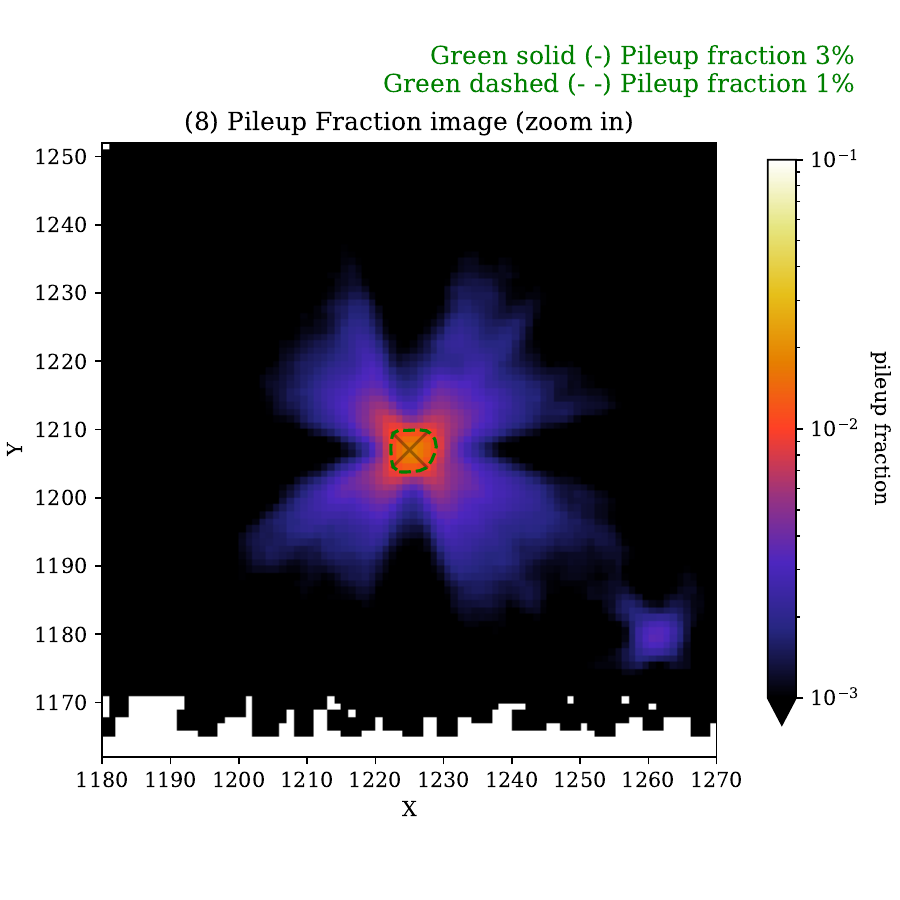}
    \includegraphics[clip,trim=0cm 0cm 0cm 0.6cm,width=0.44\textwidth]{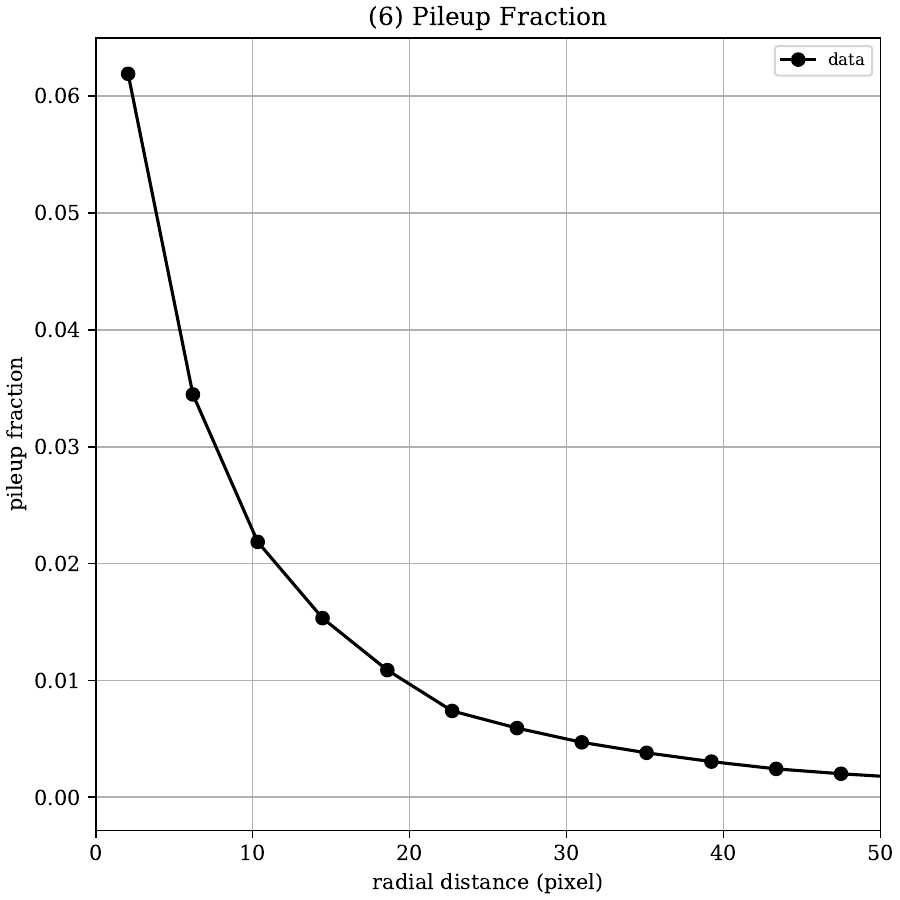}
    \vspace{-0.1em}
    \includegraphics[clip,trim=0cm 1.4cm 0cm 2.35cm,width=0.56\textwidth]{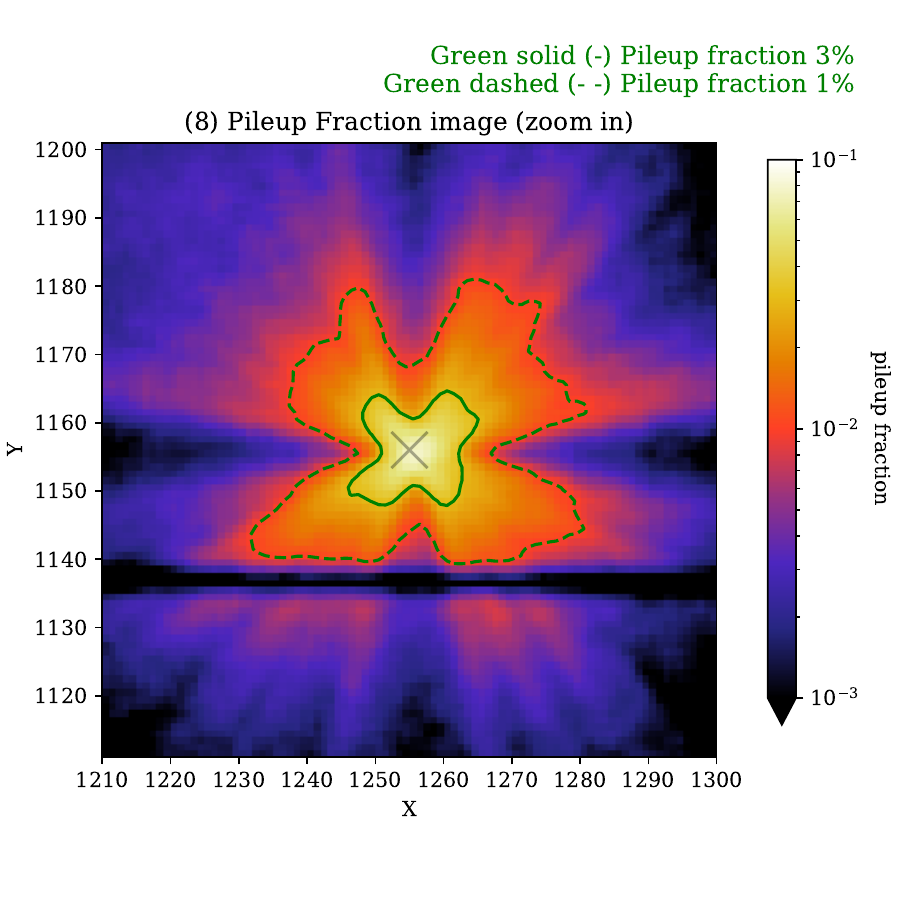}
    \includegraphics[clip,trim=6.5cm 13.6cm 0.7cm 0.5cm,width=0.3\textwidth,right]{xa300044010xtd_p030000010_cl_radialprofile_xcenter1255_ycenter1156_p8.pdf}
    \caption{Xtend radial pile-up fraction evolution \textbf{(left)} and pile-up fraction image \textbf{(right) }for the MAXI J1744-294 PSF in the burst mode DDT observation \textbf{(top)}, and the AX J1745.6-2901 PSF in the full window PV phase observation \textbf{(bottom)}.}
    \label{fig:Xtend_pileup}
\end{figure*}

\clearpage

\section{Diffuse emission models}\label{app:tables}

\subsection{Resolve}\label{app:diffuse_Resolve}

\begin{figure*}[h!]
\centering
\vspace{-1em}
    \includegraphics[clip,trim=0.5cm 0.4cm 0.4cm 0.4cm,width=0.49\textwidth]{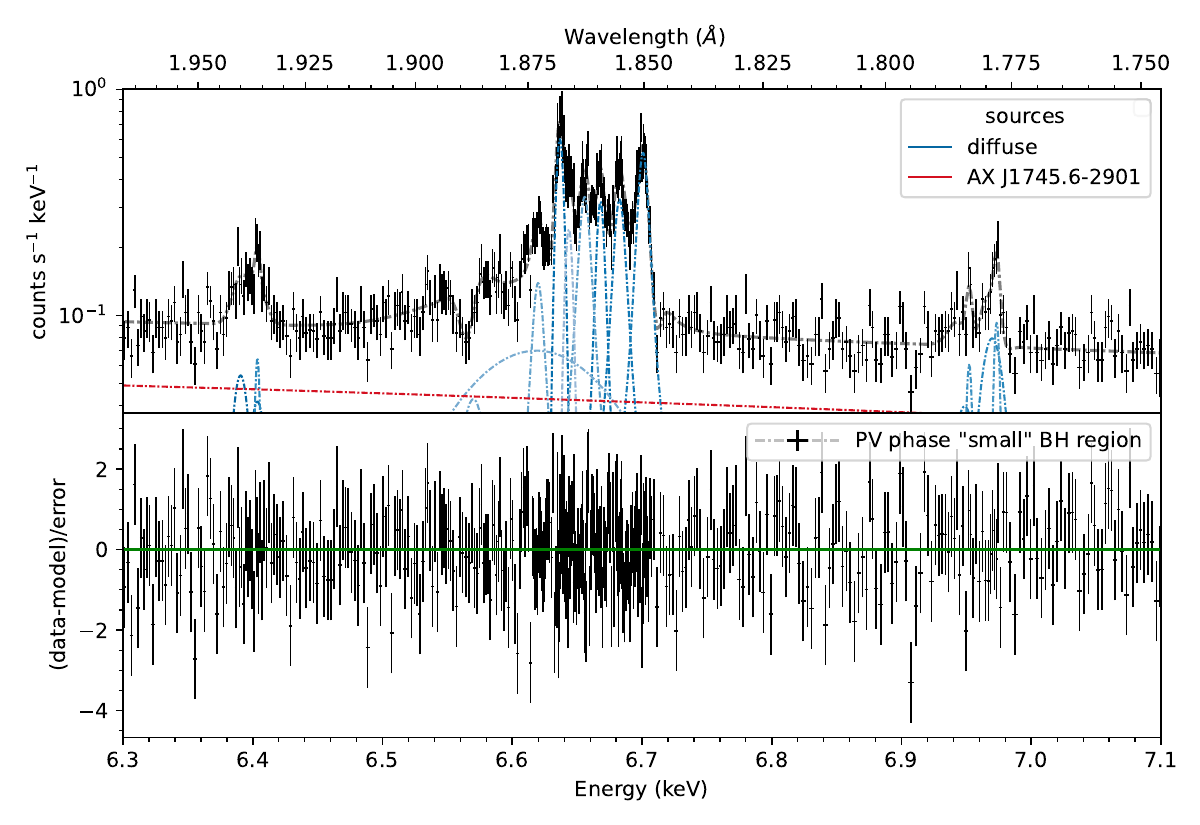}
    \includegraphics[clip,trim=0.4cm 0.4cm 0.4cm 0.4cm,width=0.49\textwidth]{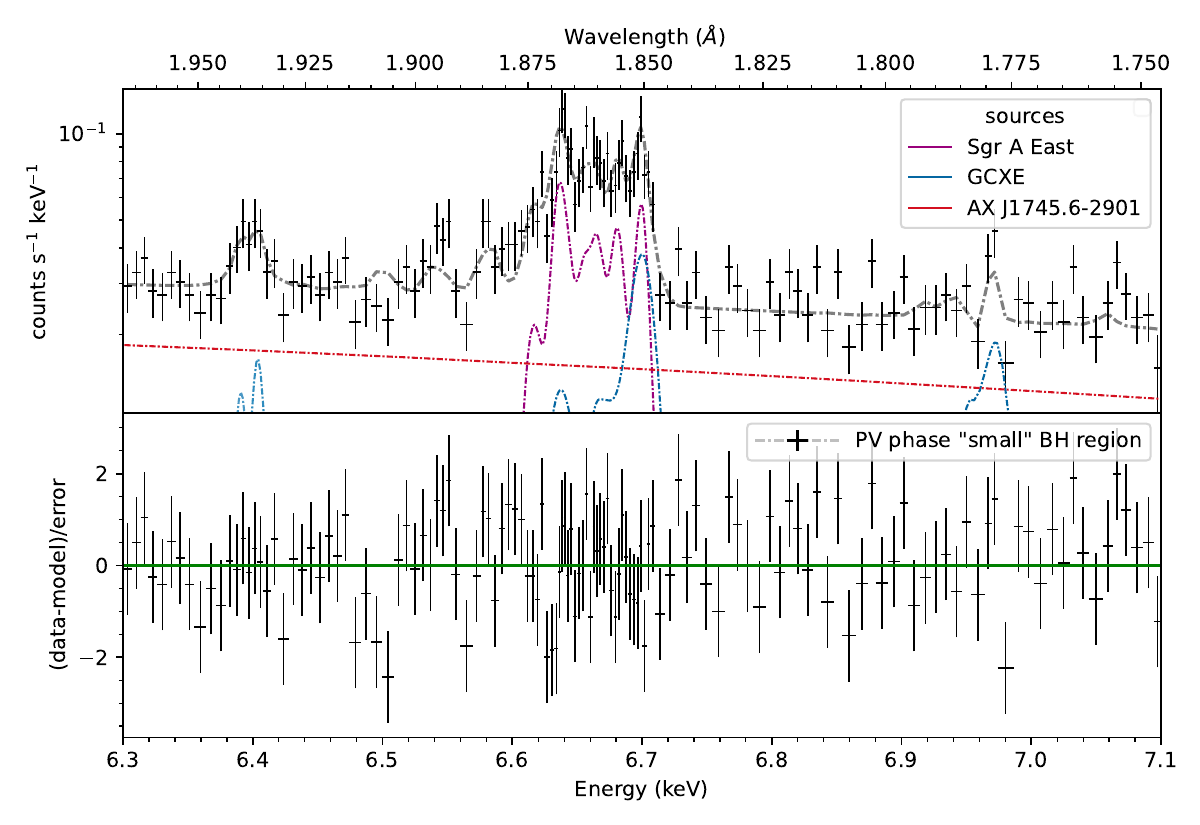}
    \caption{Zoomed spectrum and residuals in the band of the diffuse emission in the PV phase observation, using the "big pixel" region and empirical modeling \textbf{(left)}, and the "small pixel" region and physical modeling \textbf{(right)}. Both spectra were rebinned at a 5$\sigma$ significance level for visibility, and the components at 3$\sigma$.}
    \label{fig:PV_diffuse_resid_zoom}
\end{figure*}

\begin{figure*}[h!]
\centering
\vspace{-2.em}
    \includegraphics[clip,width=0.49\textwidth]{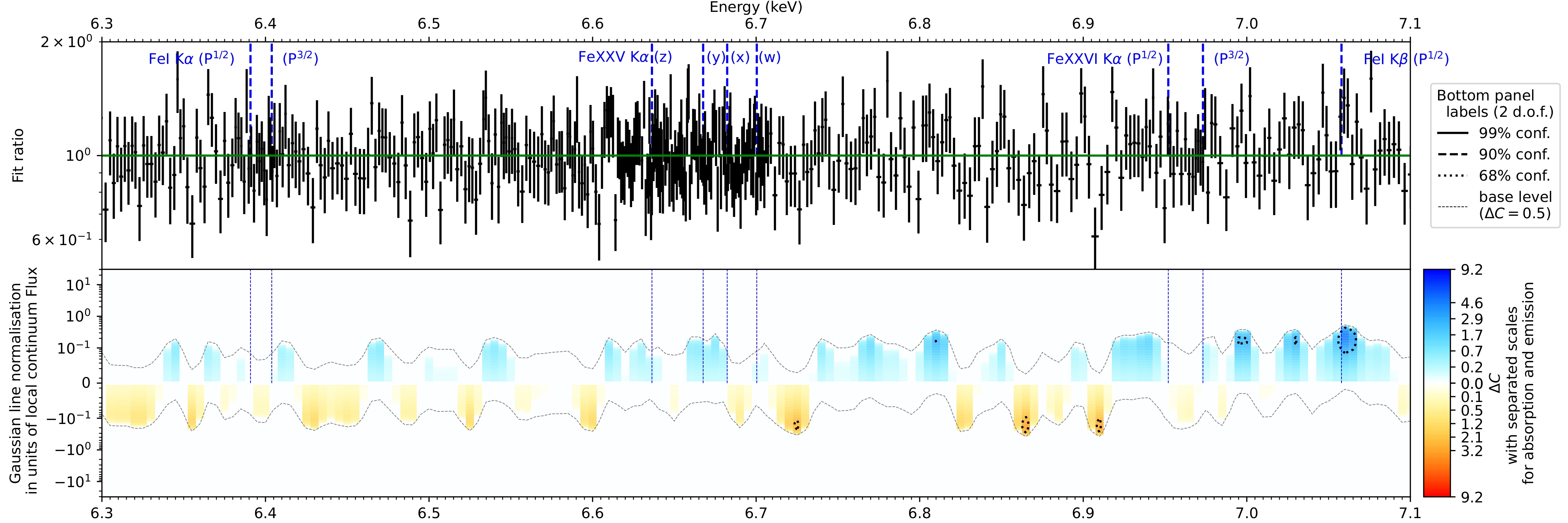}
    \includegraphics[clip,width=0.49\textwidth]{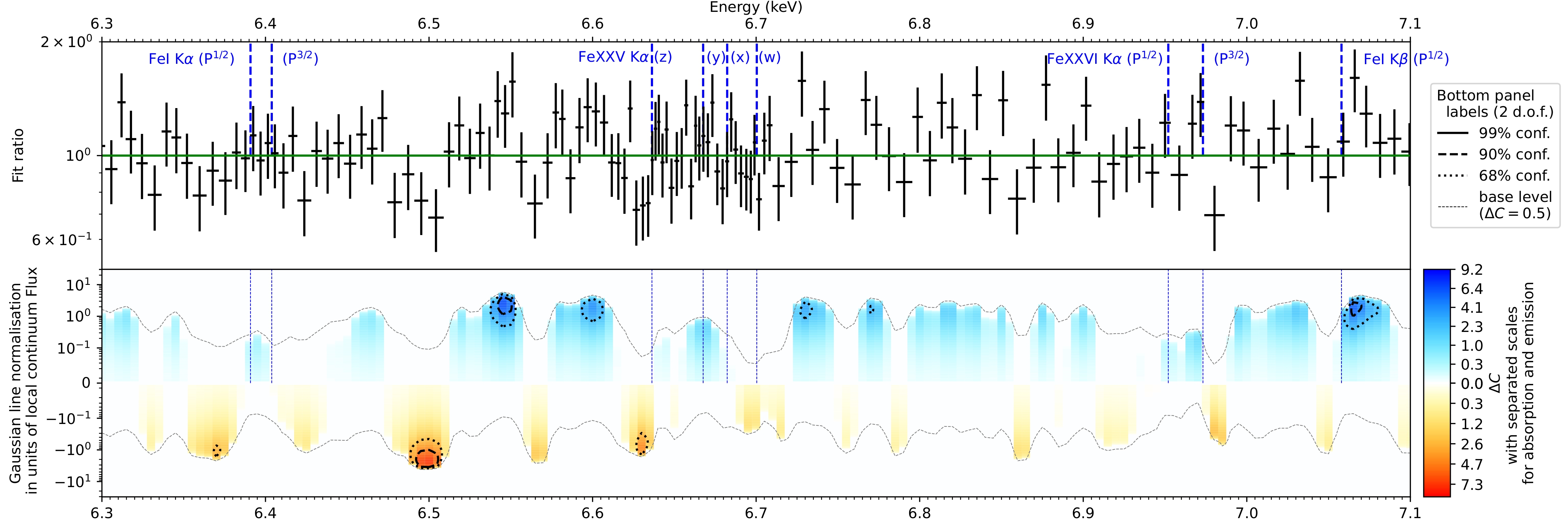}
    \includegraphics[clip,width=0.49\textwidth]{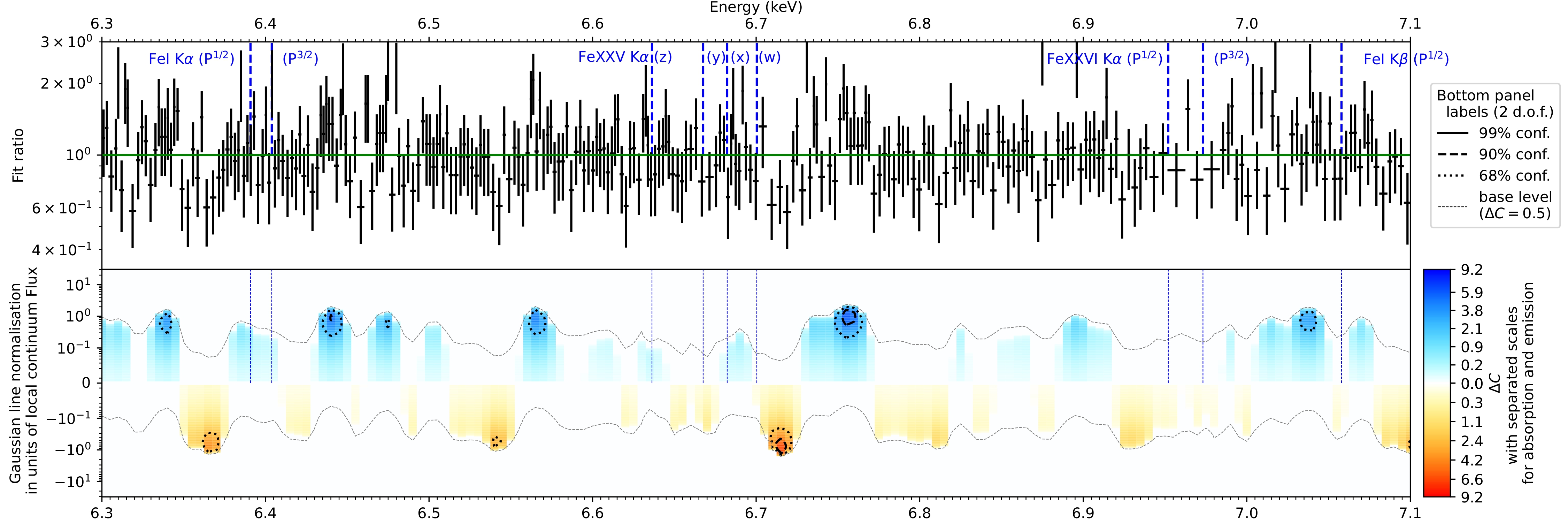}
    \includegraphics[clip,width=0.49\textwidth]{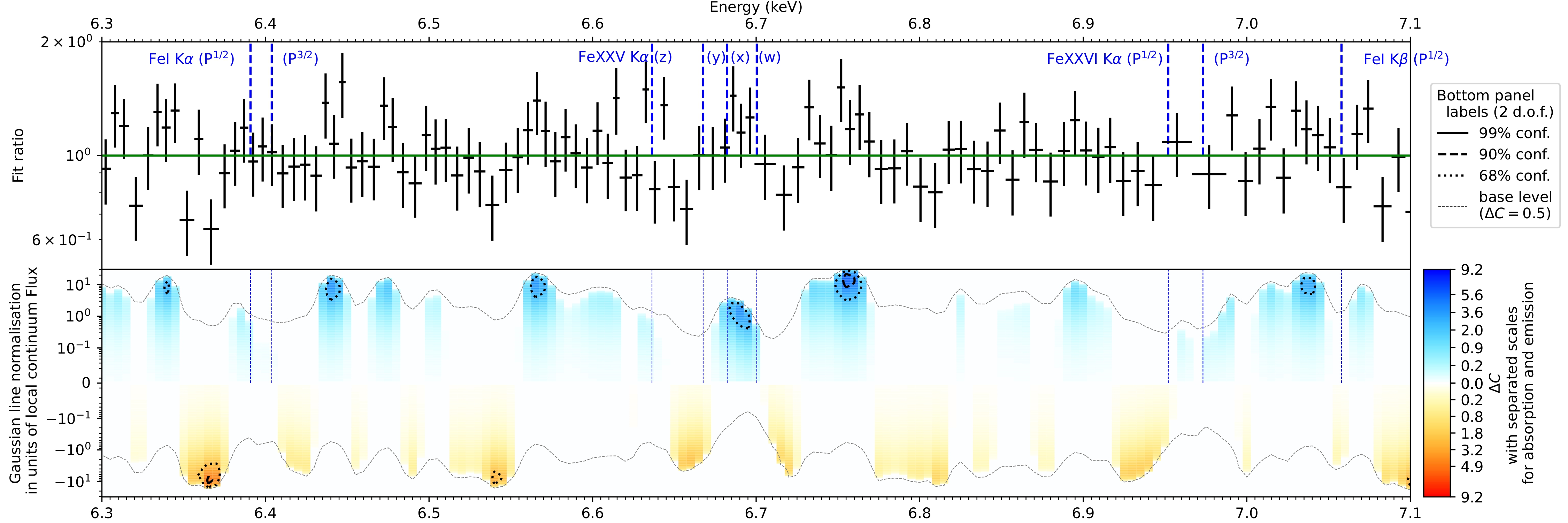}
    \caption{Blind searches for narrow line features in the 6.3-7.1 keV range, performed on the MAXI J1744-294 diffuse emission region \textbf{(top)} and the AX J1745.6-2901 pixel 11 spectrum \textbf{(bottom)} in the PV phase data, using the "big pixel" region and empirical modeling \textbf{(left)}, and 
    the "small pixel" region and physical modeling \textbf{(right)}. In the left panels, the M1744 spectrum was rebinned at a 5$\sigma$ significance level, and the NS spectrum at a 3$\sigma$ significance level. In the right panels, both spectra were rebinned at the 8$\sigma$ level.}
    \label{fig:blind_search_PV_Resolve}
\end{figure*}

\begin{table*}[h!]
\centering
\caption{Continuum parameters and spectral lines identified with empirical background modeling of the  "big" MAXI J1744-294 Resolve region in the PV observation.}\vspace{-1em}
\label{tab:comp_param_diffuse_bigpix_PV}

\begin{tabular}{lccccccc}
\cline{1-8}
\multicolumn{8}{c}{Continuum} \T\B\\
&Parameter & NH & $\Gamma$ & F$_{abs,2-10}$$^\star$ & C-stat/d.o.f.$^\star$ \T\B\\
&Unit  & 10$^{22}$ cm$^{-2}$ &  & 10$^{-11}$ cgs & \T\B\\
\cline{1-8}
&Value & $16.1\pm 0.5$ & $3.3\pm 0.1$ & $8.2_{-0.2}^{+0.1}$ & 33214/31850\T\B\\
\cline{1-8}
\cline{1-8}
\multicolumn{8}{c}{Spectral lines}\T\B\\
line complex & Line ID & $E_{\rm rest}$ (eV) & $v_{raw}$ (km s$^{-1}$)$^\ddagger$ & $\sigma$ (eV) & norm ($10^{-5}$) & EW (eV) & $\Delta$ C-stat \T\B \\

\cline{1-8}

\multirow{3}{*}{\Sxv{} He$\alpha$} & $z$ & $2431.3_{-0.6}^{+0.9}$ & \multirow{3}{*}{/} & $1.4_{-0.7}^{+0.6}$ & $133_{-28}^{+30}$ & $39_{-10}^{+9}$ & 97\T\B \\
& $x+y$ & $2450.6_{-4.4}^{+4.1}$ &  & $8.1_{-4.3}^{+4.6}$ & $129_{-45}^{+60}$ & $39_{-14}^{+19}$ & 40\T\B \\
& $w$ & $2461.2_{-1.0}^{+0.5}$ &  & $1.4_{-0.7}^{+0.6}$ & $131_{-23}^{+26}$ & $42_{-10}^{+8}$ & 106\T\B \\
\cline{1-8}
\Sxvi{} Ly$\alpha$ & 1/2+3/2 & $2623.1_{-1.9}^{+1.4}$ & / & $2.8_{-1.1}^{+2.4}$ & $42_{-14}^{+19}$ & $23\pm9$ & 38\T\B \\
\cline{1-8}
\Sxv{} He$\beta$ & 1/2+3/2 & $2884.0_{-1.3}^{+1.5}$ & / & $1.3_{-1.3}^{+1.7}$ & $14\pm6$ & $10_{-5}^{+6}$ & 21\T\B \\
\cline{1-8}
\multirow{3}{*}{Ar XVII He$\alpha$} & $z$ & $3106.2_{-0.8}^{+0.6}$ & \multirow{3}{*}{/} & $2.2_{-1.0}^{+0.8}$ & $26_{-4}^{+5}$ & $23\pm5$ & 149\T\B \\
 & $x+y$ & $3124.4_{-3.3}^{+4.1}$ &  & $4.7_{-3.5}^{+5.0}$ & $10_{-5}^{+3}$ & $9_{-6}^{+5}$ & 18\T\B \\
 & $w$ & $3139.8\pm 0.6$ &  & $1.1_{-1.1}^{+0.9}$ & $17_{-3}^{+4}$ & $15\pm4$ & 96\T\B \\
\cline{1-8}
%
Ca XVIII K$\alpha$  & sat & $3872.1_{-2.6}^{+2.5}$  & / & $3.8_{-2.0}^{+2.2}$ & $5.0_{-2.3}^{+1.9}$ & $8_{-3}^{+5}$ & 18\T\B \\
\cline{1-8}
\multirow{3}{*}{\Caxix{} He$\alpha$} & z & $3861.9_{-0.9}^{+1.3}$ & \multirow{3}{*}{/} & \multirow{3}{*}{$2.9_{-1.0}^{+0.5}$} & $9.2_{-1.7}^{+1.8}$ & $16\pm4$ & 18\T\B \\
& x+y & $3886.6_{-1.1}^{+1.3}$ &  &  & $6.7_{-1.8}^{+1.5}$ &$12\pm4$ & 33\T\B \\
& w & $3901.9_{-1.3}^{+1.2}$ &  &  & $8.4_{-1.5}^{+1.9}$ & $15\pm5$ & 58\T\B \\
\cline{1-8}
\multirow{2}{*}{Fe I K$\alpha$ (broad)} & 1/2 & 6390.8 &  \multirow{2}{*}{$13_{-300}^{+347}$} & \multirow{2}{*}{$5.9_{-2.1}^{+5.2}$} & $3.0_{-1.0}^{+1.2}$ & $11_{-11}^{+7}$ & 38\T\B \\
& 3/2 & 6403.8 & & & $2.3_{-1.0}^{+1.1}$ & $8\pm4$ & 19\T\B \\
\cline{1-8}
\multirow{1}{*}{Fe I K$\alpha$ (narrow)} & 3/2 & 6403.8 & $10_{-83}^{+74}$ & $0^{+5}$ & $1.1_{-0.5}^{+0.6}$ & $4\pm2$ & 6\T\B \\
\cline{1-8}
%
\multirow{8}{*}{6.6-6.7 keV satellites} & abs & 6565.6$^\dagger$ & \multirow{8}{*}{/} & $6.2_{-1.4}^{+1.7}$ & $-4.7_{-0.9}^{+0.3}$ & $-17_{-3}^{+4}$ & 76 \T\B \\
 &  & 6570.0$^\dagger$ &  & $8.0_{-3.1}^{+5.4}$ & $3.1_{-1.3}^{+2.1}$ & $11_{-5}^{+7}$ & 52\T\B \\
  &  & $6612.9_{-8.9}^{+11.8}$ &  & $59_{-7}^{+8}$ & $28\pm4$ & $124_{-21}^{+19}$ & 491\T\B \\
 &  & $6620.0_{-1.3}^{+1.6}$ &  & 8$^\dagger$ & $5.9_{-1.1}^{+0.8}$ & $22_{-4}^{+6}$ & 80\T\B \\
  &  & $6643.5_{-0.9}^{+0.8}$ &  & $2.5\pm 1.0$ & $6.2_{-1.2}^{+1.4}$ & $-23_{-4}^{+5}$ & 138\T\B \\
  &  & $6656.2\pm 0.7$ &  & 8$^\dagger$ & $15\pm1$ & $56_{-6}^{+8}$ & 426\T\B \\
 & abs & $6707.4\pm2.3$ &  & $8.6_{-2.6}^{+1.4\dagger}$ & $-6.7_{-1.6}^{+1.7}$ & $-23\pm6$ & 77\T\B \\
\cline{1-8}
\multirow{4}{*}{\Fexxv{} He$\alpha$ (lorentz)} & $z$ & 6636.3 & \multirow{4}{*}{$-18_{-9}^{+10}$} & $2.0\pm 0.4$ & $14\pm1$ & $55\pm5$ & 947\T\B \\
& $y$ & 6667.6 &  & $6.1_{-2.2}^{+2.8}$ & $12_{-3}^{+2}$ & $43_{-8}^{+10}$ & 395\T\B \\
& $x$ & 6682.3 &  & $6.1_{-2.0}^{+2.6}$ & $13_{-2}^{+3}$ & $50_{-12}^{+10}$ & 504\T\B \\
& $w$ & 6700.4 &  & $6.7\pm0.9$ & $26\pm2$ & $108_{-7}^{+9}$ & 989\T\B \\

\cline{1-8}
%
\multirow{2}{*}{\Fexxvi{} Ly$\alpha$ (broad)} & 1/2  & 6952.0 & \multirow{2}{*}{$143_{-81}^{+88}$} & \multirow{2}{*}{$8.4_{-1.5}^{+3.1}$} & $3.0\pm 0.4$ & $13_{-2}^{+2}$ & \multirow{2}{*}{262}\T\B \\
 & 3/2 & 6973.2 &  &  & $6.0\pm 0.8$ & $27_{-2}^{+4}$ & \T\B \\
\cline{1-8}
\multirow{2}{*}{\Fexxvi{} Ly$\alpha$ (narrow)} & 1/2  & 6952.0 & \multirow{2}{*}{$-11_{-32}^{+31}$} & \multirow{2}{*}{$1^\dagger$} & $1.2\pm0.5$ & $ 5\pm2$ & 87\T\B \\
 & 3/2 & 6973.2 &  &  & $1.8_{-0.5}^{+0.6}$ & $9\pm3$ & 175\T\B \\
\cline{1-8}
\multirow{2}{*}{\Nixxvii{} He$\alpha$ (lorentz)} & $y$ & 7786.4 & \multirow{2}{*}{$157_{-91}^{+101}$} & $3.0_{-3.0}^{+3.1}$ & $1.4_{-0.7}^{+0.8}$ &$24_{-12}^{+14}$ & 23\T\B \\
& $w$ & 7805.6 &  & $7.5_{-3.6}^{+9.7}$ & $1.2_{-0.6}^{+1.0}$ &  $21_{-13}^{+17}$ & 11 \T\B \\
\cline{1-8}
\multirow{2}{*}{\Fexxvi{} Ly$\beta$} & 1/2  & 7872.0 & \multirow{2}{*}{$-47_{-27}^{+26}$} & \multirow{2}{*}{$1^\dagger$} & $1.6_{-0.4}^{+0.5}$ & $29_{-9}^{+8}$ & 76\T\B \\
 & 3/2 & 7881.0 &  &  & $1.6\pm0.5$ & $28\pm8$ & 54\T\B \\
\cline{1-8}
\cline{1-8}
\cline{1-8}
\end{tabular}\\
\raggedright
$\star$ computed from the full diffuse emission model, including the line components, and the AXJ model, both applied to the two spectra of the M1744 and AXJ regions.
$\dagger$ frozen or at the limit of the parameter space. 
For the iron lines at high energy with definitive identification, we use fixed energies $E_{rest}$ (from NIST) and the velocities $v_{raw}$ are allowed to vary. For the lines at low energy, with more satellites and overlap, E$_{rest}$ is instead free to vary, except for some lines (e.g. weak FeXX+ satellites) where the energies and widths are frozen due to the high overlap between transitions. Widths and velocities are linked within a single complex whenever possible. 
$\ddagger$ does not include the correction of -28 km/s due to the relative motion of Earth in the Solar System on the date of the observation.
\end{table*}

\clearpage

\begin{table*}[h!]
\centering
\caption{Model parameters for the physical background models of the "small" MAXI J1744-294 Resolve region in the PV observation. The GCXE model is directly imported from Uchiyama et al. (in prep.) without fitting.}
\label{tab:comp_param_diffuse_smallpix_PV}

  \begin{minipage}[c]{0.5\textwidth}
\begin{tabular}{lcc}
\cline{1-3}
\multicolumn{3}{c}{Sgr A East} \T\B\\
\cline{1-3}
Parameter & Unit & Value \T\B\\
C$_{Region}\star$ & & 0.66 \T\B\\
$v_{shift}$ & km s $^{-1}$ & 28$^\dagger$ \T\B\\
 NH & $10^{22}$ cm$^{-2}$ & 17.6$_{-1.3}^{+1.4}$\T\B\\
   \cdashline{1-3}
\multicolumn{3}{c}{\texttt{powerlaw}} \T\B\\
  \cdashline{1-3}
$\Gamma$ & & 1$^\dagger$ \T\B\\
F$_{abs,2-10}$& erg s$^{-1}$ cm$^{-2}$ & $(3.4\pm1.0)\times10^{-12}$ \T\B\\
   \cdashline{1-3}
 \multicolumn{3}{c}{low-$kT_{e}$ \texttt{bvvrnei}} \T\B\\
 \cdashline{1-3}
$kT_{e}$ & keV & 0.95$^\dagger$ \T\B\\
$kT_{init}$ & keV & 10$^\dagger$ \T\B\\
$S$ & solar & $0.31_{-0.31}^{+0.43}$\\
$Ar$ & solar & $0.57_{-0.37}^{+0.44}$\\
$Ca$ & solar & $1.0_{-0.45}^{+0.52}$\\
$Z_{base}^{\star\star}$ & solar & 1.44$^\dagger$\\
$\tau$ & s cm$^{-3}$ & $1.7\times10^{12\dagger}$\T\B\\
$\sigma$ & km s$^{-1}$ & $180_{-53}^{+64}$ \T\B\\
$n_{e}n_{H}V^+$ & cm$^{-5}$ &  $(4.8\pm0.6)\times10^{59}$ \T\B\\
  \cdashline{1-3}
 \multicolumn{3}{c}{high-$kT_{e}$ \texttt{bvvrnei}} \T\B\\
  \cdashline{1-3}
$kT_{e}$ & keV & $1.3\pm0.2$ \T\B\\
$kT_{init}$ & keV & $13.5^\dagger$ \T\B\\
$Fe$ & solar & $2.4_{-0.8}^{+1.5}$ \T\B\\
$Z_{base}^{\star\star}$ & solar & 1.$^\dagger$\\
$\tau$ & s cm$^{-3}$ & $(8.5_{-6.7}^{+9.4})\times10^{11}$ \T\B\\
$\sigma$ & km s$^{-1}$ & $180_{-53}^{+64}$ \T\B\\
$n_{e}n_{H}V^+$ & cm$^{-5}$ & $(8.6\pm1.2)\times10^{58}$\T\B\\
  \cdashline{1-3}
C-stat/d.o.f.$^\ddagger$ & & 32379/31970\T\B\\
\end{tabular}
  \end{minipage}\hfill
  \begin{minipage}[c]{0.5\textwidth}

\begin{tabular}{lcc}
\cline{1-3}
\multicolumn{3}{c}{GCXE$\dagger$} \T\B\\
\cline{1-3}
Parameter & Unit & Value \T\B\\
C$_{Region}\star$ &  & 7.9 \T\B\\
$v_{shift}$ & km s $^{-1}$ & 28 \T\B\\
 NH & $10^{22}$ cm$^{-2}$ & 7.7 \T\B\\
   \cdashline{1-3}
\multicolumn{3}{c}{\texttt{powerlaw}} \T\B\\
  \cdashline{1-3}
$\Gamma$ & & 2.17 \T\B\\
F$_{abs,2-10}$& erg s$^{-1}$ cm$^{-2}$ & $1.9\times10^{-11}$\T\B\\
   \cdashline{1-3}
 \multicolumn{3}{c}{low-$kT_{e}$ \texttt{bvvrnei}} \T\B\\
 \cdashline{1-3}
$kT_{e}$ & keV & 0.95 \T\B\\
$kT_{init}$ & keV & 10 \T\B\\
$S$ & solar & 2.2\\
$Ar$ & solar & 0.9\\
$Ca$ & solar & 0.8\\
$Ni$ & solar & 2.8\\
$Z_{base}^{\star\star}$ & solar & 1.9\\
$\tau$ & s cm$^{-3}$ & $5.0\times10^{13\dagger}$ \T\B\\
$\sigma$ & km s$^{-1}$ &  354 \T\B\\
$n_{e}n_{H}V^+$ & cm$^{-5}$ & $9.9\times10^{57}$ \T\B\\
  \cdashline{1-3}
 \multicolumn{3}{c}{high-$kT_{e}$ \texttt{bvvrnei}} \T\B\\
  \cdashline{1-3}
$kT_{e}$ & keV & 7.5 \T\B\\
$kT_{init}$ & keV & 10 \T\B\\
$S$ & solar & 2.2\\
$Ar$ & solar & 0.9\\
$Ca$ & solar & 0.8\\
$Ni$ & solar & 2.8\\
$Z_{base}^{\star\star}$ & solar & 1.9\\
$\tau$ & s cm$^{-3}$ & $5.0\times10^{13\dagger}$ \T\B\\
$\sigma$ & km s$^{-1}$ & 354 \T\B\\
$n_{e}n_{H}V^+$ & cm$^{-5}$ & $2.4\times10^{57}$\T\B\\
   \cdashline{1-3}
 \multicolumn{3}{c}{Fe I K$\alpha$ $\texttt{lorentz}$} \T\B\\
 \cdashline{1-3}
 $\sigma$ &  eV & 10.2 \T\B\\
  $E_{1/2}$ & eV & 6390.8  \T\B\\
norm$_{1/2}$ & $10^{-5}$ & 0.49 \T\B\\
 $E_{3/2}$ & eV & 6403.8 \T\B\\
norm$_{3/2}$ & $10^{-5}$ & 0.63 \T\B\\
   \cdashline{1-3}
 \multicolumn{3}{c}{Fe I K$\beta$ $\texttt{lorentz}$} \T\B\\
 \cdashline{1-3}
 $\sigma$ &  eV & 10.2 \T\B\\
  $E$ & eV & 7006.00  \T\B\\
norm & $10^{-5}$ & 0.1 \T\B\\
   \cdashline{1-3}
 \multicolumn{3}{c}{Ni I K$\alpha$ $\texttt{lorentz}$} \T\B\\
 \cdashline{1-3}
 $\sigma$ &  eV & 10.2 \T\B\\
  $E$ & eV & 7476.3  \T\B\\
norm & $10^{-5}$ & 0.04 \T\B\\
  \cdashline{1-3}
  \end{tabular}
  \end{minipage}\hfill
\\

{$\star$ scaling from the fit of \cite{XrismCol2025_GC_obs_diffuse}. $\star\star$ Abundances for all elements except those highlighted individually.
$\dagger$ frozen or at the limit of the parameter space.$\ddagger$ Global fit statistic including both the Sgr A East, GCXE, and AXJ models, applied to the "small" pixel region and the AXJ spectrum simultaneously. 
$^+$ emission measure (normalization) computed for a distance of 8.2kpc.}
\end{table*}

\clearpage
\subsection{Xtend and XMM}\label{app:diffuse_CCD}

\begin{table*}[h!]
\centering
\caption{Continuum parameters and spectral lines for the empirical Sgr A East background modeling with Xtend in the PV observation.}
\label{tab:comp_param_diffuse_xtend_PV}
\begin{tabular}{lccccc}
\cline{1-6}
\multicolumn{6}{c}{Continuum} \T\B\\
&Parameter & NH & $\Gamma$ & F$_{abs,2-10}$$^\star$ & C-stat/d.o.f.$^\star$ \T\B\\
&Unit  & 10$^{22}$ cm$^{-2}$ &  & 10$^{-11}$ cgs & \T\B\\
\cline{1-6}
&Value & $24_{-4}^{+6}$ & $4.7_{-0.8}^{+1.2}$ & $1.6_{-1.2}^{+0.1}$ & 86/79\T\B\\
\cline{1-6}
\cline{1-6}
\multicolumn{6}{c}{Spectral lines}\T\B\\
line complex  & $E_{\rm rest}$ (keV) & $\sigma$ (eV) & norm ($10^{-5}$) & EW (eV) & $\Delta$ C-stat \T\B \\

\cline{1-6}
\Sxv{} He$\alpha$ & 2.45$^\dagger$ & 0$^\dagger$ & $407_{-251}^{+902}$ & $299_{-149}^{+163}$ & 19\T\B \\
\Caxix{} He$\alpha$ & 3.9$^\dagger$ & 0$^\dagger$ & $26_{-14}^{+17}$ & $153_{-85}^{+78}$ & 10\T\B \\
\Fei{} K$\alpha$ & 6.4$^\dagger$ & 0$^\dagger$ & $8.1_{-5.4}^{+5.8}$ & $149_{-115}^{+125}$ & 6\T\B \\
\Fexxv{} He$\alpha$ & $6.64_{-0.024}^{+0.027}$ & 0$^\dagger$ & $23_{-5}^{+7}$ & $650_{-205}^{+625}$ & 49 \T\B \\
\Fexxvi{} Ly$\alpha$ & 6.97$^\dagger$ & 0$^\dagger$ & $6.7_{-4.5}^{+4.6}$ & $156_{-114}^{+534}$ &  6\T\B \\
\Nixxvii{} He$\alpha$ & 7.8$^\dagger$ & 0$^\dagger$ & $3.8_{-3.8}^{+4.3}$ & $585_{-585}^{+619}$ &  3\T\B \\

\cline{1-6}
\cline{1-6}
\end{tabular}\\
\raggedright
$\star$ computed from the full model, including the line components.
$\dagger$ frozen or at the limit of the parameter space. 
\end{table*}

\begin{table*}[h!]
\centering
\caption{Continuum parameters and spectral lines for the empirical Sgr A East background modeling with XMM in the archival observations.}
\label{tab:comp_param_diffuse_XMM_PV}
\begin{tabular}{lcccccccc}
\cline{1-9}
\multicolumn{9}{c}{Continuum} \T\B\\
Parameter & $C_{XMM,1}$ & $C_{XMM,2}$ & NH & $\Gamma$ & $kT_{in}$ & norm$_{disk}$ &F$_{abs,2-10}$$^\star$ & C-stat/d.o.f.$^\star$ \T\B\\
Unit  & & & 10$^{22}$ cm$^{-2}$ & &  & & 10$^{-11}$ cgs & \T\B\\
\cline{1-9}
Value & 1$^\dagger$ & $0.93_{-0.03}^{+0.02}$ & $16.5_{-1.6}^{+1.9}$ & $1.8_{-0.3}^{+0.5}$ & $0.76_{-0.16}^{+0.18}$ & $5.9_{-3.7}^{+12.7}$ & $1.3_{-1.1}^{+0.1}$ & 123/89\T\B\\
\cline{1-9}
\cline{1-9}
\multicolumn{9}{c}{Spectral lines}\T\B\\
&\multicolumn{2}{c}{line complex}  & $E_{\rm rest}$ (keV) & $\sigma$ (eV) & norm ($10^{-5}$) & EW (eV) & $\Delta$ C-stat \T\B \\

\cline{1-9}
&\multicolumn{2}{c}{\Sxv{} He$\alpha$} & $2.45_{-0.01}^{+0.01}$ & 0$^\dagger$ & $63_{-12}^{+18}$ & $238_{-25}^{+52}$ & 261\T\B \\
&\multicolumn{2}{c}{Ar XVII He$\alpha$} & $3.11_{-0.02}^{+0.03}$ & 0$^\dagger$ & $11_{-2.1}^{+3}$ & $76_{-17}^{+33}$ & 73\T\B \\
&\multicolumn{2}{c}{\Caxix{} He$\alpha$} & $3.88_{-0.03}^{+0.02}$ & 0$^\dagger$ & $5.9_{-1.2}^{+1.3}$ & $76_{-14}^{+32}$ & 72\T\B \\
&\multicolumn{2}{c}{\Fei{} K$\alpha$} & $6.30_{-0.12}^{+0.13}$ & 0$^\dagger$ & $1.4_{-0.6}^{+0.6}$ & $41_{-17}^{+62}$ & 20\T\B \\
&\multicolumn{2}{c}{\Fexxv{} He$\alpha$} & $6.59_{-0.02}^{+0.02}$ & 0$^\dagger$ & $11_{-1}^{+1}$ & $549_{-54}^{+170}$ & 743 \T\B \\
&\multicolumn{2}{c}{\Fexxvi{} Ly$\alpha$} & $6.89_{-0.09}^{+0.07}$ & 0$^\dagger$ & $1.3_{-0.55}^{+0.8}$ & $45_{-25}^{+60}$ & 16\T\B \\

\cline{1-9}
\cline{1-9}
\end{tabular}\\
\raggedright
$\star$ computed from the full model, including the line components.
$\dagger$ frozen or at the limit of the parameter space. 
Unlike in Xtend, the energies of most lines are kept free to account for the systematics in \xmm{}'s energy calibration.
\end{table*}

\clearpage




\end{document}